%% file: paper.tex
\documentclass[12pt]{article}
\usepackage{jheppub}

\pdfoutput=1

\usepackage[table]{xcolor}
\usepackage{amsmath,bbm,array,amsfonts,graphicx,wrapfig,lscape,float,mathtools,multirow,longtable}
\usepackage{stackrel}
\usepackage[all]{xy}
\usepackage{diagrams}

\usepackage{physics} 
\usepackage{subcaption}
\usepackage{xcolor}
\usepackage{tikz}
\usepackage{cancel}

\usetikzlibrary{decorations.pathreplacing,decorations.markings,arrows.meta}
\definecolor{yellow2}{rgb}{0.98, 0.80, 0.20}
\definecolor{blue2}{RGB}{65,128,255}

\input{pref}

\newcommand{\tbp}[1]{\Omega^{#1}(#1)[#1]}
\newcommand{\Pn}[1]{\mathbb{P}^{#1}}
\newcommand{\E}{\mbox{Ext}}
\newcommand{\Sf}{\mathcal{E}}
\newcommand{\Ol}{\mathcal{O}}
\newcommand{\cc}{\Check{C}}
\newcommand{\om}[1]{\Omega^{#1}}
\newcommand{\shf}{\mathcal{E}}
\newcommand{\tm}[4]{\tiny\begin{pmatrix} #1 & #2 \nonumber \\ #3 & #4 \end{pmatrix}}

\newcommand{\Ext}{{\rm Ext}}
\newcommand{\Hom}{{\rm Hom}}

\newcommand{\bd}{{\bf d}}

\newcommand{\floor}[1]{ \left\lfloor{ #1} \right\rfloor}

\newcommand{\CC}{\mathcal{C}}
\newcommand{\CE}{\mathcal{E}}
\newcommand{\CV}{\mathcal{V}}
\newcommand{\CO}{\mathcal{O}}

\renewcommand{\t}{\widetilde }
\renewcommand{\d}{\partial }
\renewcommand{\b}{\overline}
\newcommand{\ov}{\over}

\newcommand{\beas}{\begin{equation} \begin{aligned}} \newcommand{\eeas}{\end{aligned} \end{equation}}

\newcommand{\Z}{\mathbb{Z}}
\newcommand{\C}{\mathbb{C}}
\newcommand{\R}{\mathbb{R}}
\newcommand{\N}{\mathbb{N}}

\definecolor{darkspringgreen}{rgb}{0.09, 0.45, 0.27}
\definecolor{forestgreen}{rgb}{0.13, 0.55, 0.13}

\usepackage{array}

\newcolumntype{C}[1]{>{\centering\let\newline\\\arraybackslash\hspace{0pt}}m{#1}}

\title{Graded quivers and B-branes \\ at Calabi-Yau singularities} 

\author[a]{Cyril Closset,} 
\author[b,c]{Sebasti\'an Franco,} 
\author[d]{Jirui Guo,} 
\author[b,c]{Azeem Hasan} 

\affiliation[a]{
 Mathematical Institute, University of Oxford \\
Woodstock Road, Oxford, OX2 6GG, United Kingdom }

\affiliation[b]{
Physics Department, The City College of the CUNY \\
160 Convent Avenue, New York, NY 10031, USA}

\affiliation[c]{The Graduate School and University Center, The City University of New York  \\
365 Fifth Avenue, New York NY 10016, USA}

\affiliation[d]{
Department of Physics and Center for Field Theory and Particle Physics \\
Fudan University, 220 Handan Road, 200433 Shanghai, China}

\emailAdd{cyril.closset@maths.ox.ac.uk}
\emailAdd{sfranco@ccny.cuny.edu}
\emailAdd{jrguo@fudan.edu.cn}
\emailAdd{ahasan@gradcenter.cuny.edu}

\abstract{A graded quiver with superpotential is a quiver whose arrows are assigned degrees $c\in \{0, 1, \cdots, m\}$, for some integer $m \geq 0$, with relations generated by a superpotential of degree $m-1$. Ordinary quivers ($m=1)$ often describe the open string sector of D-brane systems; in particular, they capture the physics of D3-branes at local Calabi-Yau (CY) 3-fold singularities in type IIB string theory, in the guise of 4d $\mathcal{N}=1$ supersymmetric quiver gauge theories. It was pointed out recently that graded quivers with $m=2$ and $m=3$ similarly describe systems of D-branes at CY 4-fold and 5-fold singularities, as 2d $\mathcal{N}=(0,2)$ and 0d $\mathcal{N}=1$ gauge theories, respectively. 
In this work, we further explore the correspondence between $m$-graded quivers with superpotential, $Q_{(m)}$, and CY $(m+2)$-fold singularities, ${\mathbf X}_{m+2}$. For any $m$, the open string sector of the topological B-model on ${\mathbf X}_{m+2}$ can be described in terms of a graded quiver.
We illustrate this correspondence explicitly with a few infinite families of toric singularities indexed by $m \in \mathbb{N}$, for which we derive ``toric'' graded quivers associated to the geometry, using several complementary perspectives. Many interesting aspects of supersymmetric quiver gauge theories can be formally extended to any $m$; for instance, for one family of singularities, dubbed $C(Y^{1,0}(\mathbb{P}^m))$, that generalizes the conifold singularity to $m>1$, we point out the existence of a formal ``duality cascade'' for the corresponding graded quivers.
}

\begin{document}

\maketitle

\section{Introduction}

The mathematical concept of a quiver---that is, a directed graph consisting of nodes and arrows between nodes---has proven very fruitful in string theory and in supersymmetric field theory, starting with the seminal work of Douglas and Moore \cite{Douglas:1996sw}.  Broadly speaking, ``ordinary'' quivers are often used to describe the structure of half-BPS states in theories with $8$ real supersymmetries. In particular, they can conveniently describe half-BPS systems of D-branes in type II string theory; schematically, the quiver nodes represent a set of mutually supersymmetric D-brane, and the arrows between nodes represent the supersymmetry-protected open string modes. 

A rich class of quivers arises from considering D3-branes probing Calabi-Yau (CY) 3-fold singularities in type IIB \cite{Morrison:1998cs,Beasley:1999uz,Feng:2000mi,Beasley:2001zp,Feng:2001xr,Feng:2001bn,Feng:2002zw,Wijnholt:2002qz,Benvenuti:2004dy,Franco:2005rj,Benvenuti:2005ja,Franco:2005sm,Butti:2005sw}. More generally, we may consider D$p$-branes transverse to CY $(m+2)$-fold singularities, with $p=5-2m$. That is, we consider a IIB background:
\be\label{IIB setup X}
\R^{6-2m} \times {\bf X}_{m+2}~,
\ee 
with ${\bf X}_{m+2}$ a local CY$_{m+2}$ singularity, and with D$(5-2m)$-branes along the transverse space, which sit at the singularity---from the point of view of ${\bf X}_{m+2}$, those branes are point-like probes. For $m=1$, the low-energy theory on the four-dimensional D3-brane worldvolume is described by a 4d $\CN=1$ supersymmetric gauge theory. More generally, if we consider $m=0, 1, 2, 3$, we obtain gauge theories in dimension $d=6,4,2,0$ with the following amounts of supersymmetry:
\be
\begin{tabular}{l|cccc}
$m$ & $0$ &$1$& $2$ & $3$  \\
 \hline
${\bf X}_{m+2}$    &CY$_2$ &CY$_3$ & CY$_4$& CY$_5$\\
SUSY & 6d  $\CN=(0,1)$ & 4d $\CN=1$ & 2d $\CN=(0,2)$ & 0d $\CN=1$ 
 \end{tabular}
\ee
The low-energy field theories have $2^{3-m}$ real supercharges.

\subsection{Graded quiver gauge theories} 

While a set of $N$ transverse D-branes at a smooth point of ${\bf X}_{m+2}$ would give rise to a $U(N)$ gauge theory on its worldvolume, the D-branes at the singularity ``fractionate'' into marginally-bound constituents, the so-called {\it fractional branes.} Each type of fractional brane supports its own gauge group. For our purpose, a {\it quiver gauge theory} is a gauge theory with a gauge group:
\be\label{gauge group Q}
 U(N_1) \times U(N_2) \times \cdots \times U(N_n)~.
\ee
We assign a gauge group $U(N_i)$ to each node $i$ of an abstract quiver; the $(6-2m)$-dimensional gauge fields $A_{\mu, i}$ sit in vector multiplets $\CV_i$ of the appropriate supersymmetry algebra. 
Open strings stretched between fractional branes give rise to matter fields in the quiver gauge theory, in adjoint or bifundamental representations of the unitary gauge groups in \eqref{gauge group Q}. For $m=0$, the matter fields sit in hypermultiplets of 6d $\CN=(0,1)$ supersymmetry, and the corresponding quiver arrows are unoriented; in this case, ${\bf X}_{2}$ is an ADE singularity, and the corresponding quivers are affine ADE quivers \cite{Douglas:1996sw}. For $m=1$, we have a 3-fold ${\bf X}_3$ and matter fields are in chiral multiplets of 4d $\CN=1$ supersymmetry, corresponding to oriented arrows of an ``ordinary'' quiver. For $m=2$ and $m=3$, the matter fields can sit in either chiral or fermi multiplets of 2d $\CN=(0,2)$ and 0d $\CN=1$ supersymmetry, respectively. For $m=2$, the chiral multiplets give rise to oriented arrows, while the fermi multiplets give rise to unoriented arrows. For $m=3$, both the chiral and fermi multiplets correspond to oriented arrows. 

The 2d and 0d gauge theories are conveniently described within the larger framework of {\it graded quivers (with superpotential)}. A graded quiver is a quiver together with a grading of the arrows by a ``quiver degree:''
\be
c \in \{ 0, 1, \cdots, m\}~.
\ee
The grading simply keeps track of the different types of matter fields.
We denote the various arrows, or ``fields,'' by:
\be\label{arrows fields}
\Phi_{ij}^{(c)} \; : i \longrightarrow j~, \qquad c=0, 1, \cdots, n_c-1~, \qquad n_c \equiv \floor{m+2\ov 2}~,
\ee
When $m$ is even, the arrows of maximal degree, $n_c-1={m\ov 2}$, are unoriented. All other arrows are oriented. For every arrow of the form \eqref{arrows fields}, we posit a ``conjugate'' arrow of degree $m-c$ and opposite orientation, denoted by:
 \be\label{Phi opp intro}
 \b\Phi_{ji}^{(m-c)}\equiv \overline{(\Phi_{ij}^{(c)})}~.
\ee
This is interpreted as the CPT conjugate fields in the supersymmetric gauge theory.

Importantly, the graded quivers can have a superpotential, which encodes interactions amongst matter fields in the gauge theory. We will come back to that important point later on. This perspective on supersymmetric quiver gauge theories was recently developed in \cite{Franco:2017lpa}. Related works include \cite{Closset:2017yte,Closset:2017xsc, Eager:2018oww}.

Gauge theory quivers have been most studied in the case of ${\bf X}_{m+2}$ a {\it toric} local CY (see e.g. \cite{Morrison:1998cs,Beasley:1999uz,Feng:2000mi,Beasley:2001zp,Feng:2001xr,Feng:2001bn,Feng:2002zw,Benvenuti:2004dy,Franco:2005rj,Benvenuti:2005ja,Franco:2005sm,Butti:2005sw,GarciaCompean:1998kh,Franco:2015tna,Franco:2015tya,Franco:2016nwv,Franco:2016fxm,Franco:2016tcm,Franco:2018qsc}). Various powerful tools become available in this case. We will review them in \sref{section_toric_rescue}.

As far as the D-brane setup \eqref{IIB setup X} goes, we are limited to $m \leq 3$ by the critical dimension, $d=10$, of type II string theory. 
From the perspective of graded quivers, however, there is no reason to stop at $m=3$.
While there is no supersymmetric field theory interpretation of general graded quivers,~\footnote{ Formally, a graded quiver with $m>3$ would correspond to a ``field theory'' in $d=6-2m<0$, with $n_c$ distinct types of matter fields, and with some ``superpotential'' interactions amongst them.} they still have a natural interpretation as describing fractional branes at a CY$_{m+2}$ singularity, as we now explain.

\subsection{From B-branes on ${\bf X}_{m+2}$ to graded quivers $Q_{(m)}$}  

By themselves, graded quivers with $m\leq 3$ do not encode the full low-energy quantum field theory on the transverse D-branes. Instead, they encode some half-BPS ``holomorphic'' information \cite{Seiberg:1994bp} which is protected by supersymmetry. In type IIB string theory, that information is preserved by the topological B-twist. 

Let us, then, focus on the B-model of the local Calabi-Yau ${\bf X}_{m+2}$.
Conveniently, this maps the problem of analyzing D-branes at a CY singularity to a purely algebraic problem, since the B-model is independent of the K{\"a}hler moduli of ${\bf X}_{m+2}$. 
The D-branes of the B-model, denoted by $\CE$, are called {\it B-branes}. They are described as objects in the bounded derived category of coherent sheaves (the B-brane category, for short) of the variety ${\bf X}_{m+2}$ \cite{Sharpe:1999qz, Douglas:2000gi, Katz:2002gh, Katz:2002jh}:
\be\label{def E in DX}
\CE \in {\bf D}^b ({\bf X}_{m+2})~.
\ee
For most purposes here, we can think of $\CE$ as a coherent sheaves with compact support. At this level of description, there is no restriction on $m$: the B-model is well-defined on any Calabi-Yau variety.

A point-like brane at a smooth point $p\in {\bf X}_{m+2}$ is described by the skyscraper sheaf $\CO_p$. When we bring $\CO_p$ to the singularity, it is expected to fractionate into marginally stable constituents:
\be
\CO_p \cong \CE_1 \oplus \cdots \oplus \CE_n~.
\ee
The B-branes $\CE_i$ are the {\it fractional branes.} They correspond to the nodes of a quiver. In the main text, we will discuss their identification in a few explicit examples, in the case of toric singularities that admit crepant resolutions.

The open strings between B-branes are described as morphism in the B-brane category. Algebraically, they are the $\Ext$ groups elements:
\be\label{x Ext element def}
\phi_{ij}^{(d)}  \in  \Ext^d_{{\bf X}_{m+2}}(\CE_j, \CE_i)~.
\ee
We review some of the necessary algebraic geometry in Appendix~\ref{App: alg geom}. Here, we just note that $\Ext$ groups are indexed by a degree:
\be
d\in \{0, 1, \cdots, m+2\}~.
\ee
The degree corresponds to the BRST charge in the B-model. 
On a Calabi-Yau $(m+2)$-fold, we have the isomorphism:
\be\label{Serre duality intro}
\Ext^d_{{\bf X}_{m+2}}(\CE_j, \CE_i) \cong \Ext^{m+2-d}_{{\bf X}_{m+2}}(\CE_i, \CE_j)~, \qquad d= 0, \cdots, m+2~,
\ee
known as Serre duality. The elements of $\Ext^0 \cong \Hom$ are identified with ``vector multiplets'' at the quiver nodes. By assumption, we must have:
\be
\Ext^0_{{\bf X}_{m+2}}(\CE_j, \CE_i)\cong \Ext^{m+2}_{{\bf X}_{m+2}}(\CE_i, \CE_j)\cong  \C  \delta_{ij}
\ee
for a consistent set of fractional branes.
The other $\Ext^{d}$ group elements \eqref{x Ext element def}, with degree $d\neq 0, m+2$, are identified with the ``matter field'' arrows in a {\it graded} quiver:
\be\label{Ext to arrows}
\phi_{ij}^{(d)} \qquad \longleftrightarrow\qquad \Phi_{ij}^{(d-1)}~.
\ee
Note that the quiver and $\Ext$ degrees are related by $c=d-1$. 

In this way, in principle, one can associate a graded quiver $Q_{(m)}$ to any local CY singularity, of any complex dimension:
\be\label{Q to X correspondence}
{\bf X}_{m+2} \qquad \longleftrightarrow \qquad Q_{(m)} ~.
\ee
The most non-trivial part of the correspondence is the identification of the ``interactions'' in either description. On the graded quiver side, there exists a quiver ``superpotential'' of degree $m-1$. On the B-brane side, this corresponds to the $A_\infty$ algebra satisfied by open string disk correlators. 

Based on the known results for $m=0, 1$ \cite{1999math......8027B, Bridgeland:2005fr}, one would expect that there exists an equivalence of derived categories between ${\bf D}^b ({\bf X}_{m+2})$ and some suitable derived category of representations of $Q_{(m)}$. This is indeed the case, as shown by Lam in \cite{lam2014calabi}.

In this paper, our goal is to flesh out the basic correspondence \eqref{Q to X correspondence} explicitly, at a ``physical'' level of rigor, in a few families of geometries $\{{\bf X}_{m+2}\}_{m\in \N}$. Given a singular CY variety ${\bf X}_{m+2}$, the procedure to obtain a graded quiver with superpotential $Q_{(m)}$ from the B-branes on ${\bf X}_{m+2}$ is as follows:
\begin{itemize}
\item[(i)] Find a consistent set of fractional branes, $\{\CE_i\}$. This gives the nodes of the quiver.
\item[(ii)]  Compute all the $\Ext$ groups \eqref{x Ext element def} between fractional branes. Using the correspondence \eqref{Ext to arrows}, draw the quiver arrows, with their quiver degrees.~\footnote{We only draw half of the arrows, as in \protect\eqref{arrows fields}. The other half of the arrows is given implicitly by the ``conjugation'' map \protect\eqref{Phi opp intro}.}
\item[(iii)]  Compute the quiver superpotential from the $A_\infty$ products between $\Ext$ group elements. (We will explain this last point in later sections.)
\end{itemize}
While the above procedure is very general and can be applied, in principle, to any singular Calabi-Yau variety, explicit computations in the B-brane category tend to be technically challenging. Moreover, the first step is problematic, since we do not have, in general, an efficient method to find a ``consistent set'' of fractional brane in the B-brane category. In fact, such sets are by no means uniquely determined by the variety ${\bf X}_{m+2}$. Different choices of fractional branes can lead to different quivers, which corresponds to ``field theory dualities'' (in particular, ``Seiberg dualities'') when $m \leq 3$. In general, we expect that any such distinct quivers for a given singularity are related by quiver mutations---see Appendix~\ref{app: mutations}  for a review of graded quiver mutations \cite{Franco:2017lpa}.

\subsection{Toric geometry to the rescue} \label{section_toric_rescue}

Fortunately, when ${\bf X}_{m+2}$ is a {\it toric} local Calabi-Yau, there exist alternative methods for associating a quiver to the singularity. We now review them briefly and point the interested reader to the references for detailed expositions.

A first approach, which is actually not restricted to toric geometries, consists of realizing ${\bf X}_{m+2}$ as a partial resolution of another geometry for which the quiver theory is easy to determine. A standard choice for such parent theory is an appropriate $\mathbb{C}^{m+2}/(\mathbb{Z}_{N_1} \times \cdots \times \mathbb{Z}_{N_{m+1}})$ orbifold. As we will elaborate in \sref{subsec: toric quivers and pms}, partial resolution translates into higgsing of the quiver. Applications of this strategy to $m=1$ and $m=2$ can be found in \cite{Morrison:1998cs,Beasley:1999uz,Feng:2000mi,Franco:2015tna}. While this method allows for a systematic derivation of the quiver theories for the desired geometries, it does not fully exploit all the structure associated to toric geometries.

The connection between toric CY$_{m+2}$'s and the corresponding quivers on D$(5-2m)$-branes, for $m=0,1,2,3$, was significantly simplified with the introduction of a class of brane configurations that are related to the original D-branes at singularities by T-duality along $m+1$ directions. For $m=1$, $2$ and $3$, these brane constructions are brane tilings \cite{Franco:2005rj,Franco:2005sm}, brane brick models \cite{Franco:2015tya,Franco:2016nwv,Franco:2016qxh} and brane hyperbrick models \cite{Franco:2016tcm}, respectively.~\footnote{The corresponding constructions for $m=0$ are the well-known elliptic models \protect\cite{Brunner:1997gf}.}  These configurations consist of stacks of D$(6-m)$-branes suspended within the voids of an NS5-brane that wraps a holomorphic hypersurface.~\footnote{For $m=3$, the suspended branes are actually Euclidean D4-branes.} This surface is $m$-complex dimensional and is defined as the vanishing locus of the Newton polynomial associated to the toric diagram, 
\beq
P(x_1,\cdots,x_{m+1})=0 ~,
\eeq
with  $x_i \in \mathbb{C}^*$, $i=1,\cdots, m+1$. Most of the non-trivial structure of these configurations lives on an $(m+1)$-torus, defined by the coamoeba projection of the $x_i$ coordinates. For many purposes, it is often sufficient to consider the ``skeletons" of these brane configurations. For brane tilings, these are bipartite graphs on $\mathbb{T}^2$; for brane brick models, they are tessellations of $\mathbb{T}^3$; and so on. In all these cases, there is a simple dictionary relating the brane setups to the corresponding quiver gauge theories.

These constructions can be formally extended to $m>3$ \cite{toappear1}.We collectively refer to them as {\it generalized dimers}. Via graph dualization, they are in one-to-one correspondence with periodic quivers on $\mathbb{T}^{m+1}$ which, likewise, fully encode both the quivers and the superpotentials of the ``field theories.''

As we will explain in \sref{subsec: toric quivers and pms}, given one of these brane setups, finding the corresponding ${\bf X}_{m+2}$ is reduced to a combinatorial problem, which is a huge simplification with respect to alternative approaches. Conversely, there are various efficient procedures for constructing generalized dimers---equivalently, quiver theories with superpotentials---starting from the corresponding toric ${\bf X}_{m+2}$. One way to do this is by using mirror symmetry. This method was developed for $m=1$ in \cite{Feng:2005gw} and for $m=2$ in \cite{Futaki:2014mpa,Franco:2016qxh}, where its extension to higher $m$ was also outlined.
 
In this paper, we focus on toric varieties. For each infinite family of examples, we present a convenient toric method to derive graded quivers with superpotential for ${\bf X}_{m+2}$, and discuss some of their interesting properties. We then proceed to check those results with an explicit B-brane computation, following the three steps above. The B-model computation provides a strong check of those recently devised toric methods. 

\vskip0.4cm
\noindent
This paper is organized as follows. In section~\ref{sec: 2}, we review the relevant aspects of graded quivers and of the B-brane category, and we spell out the relation between the two approaches. In section~\ref{sec: Cn}, we illustrate our methods in the simplest example, that of flat space $\C^{m+2}$. In section~\ref{section_family_orbifolds}, we consider an orbifold singularity, $\C^{m+2}/\Z_{m+2}$. In section~\ref{sec: Y10Pm}, we consider a family of singularities, dubbed $Y^{1,0}(\mathbb{P}^m)$, which reduces to the conifold singularity for $m=1$. In section~\ref{sec: F0m}, we consider a third family of singularities, dubbed $\mathbb{F}_0^{(m)}$, which reduces to an orbifold of the conifold for $m=1$. Appendix~\ref{App: alg geom} contains a pedagogical summary of the algebraic geometry techniques that we will need for our B-model computations. Appendix~\ref{app: mutations} reviews order $m + 1$ mutations of $m$-graded quivers.

\section{Graded quivers and B-branes}
\label{sec: 2}
In this section, we first review the concept of a graded quiver with superpotential, as developed in \cite{Franco:2017lpa}, building on mathematical ideas in \cite{MR3590528,Buan08}. We then discuss the relation between so-called ``toric'' quivers and toric singularities (while referring to \cite{toappear1} for further discussion).~\footnote{Throughout the paper, we will use the term toric quiver as a synonym of what is usually referred to as a {\it toric phase}. Toric phases are those that can be fully captured by periodic quivers on $\mathbb{T}^{m+1}$.} Finally, we discuss the derivation of the graded quiver from the B-model on the CY singularity.

\subsection{Graded quiver algebra}
A graded quiver $Q_{(m)}=(Q_0, Q_1)$ consist of a set of nodes indexed by some integers $i$, and of arrows $\Phi$ between nodes:
\be
Q_0 =\{i\}= \{1, \cdots, n\}~, \qquad \qquad  Q_1 =\{ \Phi\}~.
\ee
 Each arrow is assigned a quiver degree:
\be
c\in \{0, \cdots, m\}~,
\ee
for some integer $m \in \N$.
We denote an arrow from $i$ to $j$, of degree $c$, by:
\be
\Phi_{ij}^{(c)} \; : i \rightarrow j~.
\ee
The product of arrows is given by concatenation:
\be
\Phi_{ij} \Phi_{jk} \Phi_{kl} \cdots
\ee
Here the arrow degrees are left implicit.  A closed path is a product of arrows that comes back to itself, in the obvious way. The degree of a path is the sum of the degrees of its component arrows.
We call the degree-zero arrows the ``chiral fields,'' since they correspond to chiral multiplets in supersymmetric quiver gauge theories (when $m \leq 3$). A path of chiral fields has degree zero.

The {\it path algebra} is the algebra of paths generated by arrows, with the above product and the obvious formal sum. The freely-generated path algebra is denoted by $\C Q$. We will soon introduce relations amongst paths.

\paragraph{CPT invariance.} 
We restrict ourselves to a particular kind of graded quiver, such that every arrow $\Phi$ of degree $d$ has an ``opposite'' or ``conjugate,'' $\Phi_{\rm op}\equiv \b \Phi$, of degree $m-d$ and opposite orientation, as anticipated in \eqref{Phi opp intro}. We can then pair all the arrows according to:
\be\label{double arrows gen}
\left(\Phi^{(c)}_{ij}~, \,  \b\Phi^{(m-c)}_{ji}\right)~,\qquad  \b\Phi^{(m-c)}_{ji} \equiv \overline{( \Phi^{(c)}_{ij})}~.
\ee
This is a choice of {\it polarization} of the path algebra. A very convenient choice of polarization, which we use when drawing quivers explicitly, is to choose $\Phi^{(c)}$ for the arrows of degrees $c=0, \cdots, n_c-1$, with:
\be\label{def nc}
n_c = \floor{m+2\ov 2}~,
\ee
and $\b\Phi^{(m-c)}$ for their conjugate. In that case, one draws quivers with arrows of degrees $0$ to $n_c-1$ only. The number \eqref{def nc} is the number of ``arrow types'' in the graded quiver, also called the ``arrow colors'' \cite{Buan08}.

We may call the arrows of degree $c\in \{0, \cdots, m\}$ the ``matter fields.'' The requirement that every arrow has a conjugate corresponds to CPT invariance in quiver gauge theories.~\footnote{Conjugate arrows will always be implicit in the quiver diagrams that we will present. They are not independent objects, but can be derived from the corresponding unconjugated ones.} Note that, when $m$ is even, the arrows of degree $n_c-1={m\ov 2}$ are ``self-conjugate,'' and the choice of polarization into arrows $\Phi$ and $\b\Phi$, namely:
\be
\left(\Phi^{({m\ov 2})}_{ij}~, \, \b\Phi^{({m\ov 2})}_{ji}\right)~, 
\ee
 is arbitrary.  For $m=0$ and $m=2$, this corresponds to the fact that the 6d hypermultiplets and the 2d fermi multiplets, respectively, are self-conjugate.

\paragraph{Gauge fields.} 
Let us also introduce arrows from a node to itself:
\be
e_i : \, i \rightarrow i~, \qquad\qquad \bar e_i : \, i \rightarrow i~,
\ee
for each node, of degree $-1$ and $m+1$, respectively.~\footnote{The arrow $e_i$ is denoted by $l_i$ in \cite{Franco:2017lpa}, and its ``opposite" $\b e_i$ is introduced here for future convenience.} We may call $e_i$ and $\b e_i$ the ``gauge fields"---they are identified with vector multiplets in quiver gauge theories.

\paragraph{Superpotential relations.} 
We introduce relations on the path algebra through a ``graded quiver superpotential:''
\be
W= W(\Phi)~,\qquad\qquad  {\rm deg}(W)= m-1~.
\ee
This imposes relations on the path algebra, of the form $\d_\Phi W=0$.
The superpotential is a linear function of closed paths of matter fields, of degree $m-1$. It is clear from the grading that, for any fixed $m$, there can only be a finite number of arrows of degree $c>0$ in each closed path. On the other hand, the number of chiral multiplets $\Phi^{(0)}$ is unbounded,  {\it a priori}. For instance, at low $m$ we have:
\beas
&m=1\; :\quad  &&W= W(\Phi^{(0)})~, \cr 
&m=2\; :\quad  &&W= \Phi^{(1)}J(\Phi^{(0)})+ \b\Phi^{(1)} E(\Phi^{(0)})~, \cr 
&m=3\; :\quad  &&W= \Phi^{(1)}\Phi^{(1)} H(\Phi^{(0)})+ \Phi^{(2)} F(\Phi^{(0)})~, 
\eeas
schematically. The functions $W(\Phi^{(0)})$, $J(\Phi^{(0)}), E(\Phi^{(0)})$ and $H(\Phi^{(0)}), F(\Phi^{(0)})$ are holomorphic functions of the chiral fields. They correspond to the 4d $\CN=1$, 2d $\CN=(0,2)$, and 0d $\CN=1$ superpotentials, respectively. This obviously generalizes to any $m$:
\be
W= \Phi^{(c_1)} \cdots \Phi^{(c_k)} F_{c_1, \cdots, c_k}(\Phi^{(0)})~, \qquad c_1+ \cdots+ c_k= m-1~,
\ee
schematically,~\footnote{In general, we can have distinct paths of degree-zero chiral fields connecting each field of higher degree in the closed loop.} though there is no supersymmetric field theory interpretation for $m{>}3$.

\paragraph{Kontsevitch bracket condition.} 
There is an important condition we should impose on $W$, which can be written as:
 \be \label{WWzero}
\{ W, W\}=0~, \qquad \Leftrightarrow \qquad \sum_\Phi {\d W\ov \d \Phi}  {\d W\ov \d \b \Phi}=0~,
\ee
where the sum is over all the fields $\Phi$, for a given polarization \eqref{double arrows gen}. 
Here, $\{f, g\}$ denotes the Kontsevitch bracket on the path algebra. It is defined as:
\be
\{ f, g \}= \sum_\Phi \left( {\d f\ov\d \Phi}{\d g\ov \d\b\Phi} +(-1)^{(|f|+1)|\b\Phi|+(|g|+1)|\Phi|+ |\Phi||\b\Phi|+1} {\d f\ov\d \b\Phi}{\d g\ov\d \Phi} \right)~.
\ee
Let us note that the condition \eqref{WWzero} holds for any choice of polarization.
The Kontsevitch bracket is a natural generalization of the Poisson bracket on a graded path algebra that admits a polarization. 

\paragraph{Differential and superpotential.}
Given the superpotential above, one can define a differential, $\bd$,
of degree $-1$, acting on paths. We have the Leibniz rule:
\be
\bd (f g) = (\bd f)  g + (-1)^{|f||g|} f  \bd g~,
\ee
with $|f|$ denoting the degree of the path $f$.
The differential is given explicitly on the quiver fields by:
\beas\label{bd explicit on X}
&\bd e = - e \otimes e~, \cr
& \bd \Phi = {\d  W\ov \d \b  \Phi} + (-1)^{|\Phi|}  \Phi \otimes e- e\otimes  \Phi~, \cr
& \bd \b  \Phi = {\d  W\ov \d   \Phi} + (-1)^{|\b  \Phi|} \b  \Phi \otimes e- e\otimes \b  \Phi~, \cr
& \bd \b e=\sum_ \Phi (-1)^{|\b  \Phi|} \left(\b  \Phi \otimes  \Phi -  \Phi \otimes \b  \Phi  \right)+  (-1)^{m+1} \b e\otimes e - e\otimes \b e~.
\eeas
This is obviously of degree $-1$ since $W$ has degree $m-1$ and $|\b  \Phi|= m- | \Phi|$.
One can check that this is a differential:
\be\label{d2=0 rel}
\bd^2 =0~,
\ee
provided that \eqref{WWzero} is satisfied.

\paragraph{Representations of the quiver algebra and anomaly-free constraint.} 
Given a quiver algebra, we may want to study its representations. Recall that a quiver representation consists of a vector space $V_i \cong \C^{N_i}$ assigned to each node $i$, and of explicit homomorphisms $\Phi^{(0)}_{ij} \; : V_i \rightarrow V_j$ (that is, fixed $N_i \times N_j$ matrices such that all the quiver relations are satisfied).

In physics, the positive integers $N_i$ are the ranks of the unitary gauge group \eqref{gauge group Q} in a quiver gauge theory. The choice of homomorphism $\Phi^{(0)}$ is a choice of ``vacuum expectation values (VEVs)'' for the chiral multiplets. 
Not every choice of rank is physically acceptable. 
There are certain constraints on the allowed choices of ranks, the {\it generalized anomaly cancellation} conditions \cite{Franco:2017lpa}, which we will review in section \sref{subsec: anomaly free} below. 

It is always a good idea to distinguish between the algebra and its representations. In this work, most of our discussion will be focused on the general ``abstract'' quiver, not on a particular representation. In the B-model, a particular quiver representation corresponds to a particular bound state of D-branes, and the anomaly cancellation condition is a tadpole cancellation condition for the RR flux (at least in the physical setup with $m \leq 3$).

\subsection{Toric graded quivers and toric singularities}\label{subsec: toric quivers and pms}
A central theme of this paper is the connection between $m$-graded quivers and CY$_{m+2}$ singularities. This connection goes in both directions and can be addressed from multiple viewpoints.

The CY$_{m+2}$ variety arises from the quiver as its {\it classical moduli space}. Generalizing the $m\leq 3$ cases, for which the quivers have a gauge theory interpretation, we define the classical moduli space as the center of the Jacobian algebra with respect to fields of degree $m-1$, {\it i.e.} of next to maximal degree. The mathematical results in \cite{Ladkani} imply that it is sufficient to consider the algebra obtained by quotienting only by the corresponding relations:
\beq
{\partial W \over \partial \Phi^{(m-1)}}=0~, \ \ \ \ \ \forall \, \Phi^{(m-1)}~.
\label{relations}
\eeq
Note that, in the special case $m=2$, the field $\Phi^{(1)}$ here denotes both $\Phi^{(1)}$ and $\b\Phi^{(1)}$; they are the fermi and anti-fermi multiplets, in the 2d $\CN=(0,2)$ gauge theory.

Since the superpotential has degree $m-1$, the terms which are relevant for the relations in \eref{relations} are gauge invariants of the generic form $\Phi^{(m-1)} P(\Phi^{(0)})$, with $P(\Phi^{(0)}$ a holomorphic function of chiral fields. Borrowing the nomenclature used in the $m=2$ and $3$ cases, we refer to these terms as $J$-terms.~\footnote{Strictly speaking, $J$-term usually refers to the holomorphic function $P(\Phi^{(0)})$. We will use the name for the entire $\Phi^{(m-1)} P(\Phi^{(0)})$ term in the superpotential. For $m=1$, this corresponds to standard superpotential terms.} Therefore, the relations \eref{relations} consist entirely of chiral fields. For $m\leq 3$, chiral fields are the only superfields with scalar components, hence their relevance for the moduli space. Focusing on the center of the algebra corresponds to considering closed loops---in the gauge theory language, this is the restriction to gauge invariant fields.

\paragraph{Toric CY singularities.}  
In this paper, we focus on toric Calabi-Yau singularities, and their toric partial resolutions. A toric CY singularity ${\bf X}_{m+2}$ can be described in terms of its {\it toric diagram} $\Gamma$, a convex polytope in $\Z^{m+1}$. Let us denote the points of the toric diagram by:
\be
\{ v_1~, \cdots, v_d\} \in \Gamma \subset \Z^{m+1}~.
\ee
This includes internal points---points inside the polytope. Including all the internal points allows us to discuss toric resolutions straightforwardly. Recall that, given the toric diagram, the {\it toric fan} is the set of vectors $w_i = (v_i, 1) \in \Z^{m+2}$. The K{\"a}hler quotient description of the singularity (also known as GLSM \cite{Witten:1993yc}) is given by:
\be\label{X toric KQ}
{\bf X}_{m+2} \cong \C^d// U(1)^{d-m-2}~, \qquad z_i \sim e^{i \sum_a \alpha_a Q^a_i} z_i~, \qquad (Q^a)= {\rm ker}(w_1, \cdots, w_d)~,
\ee
with $(w_1, \cdots, w_d)$ seen as $(m+2)\times d$ matrix---here, $(z_i)\in \C^d$, $i=1, \cdots, d$, are the ``GLSM fields,'' and $a=1, \cdots, d-m-2$ runs over the ``GLSM gauge group.''

\paragraph{Toric superpotential condition.} 
To any given toric CY$_{m+2}$ singularity, we can associate a graded quiver $Q_{(m)}$ that satisfies an additional {\it toric condition}, generalizing the well-known $m=1$ and $m=2$ cases \cite{Feng:2002zw,Franco:2015tna}. More precisely, there always exists at least one such ``toric quiver,'' and other quivers are expected to be related to it by mutations. The toric condition is a condition on the superpotential:  every field $\Phi^{(m-1)}$ of degree $m-1$ should appear in exactly two $J$-terms, with opposite signs. Namely,
\beq
W= \Phi^{(m-1)} P(\Phi^{(0)})- \Phi^{(m-1)} Q(\Phi^{(0)}) + \ldots~,
\eeq
where the dots indicate terms that do not contain $\Phi^{(m-1)}$. In other words, the ``vacuum equations'' \eqref{relations} take a simple form (path$_1$)=(path$_2$). This form of the superpotential underlies the relationship between these theories and toric geometries. 

Concepts such as periodic quivers on $\mathbb{T}^{m+1}$ (and their dual brane tilings, brane brick models, and higher dimensional generalizations), perfect matchings, etc., can be generalized to arbitrary $m$. These issues will be studied in detail in a forthcoming paper \cite{toappear1}. Here, let us just quote one of the results, which we will exploit for computing moduli spaces.

Given a toric graded quiver $Q_{(m)}$ with superpotential $W$, we can define {\it perfect matchings} for arbitrary $m$, as follows. A perfect matching $p$ is a collection of arrows in $Q_{(m)}$ satisfying two conditions:
\begin{itemize}
\item $p$ contains precisely one arrow from each term in $W$. 
\item For every arrow $\Phi^{(c)}$ in $Q_{(m)}$, either $\Phi^{(c)}$ or its conjugate $\b\Phi^{(m-c)}$ is in $p$. 
\end{itemize}
This generalizes the definition of perfect matchings for brane tilings \cite{Franco:2005rj} and of brick matchings for brane brick models \cite{Franco:2015tya}.

We can regard perfect matchings as variables in terms of which the fields in the quiver can be expressed. In particular, the map between perfect matching variables and chiral fields is given by:
\beq
\Phi^{(0)}_i=\prod_{\mu} p_\mu^{P_{i\mu}} \ \ \ \ \ \quad \mbox{with }\quad P_{i\mu}=\left\{ \begin{array}{c} 1 \mbox{ if } \Phi^{(0)}_i \in p_\mu~, \\[.15mm]
0 \mbox{ if } \Phi^{(0)}_i \notin p_\mu~,
\end{array}  \right.
\eeq
where $i$ runs over the chiral fields and $\mu$ runs over perfect matchings. The $P_{i\mu}$ can be regarded as entries in the so-called $P$-matrix. This change of variables is extremely powerful, since it trivializes the relations \eref{relations}. There is then a one-to-one correspondence between perfect matchings and ``GLSM fields'' in a (possibly redundant~\footnote{If the GLSM description is redundant, there are several perfect matchings for the same point in the toric diagram.}) toric description \eqref{X toric KQ} of the CY$_{m+2}$. Perfect matchings are therefore mapped to points in the toric diagram. The $\mathbb{Z}^{m+1}$ coordinates for each perfect matching are easily determined from the intersections between the chiral fields it contains and the fundamental cycles of the $(m+1)$-torus on which the corresponding periodic quiver lives.

In this way, the determination of the moduli space is significantly simplified, reducing to the combinatorial problem of determining perfect matchings. Moreover, efficient methods for finding perfect matchings, analogous to the Kasteleyn matrices for brane tilings, exist for all $m$ \cite{toappear1}.

\paragraph{Partial resolution and higgsing.}
Partial resolution of a toric CY$_{m+2}$ corresponds to removal of points in the toric diagram, and can be used to connect different geometries. At the level of the quiver theory, this process maps to ``higgsing" by non-zero ``VEVs" for certain chiral fields, where we have extended the physical nomenclature used for low $m$ in the obvious way.

The map between chiral and GLSM fields, encoded by the $P$-matrix, provides a systematic procedure for identifying the chiral fields that acquire non-zero VEVs in order to achieve a desired partial resolution. In general, given a partial resolution, the choice of VEVs that realize it might not be unique. This procedure is a straightforward generalization of the one for CY$_3$ and CY$_4$ cases. We refer the reader to \cite{Morrison:1998cs,Beasley:1999uz,Feng:2000mi,Franco:2015tna} and references therein for in depth discussions of these cases. Later in the paper, we will investigate the connection between infinite families of geometries and the associated quiver theories via partial resolution.

\subsection{B-branes, Ext groups and $A_\infty$ algebra}

Let us now consider the B-model on a local CY$_{m+2}$ singularity ${\bf X}_{m+2}$. The {\it B-branes} are objects in the derived category of coherent sheaves on ${\bf X}_{m+2}$, as in \eqref{def E in DX}. 
In all the examples that we consider, there will exist a crepant resolution of the singularity:
\be
\pi \; : \; \t{\bf X}_{m+2} \rightarrow {\bf X}_{m+2}~,
\ee
with $\t{\bf X}_{m+2}$ a {\it smooth} local Calabi-Yau. Then, all the B-branes of interest will be coherent sheaves with compact support on complex submanifolds of $\t{\bf X}_{m+2}$. Intuitively, we simply have D-branes wrapping all possible closed complex cycles. 

Since the B-model is independent of K{\"a}hler deformations, the B-brane category on $\t{\bf X}_{m+2}$ must be equivalent to the B-brane category on the singularity ${\bf X}_{m+2}$, but the former is generally much simpler to describe.  In all our examples, the smooth resolution is the total space of a vector bundle $E$:
\be\label{res t X as bundle over B}
\t{\bf X}_{m+2}  \cong {\rm Tot}\left(E \rightarrow B_{m+2-r} \right)~, \qquad\qquad  r= {\rm rank}(E)~,
\ee
over $B_{m+2-r}$, a compact K{\"a}hler surface of complex dimension $m+2-r$; in the simplest case, we have the canonical line bundle over $B_{m+1}$. Then, the B-branes on  $\t{\bf X}_{m+2}$ can be described more simply in terms of sheaves on $B_{m+2-r}$.

The ``fractional branes,'' denoted by: 
\be
\{\CE_i\}_{i=1}^n~,
\ee
are distinguished B-branes which ``generate'' the derived category $D^b(\t{\bf X}_{m+2} )$, in some physical sense.~\footnote{Here we are being voluntarily vague. A better definition of fractional branes can be given if we are provided with a stability structure on $D^b(\t{\bf X}_{m+2})$, which does depend on the K{\"a}hler moduli (in physics, that is the central charge of the D-branes). The fractional branes are obtained by marginal decay of the point-like brane $\CO_p$ at the singularity.}
 In the setup \eqref{res t X as bundle over B}, a good set of fractional brane can be obtained from any strongly exceptional collection of sheaves on $B_{m+2-r}$ \cite{Cachazo:2001sg, Herzog:2003zc, Aspinwall:2004vm, Hanany:2006nm, Herzog:2006bu, Closset:2017yte}.
The open string states between two B-branes $\CE$ and $\CF$ are identified with the generators of the $\Ext$ groups \cite{Sharpe:1999qz, Douglas:2000gi, Katz:2002gh, Katz:2002jh}:
\be
\Ext_{\t{\bf X}_{m+2} }^d(\CE, \CF)~, \qquad d=0, \cdots, m+2~.
\ee
The interactions amongst these open string modes are encoded in a $A_\infty$ algebra. Let us define the graded vector space:
\be
A\cong  \oplus_{i, j} \oplus_{d=0}^{m+2} \Ext_{\t{\bf X}_{m+2} }^{d}(\CE_j, \CE_i)~,
\ee
of all the $\Ext$ groups elements amongst the fractional branes. One can define the multi-products $m_k$ on the $\Ext$ algebra $A$:
\be
m_k \; : \; A^{\otimes k} \rightarrow A~,
\ee
of degree $2-k$.
They satisfy the $A_\infty$ relations \cite{Herbst:2004jp}:
\be\label{Ainfinty rel}
\sum_{p+q+r=k} (-1)^{r + pq} m_{k+1-p}({\bf 1}^{\otimes r}\otimes m_p \otimes {\bf 1}^{\otimes q})=0~, \qquad \forall k>0~,
\ee
Note that, in particular, $m_1$ is a differential---that is, $(m_1)^2=0$, and $m_2$ is an associative product. The $\Ext$ algebra $A$ is a {\it minimal} $A_\infty$ algebra, meaning that $m_1=0$ identically.
There also exists a natural trace map:
\be
\gamma: A\rightarrow\C~,
\ee
of degree $-m-2$. This is used, in particular, to map to top $\Ext$ elements of degree $m+2$ to elements of $\Ext^0 \cong \Hom$.

The multi-products $m_k$ on the $\Ext$ algebra can be computed in the following manner \cite{Aspinwall:2004bs, Closset:2017yte}.
Given any $A_{\infty}$ algebra $\t A$, let us denote by $H^{\bullet}(\t A)$ to be
the cohomology of $m_1$.  If $\t A$ has no multiplications beyond $m_2$,
it turns out that one can define an $A_{\infty}$ structure
on $H^{\bullet}(\t A)$ in such a way that there exists an $A_{\infty}$ map \cite{kad, Aspinwall:2004bs}:
\be
f: H^{\bullet}(\t A) \rightarrow \t A~,
\ee 
with $f_1$ equal to a particular representation $H^{\bullet}(\t A) \hookrightarrow \t A$, in which cohomology classes map to (noncanonical) representatives in $\t A$, and such that $m_1=0$ in the
$A_{\infty}$ algebra on $H^{\bullet}(\t A)$. One can then use the 
consistency conditions satisfied by elements of an $A_{\infty}$ map
to solve algebraically for the higher products on $H^{\bullet}(\t A)$.

In the B-brane description, the algebra $\t A$ is the algebra of complexes of coherent sheaves, with chain maps between complexes. In that construction,  $m_1$ is identified with the BRST charge of the B-model. The ``physical'' open string states then live in the cohomology $H^\bullet(\t A)$, which gives us the derived category ${\bf D}^b({\bf X})$---see \cite{Aspinwall:2004jr} for a thorough review. The minimal $A_\infty$ algebra:
\be
A\equiv H^\bullet(\t A)
\ee
is precisely the $\Ext$ algebra.
In the examples discussed in this paper, each B-brane will correspond to a single coherent sheaf, which can be represented in the derived category by a locally-free resolution. The $\Ext$ elements can then be represented by chain maps between resolutions, modulo chain homotopies. The $m_2$ products in $A$ are given by chain map composition. The higher products can be computed by the procedure that we just outlined. 

In Appendix~\ref{App: alg geom}, we explain more thoroughly how to perform these computations explicitly.

\subsection{From $\Ext$ groups to quiver fields}
\label{section_ext_to_quiver_fields}

The relation between the quiver algebra and the $\Ext$ algebra was explained by Aspinwall and Katz in \cite{Aspinwall:2004bs}, in the physical context of D3-branes  at CY 3-folds ($m=1$). The general case is discussed by Lam \cite{lam2014calabi}, in a purely mathematical context. 

Here, we follow the physical argument of \cite{Aspinwall:2004bs}. In that language, the quiver fields $\Phi$ are {\it sources} for the open string vertex operators in the B-model. 
 Given the open string mode $\phi \in A$ of degree $|\phi|$, there is a one-form descendent $\phi^{(1)}$ of degree $|\phi|-1$. 
Then, to every $\phi \in A$, one can associate a ``spacetime field'' $\t \Phi$ of degree $|\t \Phi|= 1-|\phi|$, which acts as a source for $\phi$ in the B-model: 
\be
S \rightarrow S + \sum_\phi \t \Phi \,  \phi^{(1)}~.
\ee
Due to our choice of notation for the graded quivers $Q_{(m)}$, following  \cite{Franco:2017lpa}, we find it convenient to define the ``quiver field'' $\Phi$ of degree $|\Phi|=- |\t\Phi|$, so that:
\be
|\Phi|= |\phi|-1~.
\ee
The explains the relation between quiver fields and $\Ext$ elements given in  \eqref{Ext to arrows} in the introduction.~\footnote{As we just explained, a more natural definition of the quiver degree would be {\it minus} the degree that we use in this paper. This is the conventions used, for instance, in \protect\cite{lam2014calabi}. (Also in \protect\cite{Closset:2017yte}.) In our present conventions, the quiver degree is equal to minus the BRST degree of the B-model.}

Algebraically, the graded quiver algebra, $V$, and the $\Ext$ algebra, $A$, are related as follows \cite{Aspinwall:2004bs}.
Let $V$ denote the path algebra modulo the quiver relations, and let $\t V$ denote the same vector space but with the degrees $c$ exchanged with $-c$. (That is, $\Phi \in V$ and $\t \Phi \in \t V$. 
  Let also $\t V[1]$ denote the vector space $\t V$ with all degrees decreased by one, and let $s:  \t V\rightarrow \t V[1]$ denote the corresponding map of degree $-1$. Then,  $A$ is simply the {\it dual} of $\t V[1]$:
  \be
A= \big(\t V[1]\big)^\ast~.
\ee
Then, it turns out that the $A_\infty$ relations \eqref{Ainfinty rel} on $A$ are equivalent to the existence of the differential $\bd$, \eqref{bd explicit on X}, on $V$ \cite{Aspinwall:2004bs, lam2014calabi}.

\paragraph{Mapping nodes and arrows.} 
As anticipated in the introduction, we can assign a graded quiver $Q_{(m)}$ to a CY singularity. More precisely, we work with a particular crepant resolution $\t {\bf X}_{m+2}$. We should also insist on the fact that the quiver is really associated to {\it a particular set of fractional branes.} A different choice of fractional branes can lead to a different quiver.

Let us now spell out the B-brane-to-quiver correspondence. First of all, of course, the quiver nodes are in one-to-one correspondence with the fractional branes:
\be
\text{node}\; i\qquad \longleftrightarrow \qquad \CE_i
\ee
In the case of a singularity that admits a crepant resolution as in \eqref{res t X as bundle over B}, the number of fractional branes (and thus, the number of nodes in the quiver) is equal to $\chi(B_{m+2-r})$, the Euler character of the K{\"a}hler base $B_{m+2-r}$---physically, this is because we should have a basis of wrapped branes that generates the full even-homology lattice. 

Secondly, all the quiver arrows $\Phi$ of degree $|\Phi|=c$ correspond to $\Ext$-group elements $x$ of degree $|\phi|=c+1$:
\be\label{Ext to arrows full}
\phi_{ij}^{(d)} \in  \Ext_{\t{\bf X}_{m+2} }^{d}(\CE_j, \CE_i)  \qquad \longleftrightarrow\qquad \Phi_{ij}^{(c)}~, \quad {\rm with}\quad c= d-1 \in \{0, 1, \cdots, m\}~.
\ee
Of course, Serre duality \eqref{Serre duality intro} corresponds to the pairing \eqref{double arrows gen} of quiver arrows.
Note that we identify the arrow $\Phi_{ij}$ with the $\Ext$ element $\phi_{ij}$.~\footnote{Note that $\phi_{ij}^{(d)}$ correspond to a morphism from $\CE_j$ to $\CE_i$. While the product of arrows is by concatenation, the product of two $\Ext$ elements correspond to the composition of maps. In our conventions, we then have the convenient relations: $$
\Phi_{ij}\Phi_{jk}\qquad \longleftrightarrow \qquad  m_2(\phi_{ij}, \phi_{jk}) \equiv \phi_{ij}  \circ \phi_{jk}~.$$}
The quiver algebra elements of quiver degrees $-1$ and $m+1$ correspond to $e$ and $\b e$, respectively. The fact that each element is a loop attached to a single node is a property that we assume of any ``allowed fractional branes,'' namely:
\be
\Ext^0(\CE_i, \CE_j)=\Ext^{m+2}(\CE_j, \CE_i) = \delta_{ij} \C~.
\ee
These groups are identified with the ``vector multiplets'' in supersymmetric quiver gauge theories.

\paragraph{The quiver superpotential.} 
The graded quiver superpotential takes the general form:
\be\label{W from Bmod 1}
W= \sum_{\text{closed paths} \; p} \alpha_p \, \Phi_{i_1 i_2}^{(c_1)}  \Phi_{i_2 i_3}^{(c_2)} \cdots  \Phi_{i_s i_1}^{(c_1)}~,
\ee
The sum is over all closed paths, 
\be
p = \Phi_{i_1 i_2}^{(c_1)}  \Phi_{i_2 i_3}^{(c_2)} \cdots  \Phi_{i_s i_1}^{(c_s)} \qquad {\rm with} \quad \sum_{l=1}^s  c_l =m-1~,
\ee
 which consists of $s$ concatenated arrows of any degrees $c_l\in \{0, \cdots, m\}$, subject to the above constraint---that is, here $\Phi$ denotes both the fields $\Phi$ and their ``conjugates'' $\b\Phi$.~\footnote{Notice that while the sum in \eref{W from Bmod 1} is formally over all closed paths of degree $m-1$, not all of them are necessarily in the superpotential since the corresponding coefficients $\alpha_p$ may vanish.}
The superpotential couplings are given by open string disk correlators:
\be
\alpha_p = \left\langle \phi_{i_1 i_2}^{(c_1+1)}  \phi_{i_2 i_3}^{(c_2+1)}\cdots  \phi_{i_s i_1}^{(c_s+1)} \right\rangle~.
\ee
More explicitly, they are given in terms of the multi-products on $A$, according to:
\be\label{W from Bmod 2}
\alpha_p = \gamma\Big(m_2\big( \phi_{i_1 i_2}^{(c_1+1)}~,\,  m_{s-1}( \phi_{i_2 i_3}^{(c_2+1)}~,\, \cdots~,\,  \phi_{i_s i_1}^{(c_s+1)})\big)\Big)~.
\ee
Note that $\alpha_p$ has degree $0$, by construction.

\subsection{Anomaly-free conditions on the quiver ranks}\label{subsec: anomaly free}
To conclude this section, let us state the anomaly-free condition, alluded to above, in full generality  \cite{Franco:2017lpa}. Consider a graded quiver $Q_{(m)}$ (not necessarily toric), with an assignment of ranks $N_i \in \N$ to the nodes $i \in Q_0$.
Let us denote by $\CN(\Phi_{ij}^{(c)})$ the number of arrows from $i$ to $j$ of degree $c$. Then, the generalized anomaly-free conditions for $m$ odd are:
\be\label{af cond i}
\sum_j N_j \sum_{c=0}^{n_c-1} (-1)^c \left(\CN(\Phi_{ji}^{(c)})-\CN(\Phi_{ij}^{(c)})\right)=0~, \qquad \forall i~, \qquad {\rm if}\;\; m \in 2\Z+1~.
\ee
Here, for each fixed $i$, the sum over $j$ is over all nodes in the quiver (including $i$), and $n_c$ was defined in \eqref{def nc}.
For $m$ even, instead, we have the conditions:
\be\label{af cond ii}
\sum_j N_j \sum_{c=0}^{n_c-1}(-1)^c \left(\CN(\Phi_{ji}^{(c)})+\CN(\Phi_{ij}^{(c)})\right)=2N_i~, \qquad \forall i~, \qquad {\rm if}\;\; m \in 2\Z~.
\ee
For $m=0,1,2,3$, these conditions coincide with the cancellation of non-abelian anomalies for the corresponding $d=6,4,2,0$ gauge theories with gauge group $\prod_i U(N_i)$. 

Using the correspondence between quiver arrows and $\Ext$ group generators, the anomaly-free conditions have a simple expression in the B-brane language. Namely, for a configuration of $N_i$ fractional branes of each type $\CE_i$, we should impose \cite{Closset:2017yte}:
\be\label{tadpole gen}
\sum_j  N_j \sum_{d=0}^{m+2} (-1)^d {\rm dim}\, \Ext^d_{\t{\bf X}_{m+2}}(\CE_i, \CE_j)=0~, \quad \forall i~.
\ee
This is interpreted as a ``generalized tadpole cancellation condition'' for a given set of fractional branes.

In the special case of toric quivers, we always have the ``regular branes'' with rank assignment $N_i=N$, $\forall i$. In that case, a factor of $N$ factorizes out of the anomaly-free condition, and \eqref{tadpole gen} becomes a statement about the set of fractional branes. All the examples that we will consider below satisfy those conditions with $N_i=N$.

\section{Flat space: the $\C^{m+2}$ graded quiver}\label{sec: Cn}
\begin{figure}[t]
	\centering
	\includegraphics[width=12cm]{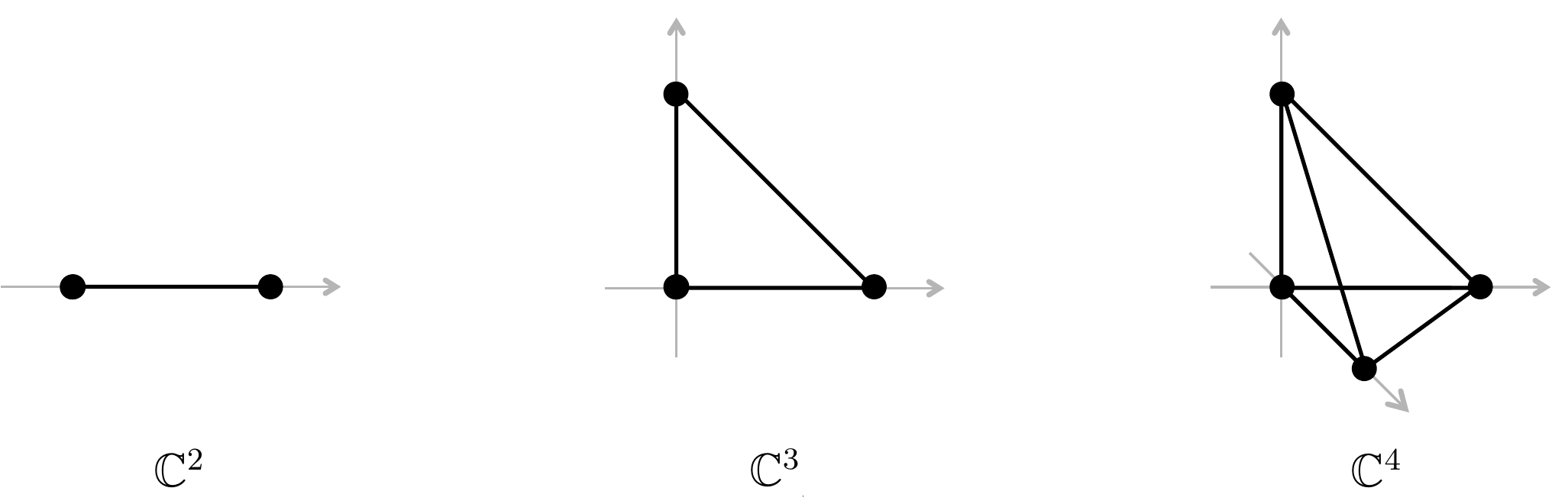}
\caption{Toric diagrams for $\mathbb{C}^{m+2}$ with $m=0,1,2$.}
	\label{toric_diagrams_Cn}
\end{figure}

The simplest local Calabi-Yau $(m+2)$-fold is flat space, $\C^{m+2}$. Its toric diagram is the minimal simplex in $\Z^{m+1}$, namely:
\beas
&v_0=(0,\ldots,0)~, \cr
&v_1= (1,0,0,\ldots,0)~, \quad
 v_2=(0,1,0,\ldots,0)~, \quad \ldots~, \quad
 v_{m+1}=(0,0,\ldots,0, 1)~.
 \eeas
 The toric diagrams for $m\leq 3$ are shown in \fref{toric_diagrams_Cn}.
As a warm up exercise, we consider the graded quiver associated to $\C^{m+2}$. We first derive it using the {\it algebraic dimensional reduction} procedure introduced in \cite{Franco:2017lpa}. We then verify this result by a direct B-brane computation.

\subsection{Algebraic dimensional reduction}

Let us quickly review algebraic dimension reduction. This corresponds to replacing the underlying CY singularity ${\bf X}_{m+2}$ by a product space of the form:
\beq
{\bf X}_{m+2} \quad\to\quad {\bf X}_{m+3} = {\bf X}_{m+2}\times \mathbb{C} ~ .
\label{dimensional_reduction}
\eeq
The effect on the corresponding graded quiver,
\be
Q_{(m)}  \quad\to\quad Q_{(m+1)}~,
\ee
 is a generalization of the $\mathbb{T}^2$ dimensional reduction of supersymmetric gauge theories. 
The quiver diagram transforms as follows:
\beq 
\begin{array}{ccc}
m & & m+1 \\ \hline \\[-.4 cm]
\mbox{node}_i & \ \ \ \ \to \ \ \ \ &  \mbox{node}_i + \mbox{adjoint chiral } \Psi^{(0)}_{ii} \\[.12 cm]
\Phi^{(c)}_{ij} & \ \ \ \ \to \ \ \ \ & \Psi^{(c)}_{ij} + \tilde{\Psi}^{(c+1)}_{ij} 
\end{array}
\label{dim_red_quiver}
\eeq
where $0\leq c\leq \floor{m\ov 2}$. This table also applies when $i=j$, namely when the theory we start with contains adjoint fields. 
It is interesting to consider more carefully what \eref{dim_red_quiver} implies for the undirected fields of degree ${m\ov 2}$ that can be present in theories with even $m$:
\beq
\begin{array}{ccc}
\mbox{even } m & & m+1 \\ \hline \\[-.3 cm]
\Phi^{({m\ov 2})}_{ij} & \ \ \ \ \to \ \ \ \ & \Psi^{({m\ov 2})}_{ij} +\tilde{\Psi}^{({m\ov 2}+1)}_{ij} = \Psi^{({m\ov 2})}_{ij}+\tilde{\Psi}^{({m\ov 2})}_{ji} 
\end{array}
\eeq
Thus, for each conjugate pair of arrows of degree ${m\ov 2}$ in $Q_{(m)}$, we get two pairs of arrows of degree ${m\ov 2}$ in $Q_{(m+1)}$. (For instance, for $m=0$, one 6d hypermultiplet gives rise to one 4d hypermultiplet, which is equivalent to two chiral multiplet arrows of opposite orientations.)
 
Let $W_{(m)}$ denote the original superpotential of $Q_{(m)}$,  and let $W_{(m+1)}$ be the one for the dimensionally reduced quiver $Q_{(m+1)}$. There are two types of contributions to $W_{m+1}$:
 \begin{itemize}
 \item[{\bf 1)}] {\bf Dimensional reduction of terms in $W_m$.} 
 Schematically, for any term in $W_m$ we have a series of terms in $W_{m+1}$ of the form:
\beq 
\begin{array}{ccc}
m & & m+1 \\ \hline \\[-.3 cm]
\Phi^{(c_1)}_{i_1 i_2} \Phi^{(c_2)}_{i_2 i_3} \ldots \Phi^{(c_k)}_{i_k i_1} & \ \  \to \ \  & 
\begin{array}{cccl}
& \tilde{\Psi}^{(c_1+1)}_{i_1 i_2} \Psi^{(c_2)}_{i_2 i_3} \ldots \Psi^{(c_k)}_{i_k i_1}& + & \Psi^{(c_1)}_{i_1 i_2} \tilde{\Psi}^{(c_2+1)}_{i_2 i_3} \ldots \Psi^{(c_k)}_{i_k i_1} \\[.12 cm] 
+ & \ldots & + & \Psi^{(c_1)}_{i_1 i_2} \Psi^{(c_2)}_{i_2 i_3} \ldots \tilde{\Psi}^{(c_k+1)}_{i_k i_1}.
\end{array}
\end{array}
\label{dim_red_potential}
\eeq
 \item[{\bf 2)}]  {\bf New terms involving adjoints.} 
In addition, $W_{m+1}$ contains a new class of terms. For every arrow $\Phi^{(c)}_{ij}$ in the original quiver, we introduce the following pair of superpotential terms in the dimensionally reduced one:
 \be
\Psi^{(0)}_{ii} \Psi^{(c)}_{ij} \tilde{\Psi}^{(m-c-1)}_{ji} - \tilde{\Psi}^{(m-c-1)}_{ji} \Psi^{(c)}_{ij} \Psi^{(0)}_{jj} ~ .
\label{dim_red_potential_adjoints}
\eeq

\end{itemize}
These rules fully determine the ``dimensionally reduced'' quiver with superpotential, $Q_{(m+1)}$.

\subsection{The graded quivers}\label{subsec: Cn quiver}
Using dimensional reduction, we can construct the field content and superpotential for $\mathbb{C}^{m+2}$ starting from $\mathbb{C}^2$, which has a single node with a single unoriented arrow from the node to itself and no superpotential.

   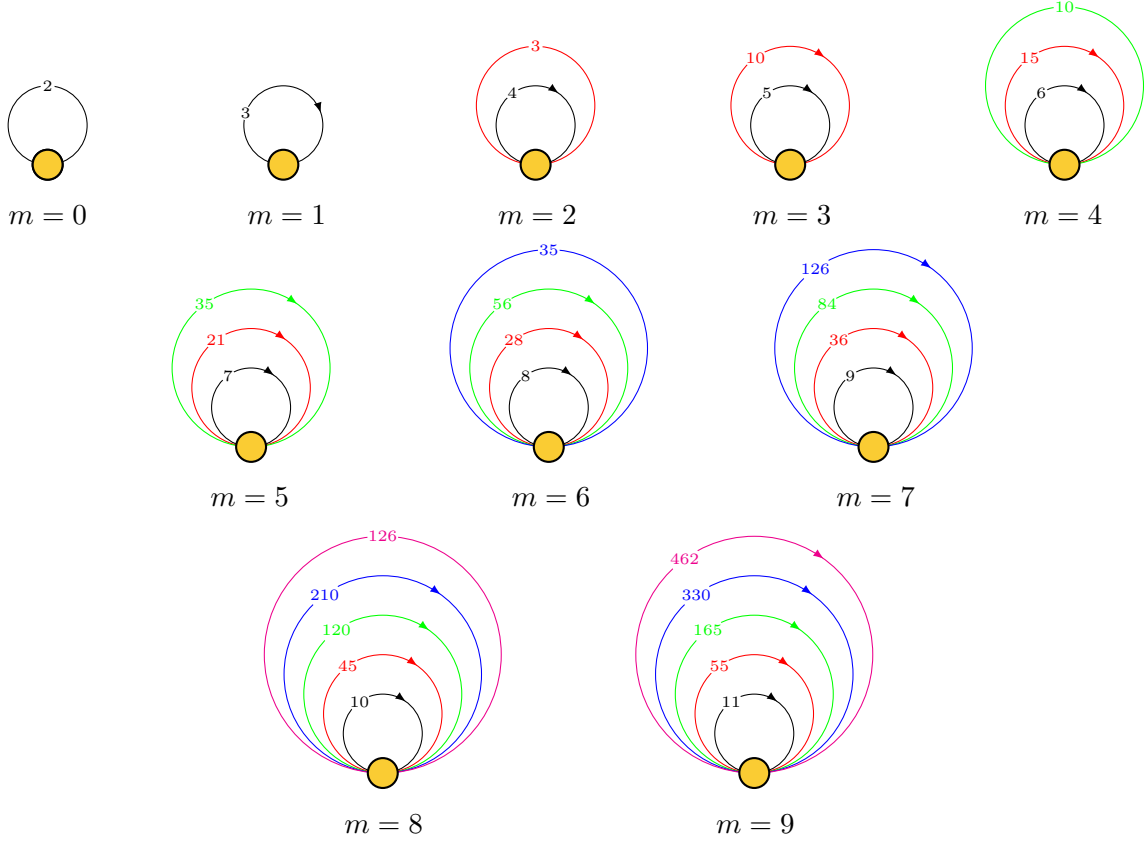
\begin{figure}
\captionsetup[subfigure]{labelformat=empty}
        \centering
        \begin{subfigure}[b]{0.11\textwidth}
             \newcommand{\pos}{0.5}
             \newcommand{\arrowHeadPosition}{0.7} 
            \begin{tikzpicture}[scale=1.75 , decoration={markings,mark=at position \arrowHeadPosition with {\arrow{latex}}}]
                \tikzstyle{every node}=[circle,thick,fill=yellow2,draw,inner sep=4pt,font=\tiny]
                \draw (0,0) node (A) {};
                    \draw[black] (0,0) arc(270:-90:0.3) node[pos = \pos, draw = none , fill = white , inner sep = 0]{$2$};
                \draw (0,0) node (A) {};
            \end{tikzpicture} 
            \subcaption{$m = 0$ \ \ \ \ \ }        
        \end{subfigure}
        \hspace{0.07\textwidth}
        \begin{subfigure}[b]{0.115\textwidth}
            \newcommand{\pos}{0.3}
            \newcommand{\arrowHeadPosition}{0.7} 
            \begin{tikzpicture}[scale=1.75 , decoration={markings,mark=at position \arrowHeadPosition with {\arrow{latex}}}]
                \tikzstyle{every node}=[circle,thick,fill=yellow2,draw,inner sep=4pt,font=\tiny]
                    \draw[postaction={decorate}, black] (0,0) arc(270:-90:0.3) node[pos = \pos, draw = none , fill = white , inner sep = 0]{$3$};
                \draw (0,0) node (A) {};
            \end{tikzpicture} 
             \subcaption{$m = 1$ \ \ \ \ }
        \end{subfigure}
        \hspace{0.07\textwidth}
        \begin{subfigure}[b]{0.13\textwidth}
            \newcommand{\pos}{0.4}
            \newcommand{\arrowHeadPosition}{0.6} 
            \begin{tikzpicture}[scale=1.75 , decoration={markings,mark=at position \arrowHeadPosition with {\arrow{latex}}}] 
                \tikzstyle{every node}=[circle,thick,fill=yellow2,draw,inner sep=4pt,font=\tiny]
                \draw[postaction={decorate}, black] (0,0) arc(270:-90:0.3) node[pos = \pos, draw = none , fill = white , inner sep = 0]{$4$};
                \draw[red] (0,0) arc(270:-90:0.44999999999999996) node[pos = 0.5, draw = none , fill = white , inner sep = 0]{$3$};
                \draw (0,0) node (A) {};
            \end{tikzpicture} 
             \subcaption{$m = 2$ \ \ \ }
        \end{subfigure}
        \hspace{0.07\textwidth}
        \begin{subfigure}[b]{0.13\textwidth}
            \newcommand{\pos}{0.4}
            \newcommand{\arrowHeadPosition}{0.6} 
            \begin{tikzpicture}[scale=1.75 , decoration={markings,mark=at position \arrowHeadPosition with {\arrow{latex}}}] 
                \tikzstyle{every node}=[circle,thick,fill=yellow2,draw,inner sep=4pt,font=\tiny]
                \draw[postaction={decorate}, black] (0,0) arc(270:-90:0.3) node[pos = \pos, draw = none , fill = white , inner sep = 0]{$5$};
                \draw[postaction={decorate}, red] (0,0) arc(270:-90:0.44999999999999996) node[pos = \pos, draw = none , fill = white , inner sep = 0]{$10$};
                \draw (0,0) node (A) {};
            \end{tikzpicture} 
             \subcaption{$m = 3$ \ \ \ }
        \end{subfigure}
        \hspace{0.07\textwidth}
        \begin{subfigure}[b]{0.15\textwidth}
            \newcommand{\pos}{0.4}
            \newcommand{\arrowHeadPosition}{0.6} 
            \begin{tikzpicture}[scale=1.75 , decoration={markings,mark=at position \arrowHeadPosition with {\arrow{latex}}}]
                \tikzstyle{every node}=[circle,thick,fill=yellow2,draw,inner sep=4pt,font=\tiny]
                \draw[postaction={decorate}, black] (0,0) arc(270:-90:0.3) node[pos = \pos, draw = none , fill = white , inner sep = 0]{$6$};
                \draw[postaction={decorate}, red] (0,0) arc(270:-90:0.44999999999999996) node[pos = \pos, draw = none , fill = white , inner sep = 0]{$15$};
                \draw[green] (0,0) arc(270:-90:0.6) node[pos = 0.5, draw = none , fill = white , inner sep = 0]{$10$};
                \draw (0,0) node (A) {};
            \end{tikzpicture} 
             \subcaption{$m = 4$ \ \ }
        \end{subfigure}
        \begin{subfigure}[b]{0.15\textwidth}
            \newcommand{\pos}{0.4}
            \newcommand{\arrowHeadPosition}{0.6} 
            \begin{tikzpicture}[scale=1.75 , decoration={markings,mark=at position \arrowHeadPosition with {\arrow{latex}}}] 
                \tikzstyle{every node}=[circle,thick,fill=yellow2,draw,inner sep=4pt,font=\tiny]
                \draw[postaction={decorate}, black] (0,0) arc(270:-90:0.3) node[pos = \pos, draw = none , fill = white , inner sep = 0]{$7$};
                \draw[postaction={decorate}, red] (0,0) arc(270:-90:0.44999999999999996) node[pos = \pos, draw = none , fill = white , inner sep = 0]{$21$};
                \draw[postaction={decorate}, green] (0,0) arc(270:-90:0.6) node[pos = \pos, draw = none , fill = white , inner sep = 0]{$35$};
                \draw (0,0) node (A) {};
            \end{tikzpicture} 
             \subcaption{$m = 5$ \ \ }
        \end{subfigure}
        \hspace{0.07\textwidth}
        \begin{subfigure}[b]{0.19\textwidth}
            \newcommand{\pos}{0.4}
            \newcommand{\arrowHeadPosition}{0.6} 
            \begin{tikzpicture}[scale=1.75 , decoration={markings,mark=at position \arrowHeadPosition with {\arrow{latex}}}] 
                \tikzstyle{every node}=[circle,thick,fill=yellow2,draw,inner sep=4pt,font=\tiny]
                \draw[postaction={decorate}, black] (0,0) arc(270:-90:0.3) node[pos = \pos, draw = none , fill = white , inner sep = 0]{$8$};
                \draw[postaction={decorate}, red] (0,0) arc(270:-90:0.44999999999999996) node[pos = \pos, draw = none , fill = white , inner sep = 0]{$28$};
                \draw[postaction={decorate}, green] (0,0) arc(270:-90:0.6) node[pos = \pos, draw = none , fill = white , inner sep = 0]{$56$};
                \draw[blue] (0,0) arc(270:-90:0.75) node[pos = 0.5, draw = none , fill = white , inner sep = 0]{$35$};
                \draw (0,0) node (A) {};
            \end{tikzpicture} 
             \subcaption{$m = 6$ \ \ }
        \end{subfigure}
        \hspace{0.07\textwidth}
        \begin{subfigure}[b]{0.19\textwidth}
            \newcommand{\pos}{0.4}
            \newcommand{\arrowHeadPosition}{0.6} 
            \begin{tikzpicture}[scale=1.75 , decoration={markings,mark=at position \arrowHeadPosition with {\arrow{latex}}}] 
                \tikzstyle{every node}=[circle,thick,fill=yellow2,draw,inner sep=4pt,font=\tiny]
                \draw[postaction={decorate}, black] (0,0) arc(270:-90:0.3) node[pos = \pos, draw = none , fill = white , inner sep = 0]{$9$};
                \draw[postaction={decorate}, red] (0,0) arc(270:-90:0.44999999999999996) node[pos = \pos, draw = none , fill = white , inner sep = 0]{$36$};
                \draw[postaction={decorate}, green] (0,0) arc(270:-90:0.6) node[pos = \pos, draw = none , fill = white , inner sep = 0]{$84$};
                \draw[postaction={decorate}, blue] (0,0) arc(270:-90:0.75) node[pos = \pos, draw = none , fill = white , inner sep = 0]{$126$};
                \draw (0,0) node (A) {};
            \end{tikzpicture} 
             \subcaption{$m = 7$ \ \ }
        \end{subfigure}
        \hspace{0.07\textwidth}
        \begin{subfigure}[b]{0.23\textwidth}
            \newcommand{\pos}{0.4}
            \newcommand{\arrowHeadPosition}{0.6} 
            \begin{tikzpicture}[scale=1.75 , decoration={markings,mark=at position \arrowHeadPosition with {\arrow{latex}}}] 
                \tikzstyle{every node}=[circle,thick,fill=yellow2,draw,inner sep=4pt,font=\tiny]
                \draw[postaction={decorate}, black] (0,0) arc(270:-90:0.3) node[pos = \pos, draw = none , fill = white , inner sep = 0]{$10$};
                \draw[postaction={decorate}, red] (0,0) arc(270:-90:0.44999999999999996) node[pos = \pos, draw = none , fill = white , inner sep = 0]{$45$};
                \draw[postaction={decorate}, green] (0,0) arc(270:-90:0.6) node[pos = \pos, draw = none , fill = white , inner sep = 0]{$120$};
                \draw[postaction={decorate}, blue] (0,0) arc(270:-90:0.75) node[pos = \pos, draw = none , fill = white , inner sep = 0]{$210$};
                \draw[magenta] (0,0) arc(270:-90:0.8999999999999999) node[pos = 0.5, draw = none , fill = white , inner sep = 0]{$126$};
                \draw (0,0) node (A) {};
            \end{tikzpicture} 
             \subcaption{$m = 8$ \ \ \ }
        \end{subfigure}
        \hspace{0.07\textwidth}
        \begin{subfigure}[b]{0.23\textwidth}
            \newcommand{\pos}{0.4}
            \newcommand{\arrowHeadPosition}{0.6} 
            \begin{tikzpicture}[scale=1.75 , decoration={markings,mark=at position \arrowHeadPosition with {\arrow{latex}}}] 
                \tikzstyle{every node}=[circle,thick,fill=yellow2,draw,inner sep=4pt,font=\tiny]
                \draw[postaction={decorate}, black] (0,0) arc(270:-90:0.3) node[pos = \pos, draw = none , fill = white , inner sep = 0]{$11$};
                \draw[postaction={decorate}, red] (0,0) arc(270:-90:0.44999999999999996) node[pos = \pos, draw = none , fill = white , inner sep = 0]{$55$};
                \draw[postaction={decorate}, green] (0,0) arc(270:-90:0.6) node[pos = \pos, draw = none , fill = white , inner sep = 0]{$165$};
                \draw[postaction={decorate}, blue] (0,0) arc(270:-90:0.75) node[pos = \pos, draw = none , fill = white , inner sep = 0]{$330$};
                \draw[postaction={decorate}, magenta] (0,0) arc(270:-90:0.8999999999999999) node[pos = \pos, draw = none , fill = white , inner sep = 0]{$462$};
                \draw (0,0) node (A) {};
            \end{tikzpicture} 
             \subcaption{$m = 9$ \ \ \ }
        \end{subfigure}
        \caption{Quivers for $\mathbb{C}^{m+2}$. The quivers for $m=0, 1, 2, 3$ correspond to maximally supersymmetric Yang-Mills theory in $d=6,4,2,0$. The multiplicities of fields, i.e.  the dimensions of the representations for the $SU(m+2)$ global symmetry, are indicated on the arrows. For $m$ even, the multiplicity of the outmost (unoriented) line is half the dimension of the corresponding representation. 
        Black, red, green, blue and purple arrows represent fields of degree 0, 1, 2, 3 and 4, respectively.
        \label{quivers_Cn}}
    \end{figure}

\paragraph{Quiver.} 
For every $m$, the quiver is given as follows:
\begin{itemize}
\item It consists of a single node.
\item In addition, there are adjoint fields $\Phi^{(c,c+1)}$ of degree $0\leq c \leq \floor{m\ov 2}$. Here we have introduced a superindex notation in which $\Phi^{(c;k)}$ indicates an arrow with degree $c$ and transforming in the $k$ index totally antisymmetric representation of the global $SU(m+2)$ symmetry. This notation might seem excessive for these simple theories, but it will turn useful for some of the computations and more general geometries to be discussed later. Each field $\Phi^{(c,c+1)}$ thus transforms in the antisymmetric $(c+1)$-index representation of $SU(m+2)$.
\item For even $m$, the multiplicity of the unoriented degree-${m\ov 2}$ fields is half the dimension of the corresponding representation. We can regard the full representation as built out of both $\Phi^{({m\ov 2})}$ and $\bar{\Phi}^{({m\ov 2})}$, which have the same degree.
\end{itemize}
\fref{quivers_Cn} shows these quivers up to $m=9$.

\paragraph{Superpotential.} 
Following dimensional reduction, all $W$ terms are cubic. The superpotential terms are given by cubic terms of degree $m-1$ combined into $SU(m+2)$ invariants. In order to write the superpotential for general $m$, we introduce a convention in which the products of fields include the contraction $SU(m+2)$ indices and are explicitly given by 
        \begin{align}
            (A_{1}^{(c_{1},k_1)}\cdots A_{n}^{(c_{n},k_n)})^{\alpha_{k+1}\cdots\alpha_{m+2}} \equiv \frac{1}{\prod_{i}k_{i}!}\epsilon^{\alpha_{1}\cdots\alpha_{m+2}}A_{1;\alpha_{1}\cdots \alpha_{k_{1}}}^{(c_{1},k_1)}\cdots A_{n;\alpha_{k-k_{n}+1}\cdots \alpha_{k}}^{(c_{n},k_n)} ~ ,
            \label{product_contraction_indices 0}
        \end{align}
where $k=\sum_i k_i$ is the total number of $SU(m+2)$ indices before contractions. Any such term with $\sum k_{i} = m+2$ is manifestly $SU(m+2)$ invariant. The superpotential can then be compactly written as
        \begin{align}
            W = \sum_{i+j+k = m+2}\Phi^{(j-1;j)}\Phi^{(k-1;k)}\bar{\Phi}^{(m+1-j-k;m+2-j-k)} ~ .
            \label{potential_Cn}
        \end{align}  
Since we sum over terms such that $i+j+k = m+2$, the degrees of the fields in the superpotential terms are given by partitions (including 0) of $(m-1)$ into three integers.

\subsection{B-model computation}
 
We can also understand the $\C^{m+2}$ quiver in terms of B-branes, as in \cite{Closset:2017yte}. There is a single ``fractional brane'' in flat space, the skyscraper sheaf over a point $p$, $\CO_p$. Without loss of generality, we take $p$ to be the origin of $\C^{m+2}$. The Koszul resolution at point $p$ is:
    \begin{align}
        \begin{diagram}[small]
                0 & \rTo & \Omega^{m+2}  & \rTo^{f} & \Omega^{m+1} & \rTo^{f} &  \cdots &  \rTo^{f} & \Omega^{0} & \rTo^{r} & \mathcal{O}_{p} & \rTo & 0~,
        \end{diagram} \label{skyscraper_Koszul_resolution}
    \end{align}
where $\Omega$ is the cotangent bundle of flat space, and $r$ is the restriction map at the origin. Lastly, $f: \Omega^{k} \to \Omega^{k-1}$ is the vector field:
    \begin{align}
        f = \sum_{\mu} z_{\mu}\pdv{}{z_{\mu}}~,
    \end{align}
    acting by interior derivative, with $z_{\mu}$ the holomorphic coordinates of flat space.

\subsubsection{Quiver fields}
    
        The quiver fields can be computed as the chain maps between two copies of this resolution. The generators $\phi^{\mu}$ of the $\Ext^1(\CO_p, \CO_p)$ group, corresponding to chirals, are elements of $\cc^{0}(Hom^{1}(\CO_{p},\CO_{p}))$. There are $m+2$ of them, transforming in the fundamental representation of $SU(m+2)$. $\phi^{\mu}$ is explicitly given by the chain map
        \begin{diagram}[small]
             &&\Omega^{m+2}  & \rTo & \Omega^{m+1}  & \rTo& \cdots &  \rTo & \Omega^{1}  & \rTo & \Omega^{0}   \\
             &&\dTo^{\pdv{}{z_{\mu}}} &        &  \dTo^{\pdv{}{z_{\mu}}} &  &       &          & \dTo^{\pdv{}{z_{\mu}}}           \\
            \Omega^{m+2}  & \rTo & \Omega^{m+1}  & \rTo & \Omega^{m} & \rTo &  \cdots &  \rTo & \Omega^{0}  
        \end{diagram} 
        The vector field $\pdv{}{z_{\mu}}$ again acts by interior derivative.

        The generator of the other $\Ext$ groups are given by the antisymmetric composition of these basic elements. There are $\binom{m+2}{k}$ generators of $\E^{k}(\CO_{p},\CO_{p})$, given explicitly by:
        \begin{align}
            \phi^{\mu_{1}\cdots\mu_{k}} = \frac{1}{k!}\phi^{\mu_{1}}\circ \phi^{\mu_{2}}\circ \cdots \circ \phi^{\mu_{k}}~. \label{sky_gen_def}  
        \end{align} 
        If we allow $0 \le k \le m+2$, this contains both the generators $\phi$ and their Serre dual $\b\phi$. To mimic the notation that is natural for the more complicated example of later sections, we will write $\phi^{\mu_{1}\cdots\mu_{k}}$ for $k \le \frac{m+2}{2}$ and $\bar{\phi}^{\mu_{1}\cdots\mu_{k}}$ for $k \ge \frac{m+2}{2}$, including the arbitrary choice of some pairing:
        \be 
        \big(\phi^{\mu_1, \cdots, \mu_{m+2\ov2}}, \b\phi^{\mu_1, \cdots, \mu_{m+2\ov2}}\big)~,
        \ee
         when $m$ is even. In that case, the number of arrows $\phi^{({m+2\ov2})}$ is half the dimensions of the $\frac{m+2}{2}$-index representation, since the full representation is spanned by these arrows and their Serre dual arrows.
The Serre dual of ${\phi}^{\mu_{1}\cdots\mu_{k}}$ is the generator $\bar{\phi}^{\mu_{k+1}\cdots\mu_{m+2}}$, which satisfies:
        \begin{align}
             \phi^{\mu_{1}\cdots\mu_{k}}\circ \bar{\phi}^{\mu_{k+1}\cdots\mu_{m+2}} = \bar{\phi}^{\mu_{1}\cdots \mu_{m+2}}~. \label{sky_dual_def}
        \end{align}

\subsubsection{Superpotential}

            The superpotential can be computed straightforwardly. Since we defined higher $\Ext$ generators as compositions of $\E^{1}$ generators, composing them gives:
            \begin{align}
                m_{2}(\phi^{\mu_{1}\cdots\mu_{j}}, \phi^{\mu_{j+1}\cdots\mu_{k}}) &= \phi^{\mu_{1}\cdots \mu_{k}}~,  \nonumber\\
                m_{2}(\phi^{\mu_{1}\cdots\mu_{j}}, \bar{\phi}^{\mu_{j+1}\cdots\mu_{k}}) &= \bar{\phi}^{\mu_{1}\cdots \mu_{k}}~, \nonumber \\
                m_{2}(\bar{\phi}^{\mu_{1}\cdots\mu_{j}}, \bar{\phi}^{\mu_{j+1}\cdots\mu_{k}}) &= \bar{\phi}^{\mu_{1}\cdots \mu_{k}}~. \nonumber
            \end{align} 
The definition \eqref{sky_gen_def} is valid both for the \v{C}ech cohomology classes as well as for their explicit representatives, therefore all $f_{2}$ are trivially zero. Hence all higher products vanish.

Thus, all the superpotential terms present are the cubic terms we postulated before. We can compute the coefficients straightforwardly using \eqref{sky_dual_def}. They are
            \begin{align*}
                \gamma(m_{2}(m_{2}(\phi^{\mu_{1}\cdots\mu_{j}}, \phi^{\mu_{j+1}\cdots\mu_{k}}), \bar{\phi}^{\mu_{k+1}\cdots\mu_{m+2}})) = \epsilon_{\mu_{1}\cdots\mu_{m+2}}~,
            \end{align*} 
in agreement with \eqref{potential_Cn}.

\section{The $\mathbb{C}^{m+2}/\mathbb{Z}_{m+2}$ orbifolds}
\label{section_family_orbifolds}

As a first family of non-trivial CY singularities, let us consider the orbifolds $\C^{m+2}/\Z_{m+2}$, with the cyclic group acting on flat space as:
\be
z_i \sim e^{2\pi i \ov m+2} z_i~, \qquad i=1, \cdots, m+2~,\quad \qquad (z_i) \in \C^{m+2}~.
\ee
This singularity can be resolved to a local $\mathbb{P}^{m+1}$. We thus have:
\be
{\bf X}_{m+2} \cong \C^{m+2}/\Z_{m+2}~,\quad \qquad  \t{\bf X}_{m+2} \cong {\rm Tot}\big(\CO(-m-2) \rightarrow \mathbb{P}^{m+1}\big)~.
\ee
Let us first derive the quiver by toric methods. We will then discuss B-branes on the resolution $ \t{\bf X}_{m+2}$.

\subsection{The toric geometries}

\begin{figure}[t]
	\centering
	\includegraphics[width=14cm]{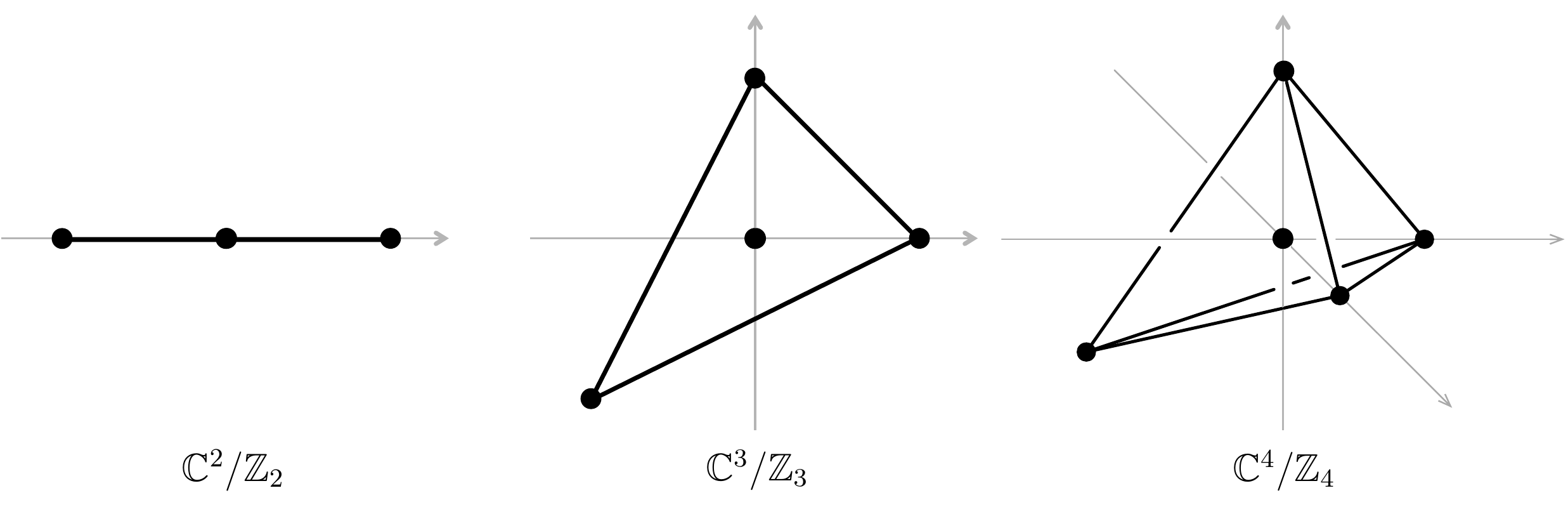}
\caption{Toric diagrams for $\mathbb{C}^{m+2}/\mathbb{Z}_{m+2}$ with $m=0,1,2$.}
	\label{toric_diagrams_Cn_Zn}
\end{figure}

The $(m+1)$-dimensional toric diagrams for these geometries contain the following $m+3$ points:
\beq
v_0=(0,\ldots,0)~, 
\qquad 
\begin{array}{l}
v_1= (1,0,0,\ldots,0)~, \\
 v_2=(0,1,0,\ldots,0)~, \\
\quad \vdots \\
 v_{m+1}=(0,0,\ldots,0, 1)~, 
\end{array} 
\qquad 
v_{m+2}=(-1,-1, \ldots, -1)~.
 \label{eq_toric_Cn_Zn}
 \eeq 
The toric diagrams for the first few values of $m$ are shown in \fref{toric_diagrams_Cn_Zn}.

\subsection{The graded quivers}\label{section_quivers_Cn_Zn}

The quivers and superpotentials can be determined by standard orbifolding~\cite{Douglas:1996sw} of the $\C^{m+2}$ quivers discussed above.

\begin{figure}
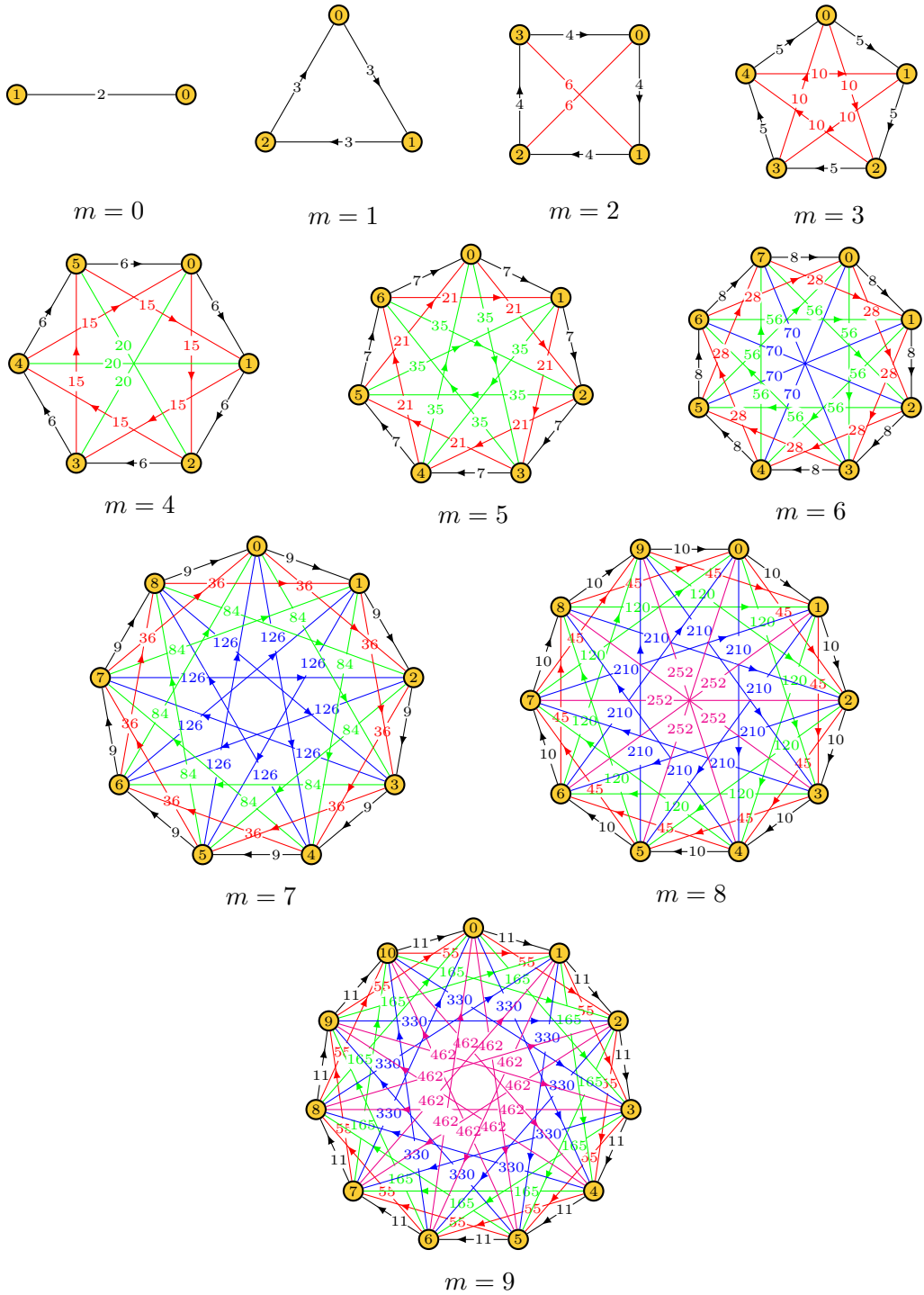

        \captionsetup[subfigure]{labelformat=empty}
\centering 
        \hspace*{.55cm}\begin{subfigure}[c]{0.22\textwidth}
            \newcommand{\pos}{0.5}
            \newcommand{\arrowHeadPosition}{0.6} 
          
             \subcaption{$m = 0$\ \ \ \ }             
         \end{subfigure}
        \hspace{.1cm}\begin{subfigure}[c]{0.22\textwidth}
            \newcommand{\pos}{0.42}
            \newcommand{\arrowHeadPosition}{0.58}
            %
            \subcaption{$m=1$\ \ \ \ \ \ \ }
        \end{subfigure}
       \begin{subfigure}[c]{0.22\textwidth} 
            \newcommand{\pos}{0.4}
            \newcommand{\arrowHeadPosition}{0.6}
            %
 
            \subcaption{$m=2$\ \ \ \ \ \ \ }
        \end{subfigure}
        \vspace*{.1cm}\begin{subfigure}[c]{0.22\textwidth} 
            \newcommand{\pos}{0.45}
            \newcommand{\arrowHeadPosition}{0.65}
            %
 
            \subcaption{$m=3$ \ \ \ \ \ }
        \end{subfigure}
         \hspace*{.35cm}\begin{subfigure}[c]{0.25\textwidth} 
            \newcommand{\pos}{0.4}
            \newcommand{\arrowHeadPosition}{0.6}
            %
 
            \subcaption{$m=4$}
            \end{subfigure}
         \hspace*{1cm}\begin{subfigure}[c]{0.25\textwidth}
            \newcommand{\pos}{0.37}
            \newcommand{\arrowHeadPosition}{0.7} 
            %
 
            \subcaption{$m = 5$ \, }
        \end{subfigure}
         \hspace*{1cm}\begin{subfigure}[c]{0.25\textwidth}
            \newcommand{\pos}{0.34}
            \newcommand{\arrowHeadPosition}{0.7} 
            %
 
            \subcaption{$m = 6$\ \ \ }
        \end{subfigure}
        \hspace*{1.4cm}\begin{subfigure}[c]{0.4\textwidth}
         \newcommand{\pos}{0.28}
         \newcommand{\arrowHeadPosition}{0.67} 
            %
 
            \subcaption{$m = 7$\ \ \ \ \ \ \ \ \ \ }
        \end{subfigure}
       \begin{subfigure}[c]{0.4\textwidth}
             \newcommand{\pos}{0.4}
             \newcommand{\arrowHeadPosition}{0.725} 
            %
 
            \subcaption{$m = 8$\ \ \ \ \ \ \ \ \ \ }
         \end{subfigure}
\hspace*{2.8cm}\begin{subfigure}[c]{0.48\textwidth}
            \newcommand{\pos}{0.35}
            \newcommand{\arrowHeadPosition}{0.75} 
            %
 
            \subcaption{$m = 9$\ \ \ \ \ \ \ \ \ \ \ \ \ \ \ \ \ \ \ }
        \end{subfigure}  
        \vspace{-.2cm}\caption{Quivers for the $\mathbb{C}^{m+2}/\mathbb{Z}_{m+2}$ orbifolds. Black, red, green, blue and purple correspond to degree 0, 1, 2, 3 and 4, respectively.}
        \label{quivers_CnZn}
    \end{figure}

\paragraph{Quiver.} 
\fref{quivers_CnZn} shows these quivers up to $m=9$.~\footnote{The first members of this family have already appeared in the literature. The $m=0$ and $1$ cases are well known. For early references on $m=2,3,4$, see \cite{GarciaCompean:1998kh,Franco:2015tna,Franco:2016tcm,Franco:2017lpa}.} For each type of field, we have indicated the corresponding $SU(m+2)$ representation. For even $m$, the multiplicities of degree ${m\ov 2}$ fields are actually half the dimension of these representations. In summary:
\begin{itemize}
\item The quiver contains $m+2$ nodes, that we will indexed by $i=0, \cdots, m+1$.
\item The quiver consists of bifundamental fields $\Phi^{(c,c+1)}_{i,i+c+1}$ of degree $0\leq c \leq \floor{m\ov 2}$, where we have used the superindex notation introduced for $\mathbb{C}^{m+2}$. The bifundamental indices are correlated with the degree. As in the unorbifolded case, $\Phi^{(c)}_{i,i+c+1}$ transforms in the antisymmetric $(c+1)$-index representation of $SU(m+2)$.
\item For even $m$, now the multiplicity of the unoriented degree ${m \ov2}$ fields is only equal to the full dimension of the corresponding representation.
\end{itemize}

\paragraph{Superpotential.} 
Using the convention for contracting $SU(m+2)$ indices introduced in \eref{product_contraction_indices 0}, the superpotential is given by
        \begin{align}
            W = \sum_{i+j+k < m+2}\Phi_{i,i+j}^{(j-1;j)}\Phi_{i+j,i+j+k}^{(k-1;k)}\bar{\Phi}_{i+j+k,i}^{(m+1-j-k;m+2-j-k)} ~ .
            \label{potential_Cn_Zn}
        \end{align}

\medskip

Below, we will perform various non-trivial checks of the proposed quiver theories. Similar tests will be presented for all the infinite families of theories considered in this paper. We will then independently derive these quiver theories using the B-model.

\subsubsection{Generalized anomaly cancellation}
Let us verify that the quivers introduced above satisfy the generalized anomaly cancellation condition discussed in \sref{subsec: anomaly free}. Let us assume that the ranks of all nodes are equal to $N$. Then, for a $\mathbb{C}^{m+2}/\mathbb{Z}_{m+2}$ orbifold, the contribution to the anomaly at any node due to the arrows in the quiver is equal to:
\beq
a_{\mbox{arrows}}=N \sum_{c=0}^{m}  (-1)^c \left(\begin{array}{c} m+2 \\ c+1 \end{array}\right) =N(1+ (-1)^m)~,
\eeq
which is precisely the condition for cancellation of anomalies. It is straightforward to show that the only solution to the anomaly cancellation conditions corresponds to equal ranks, as we have assumed. The theories considered in coming sections will exhibit a richer behavior in that respect.

\subsubsection{Kontsevich bracket}

Let us now compute the Kontsevich bracket $\{W,W\}$ for the superpotential in \eref{potential_Cn_Zn} and check that it vanishes. To do so, we need to take into account the rule for cyclic permutations of arrows. Consider a cycle $A_{i,j}^{(c;k)}B^{(d;l)}_{j,i}$, where $A^{(c;k)}_{i,j}$ and $B^{(d;l)}_{j,i}$ are monomials of arrows. Note that the difference between the number of $SU(m+2)$ indices and the degree of a monomial is equal to the number of arrows in it. The commutation relation is:
        \begin{align}
            A^{(c;k)}B^{(d;l)} = (-1)^{cd+kl} B^{(d;l)}A^{(c;k)} ~ .
            \label{commutation_relation_orbifolds}           
        \end{align}    
    The superpotential has degree $m-1$ and $m+2$ indices, so any term in it can be written as $A^{(m-1-c;m+1-c)}_{i,j}\Phi^{(c;c+1)}_{j,i}$, with $A^{(m-1-c;m+1-c)}_{i,j}$ a quadratic monomial and $\Phi^{(c;c+1)}_{j,i}$ an arrow. We then have: 
        \begin{align}
            A^{(m-1-c;m+1-c)}_{i,j}\Phi^{(c;c+1)}_{j,i} = (-1)^{m+1-c}\Phi^{(c;c+1)}_{j,i}A^{(m-1-c;m+1-c)}_{i,j} ~ .
        \end{align}
The derivatives we need for the Kontsevich bracket are
        \begin{align}
            \pdv{W}{\Phi_{i,i+j}^{(j-1;j)}} & = \sum_{k < i}(-1)^{j+k} \bar{\Phi}_{i+j,i-k}^{(m+1-j-k;m+2-j-k)}\Phi_{i-k,i}^{(k-1;k)} \nonumber \\
                & + \sum_{k< m+2-i-j}(-1)^{j+m}\Phi_{i+j,i+j+k}^{(k-1;k)}\bar{\Phi}_{i+j+k,i}^{(m+1-j-k;m+2-j-k)} 
        \end{align}
        and
        \begin{align}
            \pdv{W}{\bar{\Phi}_{i+j,i}^{(m+1-j;m+2-j)}} = \sum_{k< j}\Phi_{i,i+k}^{(j-1;j)}\Phi_{i+k,i+j}^{(k-1;k)} ~ .
        \end{align}
Using these results, we compute:
        \begin{align}
            \{W,W\} =2 \sum_{i,j|j>1;i+j < m+2}\pdv{W}{\bar{\Phi}^{(m+1-j;m+2-j)}_{i+j,i}}\pdv{W}{\Phi_{i,i+j}^{(j-1;j)}} ~ .
        \end{align}
        To simplify the resulting expression we use that fact that all terms in $\{W,W\}$ have degree $m-2$ and $m+2$ global symmetry indices. For a monomial $B_{i,j}^{(m-2-c;m+1-c)}\Phi^{(c;c+1)}_{j,i}$ in $\{W,W\}$ we then have:
        \begin{align}
            B_{i,j}^{(m-2-c;m+1-c)}\Phi^{(c;c+1)}_{j,i} = (-1)^{m+1}\Phi^{(c;c+1)}_{j,i}B_{i,j}^{(m-2-c;m+1-c)} ~ .
        \end{align}
        Using this rule, it is straightforward to verify that $\{W,W\}=0$.

\subsection{Moduli space}

We can verify that the moduli space of the quiver indeed corresponds to $\mathbb{C}^{m+2}/\mathbb{Z}_{m+2}$, using perfect matchings. Below we present the main results, namely the field content of the perfect matchings and how they are mapped to points in the toric diagram. Such detailed information not only confirms that the moduli space corresponds to the desired geometry, but can also be used, for example, to identify the graded quiver counterpart of partial resolutions. We will study examples of partial resolutions in \sref{section_Y10_partial_resolution} and \sref{section_F0_partial_resolution}. 

Let us consider how perfect matchings give rise to the toric diagram in \eref{eq_toric_Cn_Zn}. It is convenient to divide the perfect matchings according to how they transform under the global $SU(m+2)$ symmetry. We consider this approach, which is primarily based on the global symmetry, to be illuminating. It is of course also straightforward to determine the perfect matchings by direct application of their definition and to find their positions in the toric diagram from the intersections between their chiral fields and the boundaries of a unit cell in the corresponding periodic quiver.

\paragraph{Internal Point.} 
The internal point of the toric diagram, $v_0=(0,\ldots,0)$, is the only one that is invariant under $SU(m+2)$. This implies that all perfect matchings that are invariant under $SU(m+2)$ correspond to this point. We label these perfect matchings by $s_i$, $i=1,\ldots,m+2$. They are given by: 
        \begin{align}
            \renewcommand{\arraystretch}{1.3}
            \begin{array}{|c|c|c|}
                \hline
                \ \mbox{        Perfect matching       }  \ \ & \mbox{      Chirals      } & \ \mbox{      Additional fields      } \ \ \\
                \hline
                s_{0}                         & \bar{\Phi}_{m+1,0}^{(0;1)} & \bar{\Phi}_{k,j}^{(m+1-k+j;m+2-k+j)} \,\,\,(k > j)   \\
                \hline
                s_{i} \,\,\,(1 \le i \le m+1) & \Phi_{i-1,i}^{(0;1)}       & \Phi_{j,k}^{(k-j-1;k-j)} \,\,\,  (j < i \mbox{ and } j < k)  \\
                                              &                            & \bar{\Phi}^{(m+1-k+j;m+2-k+j)}_{k,j} \,\,\, (k > j \ge i)\\
                \hline
            \end{array}        
        \end{align} 
We have indicated the chiral field content separately, since it is what matters for the moduli space. From the expression of the superpotential \eref{potential_Cn_Zn}, $s_0$ is evidently a perfect matching. All the $s_{i}$ can be determined by the following simple rule. Given an unbarred field $\Phi_{j,k}^{(k-j-1;k-j)}$, it is in the perfect matching iff $j < i$; otherwise, its conjugate is in the perfect matching. It is straightforward to verify that this results in a collection of fields which covers each term in the superpotential exactly once.

\paragraph{Corners.} 
The $SU(m+2)$ symmetry permutes the corners $v_\mu$, $\mu=1,\ldots,m+2$, of the toric diagram. Thus, the perfect matching associated to any corner breaks the $SU(m+2)$ down to $SU(m+1)\times U(1)$. 
In order to find the perfect matching corresponding to a corner it is sufficient to consider how a given representation of $SU(m+2)$ decomposes under $SU(m+1)$. Since this breaking corresponds to picking a particular $SU(m+2)$ fundamental index $\mu$, this behavior is very simple: $\Phi^{(k-1;k)}_{i,i+k}$ decomposes into two representation, $\Phi^{(k-1;k;\mu)}_{i,i+k}$ and $\Phi^{(k-1;k;\cancel{\mu})}$, of $SU(m+1)$. They are in the $(k-1)-$ and $k$-index antisymmetric representations of $SU(m+1)$, respectively. Explicitly: 
\beq
\begin{array}{ll}
            (\Phi^{(k-1;k;\mu)}_{i,i+k})_{\nu_{1}\cdots\nu_{k-1}} &= (\Phi^{(c;k)}_{i,i+k})_{\mu\nu_{1}\cdots\nu_{k-1}} \\[1mm]
            (\Phi^{(k-1;k;\cancel{\mu})}_{i,i+k})_{\nu_{1}\cdots\nu_{k}} &= (\Phi^{(c;k)}_{i+k,k})_{\nu_{1}\cdots\nu_{k}} \qquad \nu_{j} \ne \mu 
\end {array}            
\eeq
        Similarly, $\bar{\Phi}_{i+k,i}^{(m+1-k;m+2-k)}$ decomposes into two representations and, in keeping with our convention of making all quantum numbers explicit, the conjugate of $\Phi^{(k-1;k;\mu)}_{i,i+k}$ is $\bar{\Phi}^{(m+1-k;m+2-k;\cancel{\mu})}$. Under this breaking, the terms in the superpotential decompose as 
\beq
\begin{array}{cr}
 & \Phi_{i,i+j}^{(j-1;j;\mu)}\Phi_{i+j,i+j+k}^{(k-1;k;\cancel{\mu})}\bar{\Phi}_{i+j+k,i}^{(m+1-j-k;m+2-j-k;\cancel{\mu})} \\[1.5mm]
\Phi_{i,i+j}^{(j-1;j)}\Phi_{i+j,i+j+k}^{(k-1;k)}\bar{\Phi}_{i+j+k,i}^{(m+1-j-k;m+2-j-k)} \to &  + \, \Phi_{i,i+j}^{(j-1;j;\cancel{\mu})}\Phi_{i+j,i+j+k}^{(k-1;k;\mu)}\bar{\Phi}_{i+j+k,i}^{(m+1-j-k;m+2-j-k;\cancel{\mu})} \\[1.5mm]
 & + \, \Phi_{i,i+j}^{(j-1;j;\cancel{\mu})}\Phi_{i+j,i+j+k}^{(k-1;k;\cancel{\mu})}\bar{\Phi}_{i+j+k,i}^{(m+1-j-k;m+2-j-k;\mu)}
 \end{array} 
\eeq
Hence we see that, for every $\mu$, we get a perfect matching $p_{\mu}$ containing the following fields:
        \begin{align}
            \renewcommand{\arraystretch}{1.3}
            \begin{array}{|c|c|c|}
                \hline
                \ \mbox{        Perfect matching       }  \ \ & \mbox{      Chirals      } & \ \mbox{      Additional fields      } \ \ \\
                \hline
                p_{\mu}                       & \Phi_{i,i+1}^{(0;1;\mu)}       & \Phi_{i,i+k}^{(k-1;k;\mu)}  \\
                                              & \bar{\Phi}_{m+1,0}^{(0;1;\mu)} & \bar{\Phi}_{i+k,i}^{(m+1-k;m+2-k;\mu)}   \\
                \hline
            \end{array}        
        \end{align} 

\medskip

In summary, the perfect matchings give rise to the toric diagrams in \eref{eq_toric_Cn_Zn}, confirming that the moduli spaces of these quiver theories are indeed the desired $\mathbb{C}^{m+2}/\mathbb{Z}_{m+2}$ orbifolds.

\subsection{B-model computation}\label{susbec: Bmod Cm+2/Zm+2}
    
Let us now consider the B-model on the $\mathbb{C}^{m+2}/\mathbb{Z}_{m+2}$ orbifold. This orbifold admits a crepant resolution as the total space of the canonical line bundle over $\mathbb{P}^{n}$:
\be\label{tX local Pmp1}
 \t {\bf X}_{m+2} = \mbox{Tot}(\Ol(-m-2) \to \Pn{m+1})~.       
\ee
The following set of sheaves form a strongly exceptional collection on $\mathbb{P}^{m+1}$:
        \begin{align}
           \big \{\Omega^{m+1}(m+1)[m+1]~,\, \Omega^{m}(m)[m]~,\, \cdots~,\, \Omega(1)[1]~,\, \mathcal{O}\big\}~.
            \label{pn_exceptional_collection}
        \end{align}
Denoting by $i$ the embedding $i : \mathbb{P}^{m+1} \to \tilde{X}_{m+2}$, the $m+2$ fractional branes on \eqref{tX local Pmp1} can be written as:~\footnote{To correctly compute the morphisms below, it is important to take into account the derived-category shifts $[j]$ in the definitions of the the fractional branes on ${\bf X}_{m+2}$. Recall that the complex $\CE^\bullet[j]$ denotes the complex $\CE^\bullet$ shifted to the left by $j$ units.}
        \begin{align}
             \big\{\shf_{j} \equiv i_{*}\tbp{j}\; \big|\;  0 \le j \le m+1\big\}~.
        \end{align}
 With these B-branes at hand, we are ready to determine the quiver. The map between $\Ext$ groups and quiver fields was discussed in \sref{section_ext_to_quiver_fields}. The $\Ext$ group elements correspond to the chain maps between the Koszul resolutions of a pair of these sheaves. A sheaf of the form $i_{*}\mathcal{F}$, with $\CF$ a sheaf on $\mathbb{P}^{m+1}$, has a Koszul resolution:
        \begin{align}
            \begin{diagram}
                0 & \rTo & \mathcal{F}(n+1) & \rTo^{v_{\mu}e_{\mu}^{m+2}} & \mathcal{F} & \rTo & i_{*}\mathcal{F} & \rTo & 0~,
                \label{Koszul_resolution_CnZn}
            \end{diagram}
        \end{align}
where $v_{\mu}$ is the  $\Ol(-m-2)$ fiber coordinate in the chart $U_{\mu}$. We refer to Appendix~\ref{App: alg geom} for an explanation of our notations, and for additional background material that will be used extensively below.

\subsubsection{Quiver fields} 
        
The simplest arrows are the generators of $\E^{1}(\shf_{i+1},\shf_{i})$. There generators, denoted by $\phi_{i,i+1}^{\mu}$, are elements of $\cc^{0}(\Hom^{1}(\shf_{i+1},\shf_{i}))$ and are explicitly given by the maps:
            \begin{diagram}
                \Omega^{i+1}(n+i+2) & \rTo & \Omega^{i+1}(i+1) \nonumber \\
                \dTo^{\varphi^{\mu}} & & \dTo^{-\varphi^{\mu}} \nonumber \\
                \Omega^{i}(n+1+i) & \rTo & \Omega^{i}(i) 
            \end{diagram}
Here, $\varphi^{\mu}$ are the global sections of $\Omega^{*}(-1)$, which are computed in Appendix \ref{App: alg geom}---see equation \eqref{global_sections}. Thus, we reproduce the chiral fields (of vanishing quiver degree) of the quiver: 
\be
\phi_{i,i+1}^{\mu}\in \E^{1}(\shf_{i+1},\shf_{i}) \qquad \longleftrightarrow \qquad \Phi^{(0; 1)}_{i, i+1}~,
\ee
in the fundamental of $SU(m+2)$.

The generators of $\E^{k}(\shf_{i+k},\shf_{i})$ take a similar form, using the global sections given in \eqref{section_composition}. The generators lie in the \v{C}ech cohomology $\cc^{0}(\Hom^{k}(\shf_{i+k},\shf_{i}))$ and can be defined to be the antisymmetric composition of $k$ generators of $\E^{1}(\shf_{i+1},\shf_{i})$:
            \begin{align}\label{arrow k antisym orb}
                \phi^{\mu_{1}\mu_{2} \cdots \mu_{k}}_{i,i+k} &= \frac{1}{k!}  \phi^{[\mu_{1}}_{i,i+1} \circ \phi^{\mu_{2}}_{i+1,i+2} \circ \cdots \circ \phi^{\mu_{k}]}_{i+k-1,i+k}
                \qquad \longleftrightarrow \qquad \Phi^{(k-1; k)}_{i, i+k}~.
            \end{align}
As expected, these arrows transform in the $k$-index antisymmetric representation of $SU(m+2)$. The B-model computation thus reproduces exactly the arrows of the  $\mathbb{C}^{m+2}/\mathbb{Z}_{m+2}$ toric quiver presented in \sref{section_quivers_Cn_Zn}.   

We now compute the Serre duals of these arrows, which correspond to the conjugate fields in the quiver.  These computations are useful for determining the superpotential, since some of the terms might involve conjugate fields. In the present case, the Serre duals can also be computed easily starting from the generators of $\E^{1}(\shf_{0},\shf_{m+1})$. They are $\bar{\phi}^{\mu}_{m+1,0} \in \cc^{m+1}(\Hom^{-m}(\shf_{0},\shf_{m+1}))$ and given by the maps:
            \begin{diagram}
                &&\mathcal{O}(m+2) & \rTo & \mathcal{O} \nonumber \\
                &&\dTo^{\bar{\varphi}^{\mu}}\nonumber \\
                \Omega^{m+1}(2m+3) & \rTo & \Omega^{m+1}(m+1) 
            \end{diagram}
 where the sections $\bar{\varphi}^{\mu}$ are given in \eqref{serre_duals}.
The Serre duals of the other arrows \eqref{arrow k antisym orb} can be found by composition of these maps with $\phi_{i,i+1}^{\mu}$. Explicitly, they are given by:
            \begin{align}
                \bar{\phi}_{i+k,i}^{\mu_{1}\cdots\mu_{m+2-k}} &= \frac{(m+1-i-k)!}{(m+2-k)!i!}\phi_{i+k,m+1}^{[\mu_{1}\cdots\mu_{m+1-i-k}}\circ \bar{\phi}^{\mu_{m+2-i-k}}_{m+1,0} \circ \phi_{0,i}^{\mu_{m+3-i-k} \cdots \mu_{m+2-k}]}~.    
            \end{align}

\subsubsection{Superpotential}

Since we have defined higher $\Ext$ groups by composition of maps used to define $\E^{1}$ groups, the product $m_{2}$ (itself given by composition) can be determined straightforwardly. We find:
            \begin{align}
                m_{2}(\phi_{i,i+j}^{\mu_{1}\cdots\mu_{j}},\phi_{k,k+l}^{\mu_{k+1}\cdots\mu_{k+l}}) = \delta_{i+j,k}\phi_{i,k+l}^{\mu_{1}\cdots\mu_{k+l}}~.
            \end{align}
Note that this relation holds not only between cohomology classes, but also between the explicit representatives we have defined. Hence, 
            \begin{align}
                f_{2}(\phi_{i,i+j}^{\mu_{1}\cdots\mu_{j}},\phi_{k,k+l}^{\mu_{k+1}\cdots\mu_{k+l}}) = 0 ~.
            \end{align}
Similarly, using our definition of Serre duals, we can compute that
            \begin{align}
                m_{2}(\bar{\phi}_{i+j,i}^{\mu_{1}\cdots\mu_{m+2-j}},\phi_{k,k+l}^{\mu_{m+1-j}\cdots\mu_{m+2+l-j}}) &= \delta_{i,k}\bar{\phi}_{i+j,i+l}^{\mu_{1}\cdots \mu_{m+2-l-j}} \nonumber \\
                m_{2}(\phi_{i,i+j}^{\mu_{1}\cdots\mu_{j}},\bar{\phi}_{k+l,k}^{\mu_{j+1}\cdots\mu_{m+2+j-l}}) &= \delta_{i+j,k+l}\phi_{i,k}^{\mu_{1}\cdots\mu_{n+1+k-i}}
            \end{align}
            These are the only non-zero $m_{2}$ products. In addition, all the $f_{2}$'s vanish, which means that there are no higher products. The last piece of information we need, in order to write down the superpotential, is the canonical pairing $\gamma$. Taking into account the $SU(m+2)$ global symmetry, it is given by
            \begin{align}
                \gamma(m_{2}(\phi^{\mu_{1}\cdots\mu_{k}}_{i,i+k},\bar{\phi}^{\mu_{k+1}\cdots \mu_{m+2}}_{i+k,i})) = \epsilon_{\mu_{1}\cdots\mu_{m+2}} ~ .
            \end{align}
Combining all this, the general prescription~\eqref{W from Bmod 1}-\eqref{W from Bmod 2} gives the quiver superpotential:
            \begin{align}
                W = \sum_{i+j+k < m+2}\frac{\epsilon_{\mu_{1}\cdots\mu_{m+2}}}{j!k!(m+2-j-k)!}\Phi_{i+j+k,i+k}^{\mu_{1}\cdots\mu_{j}}\Phi_{i+k,i}^{\mu_{j+1}\cdots\mu_{j+k}}\bar{\Phi}_{i,i+j+k}^{\mu_{j+k+1}\cdots \mu_{m+2}} ~ ,
            \end{align}
which is in perfect agreement with \eref{potential_Cn_Zn}.

\section{The $Y^{1,0}(\mathbb{P}^m)$ family}\label{sec: Y10Pm}
Our second family of singularities is a particular generalization of the conifold singularity ${\bf X}_3=\CC_0$. As we will see, the corresponding graded quivers share some rather interesting properties with the celebrated Klebanov-Witten quiver that describes D3-branes at  $\CC_0$ \cite{Klebanov:1998hh}.

\subsection{The toric geometries} \label{section_Y10_geometries}

There exist very interesting infinite families of CY$_{m+2}$ singularities given by the real cone over certain $(2m+3)$-real dimensional Sasaki-Einstein manifolds, with explicitly known metrics, known as $Y^{p,q}$, with the integers $p>0$, $0\leq q<p$ and $p, q$ mutually prime \cite{Gauntlett:2004hh}:
\beq
{\bf X}_{m+2} = C\big(Y^{p,q}(B_m)\big)~ .
\eeq
The compact manifold $Y^{p,q}$ can be understood as a certain lens space bundle over a K{\"a}hler manifold $B_m$ of complex dimension $m$. Importantly, $C(Y^{p,q}(B_m))$ is toric if $B_m$ is a compact toric variety. 

Here, we will focus on the simplest such example, $(p,q)=(1,0)$ and $B_m= \mathbb{P}^m$, namely:
\be\label{X2 Y10Pm}
{\bf X}_{m+2} =C\big(Y^{1,0}(\mathbb{P}^m)\big)~ .
\ee
The toric diagram of this singularity is given by:
\beq
v_0=(0,\ldots,0)~,
\qquad
\begin{array}{l}
v_1= (1,0,0,\ldots,0)~, \\
 v_2=(0,1,0,\ldots,0)~, \\
 \vdots \\
 v_{m+1}=(0,0,\ldots,0, 1)~, 
\end{array} 
\qquad
v_{m+2}=(1,1, \cdots, 1, 1)~.
 \label{eq_toric_Y10}
 \eeq 
These geometries possess an $SU(m+1)$ isometry, which acts on the toric diagram by permuting the points $v_1,\ldots,v_{m+1}$. We then have an $SU(m+1)$ global symmetry in the corresponding graded quivers.

Note that the points $v_0,\ldots,v_{m+1}$ in \eqref{eq_toric_Y10} give rise to the toric diagram for $\mathbb{C}^{m+2}$, which is then augmented by a single additional point $v_{m+2}$. It is hence possible to connect the quivers in this family to the ``flat-space'' quivers for $\mathbb{C}^{m+2}$. In \sref{section_Y10_partial_resolution} below, we will study this connection in detail. 

The singularity \eqref{X2 Y10Pm} has a single K{\"a}hler parameter, corresponding to a small resolution by a $\mathbb{P}^m$:
\be\label{XY01}
\t {\bf X}_{m+2}\cong {\rm Tot}\big(\mathcal{O}(-m) \oplus \mathcal{O}(-1) \longrightarrow \mathbb{P}^m \big)~.
\ee
We will use this resolution \eqref{XY01} to study B-branes in \sref{sec: Bbranes Y10}.

\subsection{The graded quivers}

Unlike $\mathbb{C}^{m+2}$ and the $\mathbb{C}^{m+2}/\mathbb{Z}_{m+2}$ orbifolds discussed in \sref{sec: Cn} and \sref{section_family_orbifolds}, determining the $Y^{1,0}(\mathbb{P}^{m})$ quivers requires a more sophisticated approach than dimensional reduction and orbifolding. Instead, it is possible to derive these quivers combining a generalization of 3d printing \cite{Franco:2018qsc} to CY$_{m+2}$'s with arbitrary $m$ \cite{toappear2}, followed by partial resolution---that is, higgsing in the quiver. Our focus is on the quiver theories in this family and their physics. See \cite{Franco:2018qsc} for a detailed presentation of 3d printing.

First of all, from the normalized volume of the toric diagram, we know that the $Y^{1,0}(\mathbb{P}^{m})$ quiver has $m+1$ nodes.~\footnote{It is also easily understood from the B-model on \protect\eqref{XY01}, since $\chi(\mathbb{P}^m)= m+1$.} In addition, the quivers have an $SU(m+1)$ global symmetry.
  
\begin{figure}[ht!]
    \centering
    \includegraphics[width=\textwidth]{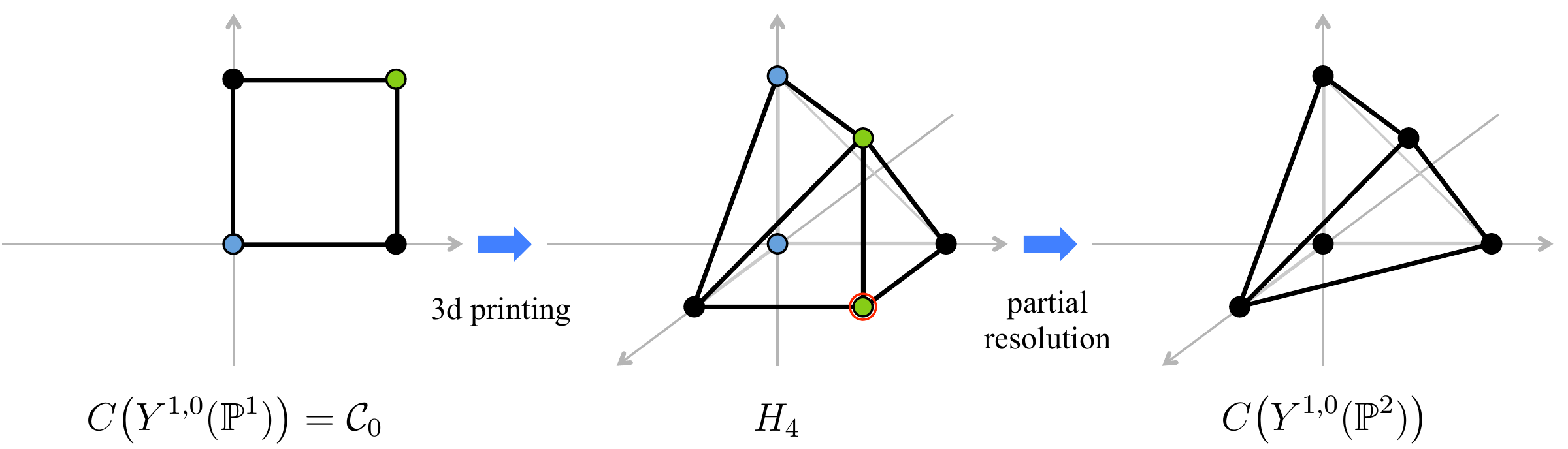}
\caption{Generation of the toric diagram for $C(Y^{1,0}(\mathbb{P}^{2}))$. Starting from the conifold, two points of the toric diagram are lifted by 3d printing. Finally, another point is removed by partial resolution.}
    \label{generating_Y10p}
\end{figure}

The entire family admits an interesting recursive construction. $C(Y^{1,0}(\mathbb{P}^{m+1}))$ can be obtained by starting from $C(Y^{1,0}(\mathbb{P}^{m}))$ and performing 3d printing to produce images of two of the points in the toric diagram, as follows:
\beq
\begin{array}{ccc}
(0,\ldots,0) & \ \ \to \ \ & (0,\ldots,0,0)+(0,\ldots,0,1) \\
(1,\ldots,1) & \ \ \to \ \ & (1,\ldots,1,0)+(1,\ldots,1,1)
\end{array}
\eeq
where the vectors in the first column are $(m+1)$-dimensional, while the ones in the second column are $(m+2)$-dimensional. Next, removing the point $(1,\ldots,1,0)$ via partial resolution, produces the toric diagram for $C(Y^{1,0}(\mathbb{P}^{m+1}))$. The field theory counterparts of these operations generates the $Y^{1,0}(\mathbb{P}^{m+1})$ quivers starting from $Y^{1,0}(\mathbb{P}^{m})$. The initial step is $Y^{1,0}(\mathbb{P}^{m})$, which has $m+1$ nodes. The 3d printing lift of two points in the toric diagram generates a quiver with $2m+2$ nodes. The final partial resolution corresponds to a higgsing with non-zero VEVs for $m$ bifundamental chiral fields, which reduces the number of quiver nodes to $m+2$ and produces the $Y^{1,0}(\mathbb{P}^{m+1})$ quiver. \fref{generating_Y10p} illustrates this process at the level of the geometry for the $Y^{1,0}(\mathbb{P}^{1}) \to Y^{1,0}(\mathbb{P}^{2})$ transition. In this case, the intermediate step corresponds to the so-called $H_4$ theory, which was studied in \cite{Franco:2017cjj,Franco:2018qsc}.
 
We can use the previous method to generate the first members of this family, up to $m=4$. This information, combined with the $SU(m+1)$ global symmetry and a few other consistency conditions (that we discuss below) is sufficient to identify the $Y^{1,0}(\mathbb{P}^{m})$ quivers for arbitrary $m$. In the following, we first present the result of the procedure we just outlined, and we then explicitly verify that these quiver theories have the correct geometry as their moduli space.

\paragraph{Quiver.}
Let us label the $m+1$ nodes with an index $i=0,\ldots, m$. The quiver contains the following arrows, which transform in representations of the global $SU(m+1)$ symmetry:  
        \begin{align}
             X_{m,0}                         & : &m &\xrightarrow[(0)]{\phantom{aabbccd}1\phantom{aabbccd}}0~, \nonumber \\
             X_{i+1,i}                       & : &i+1&\xrightarrow[(0)]{\phantom{aabbccd}1\phantom{aabbccd}}i~,        && 0 \le i \le m-1~, \nonumber\\                    
             \Lambda_{i,i+k}^{(k-1;k)}       & : &i&\xrightarrow[(k-1)]{\phantom{aabb\,\,}\binom{m+1}{k}\phantom{bbcc\,\,}} i+k~, && 0 \le i \le m-1 ; 1 \le k \le m-i~, \nonumber\\
             \Gamma_{i,i+k}^{(k+1;k+1)}      & : &  i&\xrightarrow[(k+1)]{\phantom{aabb\, \, }\binom{m+1}{k+1}\phantom{bbcc\, \, }} i+k~, && 1 \le i \le m-1 ; 0 \le k \le m-i~, 
\label{field_content_Y10m}
       \end{align} 
The subscripts, which should be taken ${\rm mod}(m+1)$, indicate the nodes connected by the arrows, which are bifundamental or adjoint depending on whether the two indices are different or the same. $X_{m,0}$ and $X_{i+1,i}$ are chirals ({\it i.e.}, of quiver degree $0$). They are also singlets under the $SU(m+1)$ global symmetry. For the rest of the arrows, we use a notation with two superindices similar to the one of \sref{subsec: Cn quiver} and \sref{section_quivers_Cn_Zn}. The first integer is the degree of the field. All of these arrows transform in the $j$-index totally antisymmetric representation of $SU(m+1)$. The second integer in the superscript is this $j$. In \eref{field_content_Y10m}, the numbers over the arrows indicate the dimension of the corresponding $SU(m+1)$ representations, and the numbers below are the degrees. Finally, in \eref{field_content_Y10m} we have allowed degrees to go over $n_c-1=\floor{m\ov 2}$, since this permits a more compact presentation of the field content. It is straightforward to restrict to fields with degree $c \leq n_c-1$ by conjugating arrows whenever necessary.

We introduce the following notation for conjugate fields, which makes all their quantum numbers explicit:
\beq
\begin{array}{rclcrcl}
\overline{\left( X_{m,0}  \right)} & = & \b{X}_{0,m}~, & \ \ \ \ \ \ \ \ & \overline{\left( \Lambda_{i,i+k}^{(k-1;k)} \right)} & = & \b{\Lambda}_{i+k,i}^{(m+1-k;m+1-k)}~,  \\[.3cm]
\overline{\left( X_{i+1,i} \right)} & = & \b{X}_{i,i+1}~, & &  \overline{\left( \Gamma_{i,i+k}^{(k+1;k+1)}  \right)} & = & \overline{\Gamma}_{i+k,i}^{(m-1-k;m-k)}~. \\[.3cm]
\end{array}
\eeq
The bifundamental indices are simply flipped. The degree $c$ transforms as $c\to m-c$. Finally the number $j$ of $SU(m+1)$ fundamental indices in the totally antisymmetric representation goes to $m+1-j$. Note that the representations with $j$ and $m+1-j$ have the same dimension and are conjugate to each other, as expected.

\begin{figure}[ht]
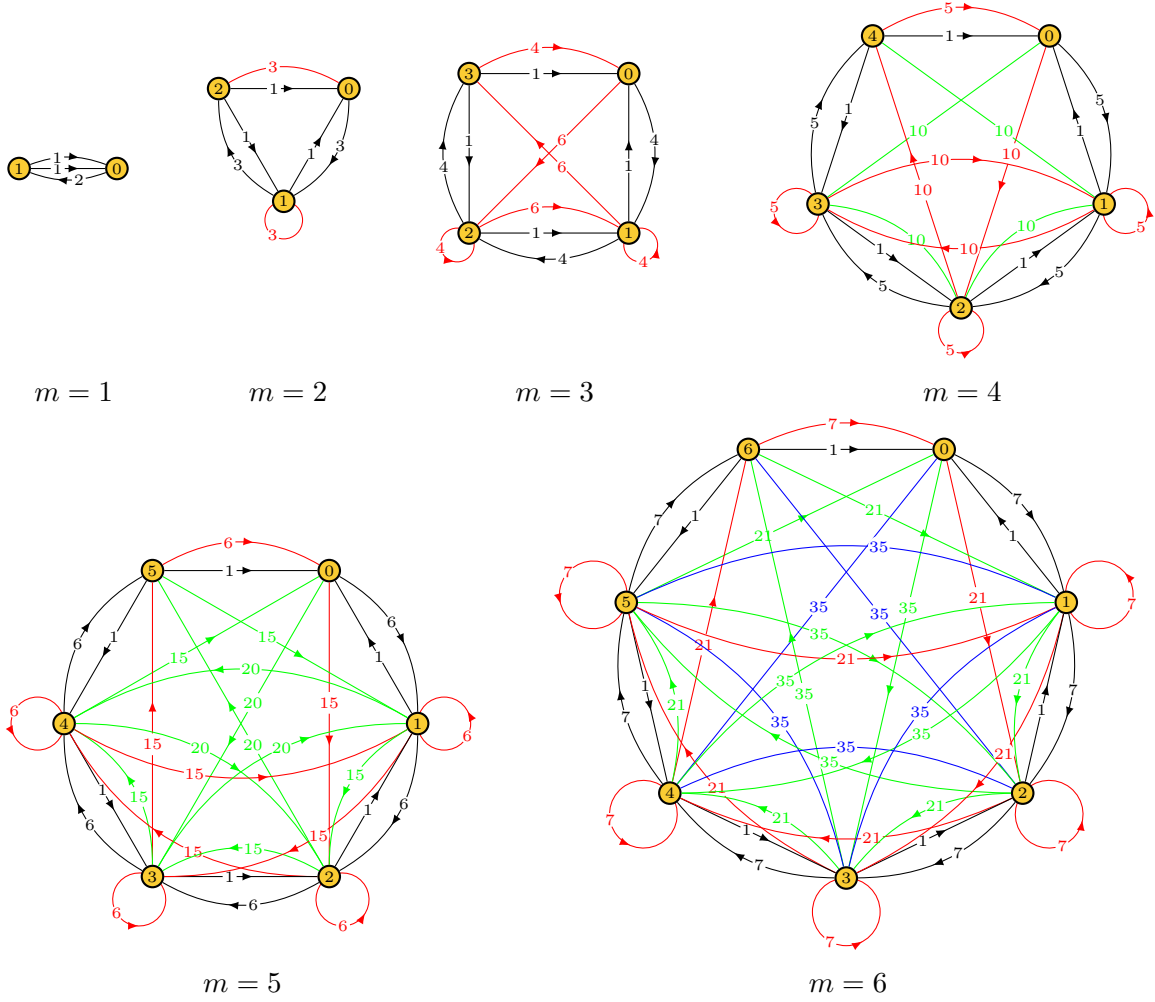

        \captionsetup[subfigure]{labelformat=empty}
            \begin{subfigure}[b]{0.13\textwidth}
                \newcommand{\pos}{0.35}
                \newcommand{\arrowHeadPosition}{0.65}

                \vspace{5.5em}
                \subcaption{$m=1$ \ \ }
            \end{subfigure}
            \hspace{.2cm}
            \begin{subfigure}[b]{0.18\textwidth} 
                \newcommand{\pos}{0.4}
                \newcommand{\arrowHeadPosition}{0.6}
                %
 
                \vspace{3.7em}
            \subcaption{$m=2$ \ \ \ \ \ }
            \end{subfigure}
            \begin{subfigure}[b]{0.29\textwidth}
                \newcommand{\pos}{0.4}
                \newcommand{\arrowHeadPosition}{0.6}
                %
 
                \vspace{2.85em}
                \subcaption{$m=3$ \ \ \ \ \ \ \ \ \ \ }
            \end{subfigure}
            \hspace{-.5cm}
            \begin{subfigure}[b]{0.35\textwidth} 
                \newcommand{\pos}{0.42}
                \newcommand{\arrowHeadPosition}{0.58}
                %
                \subcaption{$m=4$ \ \ } 
            \end{subfigure}
            \begin{subfigure}[b]{0.46\textwidth}
                \newcommand{\pos}{0.42}
                \newcommand{\arrowHeadPosition}{0.58}
                %
 
                \vspace{0.5em}
                \subcaption{$m=5$ \ \ \ \ \ \ \ }
            \end{subfigure}
            \begin{subfigure}[b]{0.5\textwidth} 
                \newcommand{\pos}{0.42}
                \newcommand{\arrowHeadPosition}{0.58}
                %
 
                \subcaption{$m=6$}
            \end{subfigure}
            \caption{Quivers for $Y^{1,0}(\mathbb{P}^{m})$ with $1\leq m \leq 6$. Black, red, green and blue arrows represent fields of degree 0, 1, 2 and 3, respectively.}
            \label{quivers_Y10m}
        \end{figure}

\fref{quivers_Y10m} shows the quivers for $1\leq m\leq 6$. In this figure we adopted the convention in which the degrees of the fields, $c$, are restricted to the range $c\leq \floor{m\ov 2}$, as explained in \sref{sec: 2}. For those fields in \eref{field_content_Y10m} with $c > \floor{m\ov 2}$, we consider their conjugates. Nodes 0 and $m$ are identical, up to conjugation of all the fields in the quiver. The rest of the nodes, 1 to $m-1$, are all equivalent. 

Let us consider the behavior of these quivers under mutations, which are reviewed in Appendix \ref{app: mutations}. Interestingly, node 0 is the only {\it toric node} of the quiver for $m>1$. By this, we mean that it is the only node with two incoming chiral arrows, which results in a toric phase when mutated. Similarly, node 1 is an {\it inverse toric node}, i.e. we obtain a toric phase when acting on it with the inverse mutation. We plan to carry out a more detailed investigation of the mutations of these quivers in future work.

For $m=1$ we have the conifold quiver. In this case, the naive $SU(2)$ global symmetry is enhanced to $SU(2)\times SU(2)$, with the two chiral fields that go from node 1 to node 0 combining to form a doublet of the new $SU(2)$. The $m=2$ quiver (with its superpotential) first appeared in the mathematical literature in \cite{lam2014calabi}; see also \cite{Eager:2018oww}.

\paragraph{Superpotential.}
Let us now consider the superpotential of this family of graded quivers. To determine it, we will again be guided by the global $SU(m+1)$ symmetry. As in \eref{product_contraction_indices 0}, we define a product of arrows in which the $SU(m+1)$ indices are contracted:
            \begin{align}
                (A_{1}^{(c_{1};k_{1})}\cdots A_{n}^{(c_{n};k_{n})})^{\alpha_{k+1}\cdots\alpha_{m+1}} \equiv \frac{1}{\prod_{i}k_{i}!}\epsilon^{\alpha_{1}\cdots\alpha_{m+1}}A_{1;\alpha_{1}\cdots \alpha_{k_{1}}}^{(c_{1};k_{1})}\cdots A_{n;\alpha_{k-k_{n}+1}\cdots \alpha_{k}}^{(c_{n};k_{n})} ~ ,
            \end{align} 
            where $k = \sum_{i}k_{i}$ and the $\alpha_\mu$'s are fundamental $SU(m+1)$ indices. With this convention, any such term with a total of $m+1$ indices is an $SU(m+1)$ invariant. All the superpotential terms we will write have this property.

The superpotential consists of cubic terms $W_{3}$ and quartic terms $W_{4}$. The cubic terms are:
{\small
        \begin{align}
            W_{3} &= 
                \sum_{i=2}^{m}\sum_{k = 0}^{i-1}s_{1}(i,k)X_{i,i-1}\bar{\Gamma}_{i-1,i-1-k}^{(m-k-1;m-k)}\Lambda_{i-1-k,i}^{(k;k+1)} \nonumber\\
            & + \sum_{i=2}^{m}\sum_{k = 1}^{m-i}s_{2}(i,k)X_{i,i-1}\Lambda_{i-1,i-1+k}^{(k-1;k)}\bar{\Gamma}_{i-1+k,i}^{(m-k;m+1-k)} \nonumber\\
            & + \sum_{i=1}^{m-1}\sum_{k=1}^{i-1}\sum_{j=k}^{m-1-i}s_{3}(i,j,k)\Lambda_{i-k,i}^{(k-1;k)}\bar{\Gamma}_{i,i-j}^{(m-j-1;m-j)}\Gamma_{i-j,i-k}^{(j-k+1,j-k+1)} \nonumber\\
            & + \sum_{i=1}^{m-1}\sum_{k=1}^{i-1}\sum_{j=0}^{m-i-1}s_{4}(i,j,k)\Lambda_{i-k,i}^{(k-1;k)}\Gamma_{i,i+j}^{(j+1;j+1)}\bar{\Gamma}_{i+j,i-k}^{(m-j-k-1;m-j-k)} \nonumber\\
            & + \sum_{i=1}^{m}\sum_{k=1}^{i-1}\sum_{j=1}^{m-i}s_{5}(i,j,k)\Lambda_{i-k,i}^{(k-1;k)}\Lambda_{i,i+j}^{(j-1;j)}\bar{\Lambda}_{i+j,i-k}^{(m+1-j-k;m+1-j-k)}~,
\label{Y10_W_cubic}
        \end{align}}
while the quartic terms are:
        \begin{align}
           W_{4} =  &\sum_{k=1}^{m}s_{6}(k)X_{k,k-1}\Lambda_{k-1,m}^{(m-k,m-k+1)}X_{m,0}\Lambda_{0,k}^{(k-1;k)} \nonumber \\
                   &\phantom{abcdef} + \sum_{k=1}^{m-1}\sum_{j=0}^{m-1-k}s_{7}(j,k)\Gamma_{k,k+j}^{(j+1;j+1)}\Lambda_{k+j,m}^{(m-k-j-1;m-k-j)}X_{m,0}\Lambda_{0,k}^{(k-1;k)}~,
 \label{Y10_W_quartic} 
        \end{align}
where $s_{7}, \cdots, s_{7}$ are signs, which we will fix momentarily by requiring that the Kontsevich bracket  $\{W,W\}$ vanishes. Note that, for $m=1$, the only non-trivial term in $W$ is the first line of $W_4$, giving us $W= X_{10} \Lambda^{(0; 1)}_{01} X_{10} \Lambda^{(0; 1)}_{01}$, which reproduces the well-known quadratic superpotential of the conifold quiver.

\subsubsection{Generalized anomaly cancellation}

Let us start by assuming that the ranks of all the nodes are equal to $N$ and check that, in this case, the quivers we propose satisfy the generalized anomaly-free conditions. We normalize all the anomalies by $N$. For node $0$ the contribution of the arrows to the anomaly is given by:
        \begin{align}
            a_{0,{\rm{arrows}}} &=   (-1)^{m}2 + \sum_{k=1}^{m}(-1)^{m-k}\binom{m+1}{k} \nonumber \\
                                    &= (-1)^{m}2 + [ 1 + (-1)^{m+1}] \nonumber \\
                                    &= 1 + (-1)^{m}         
        \end{align}
        Due to the aforementioned symmetry between nodes 0 and 1, the anomaly for node $1$ follows a very similar computation. For nodes $2$ to $m$ the contributions to the anomaly of fields of different degrees are as follows:
        \beq
            \renewcommand{\arraystretch}{1.75}
            \begin{array}{c c l}
                X_{i+1,i} , \bar{X}_{i-1,i} &: \ \ & 1 + (-1)^{m} \\
                \Lambda_{i-k,i}^{(k-1;k)} &: \ \ & \sum_{k=1}^{i}(-1)^{m+1-k}\binom{m+1}{k} = (-1)^{m} + (-1)^{m-i}\binom{m}{i} \\
                \bar{\Lambda}^{(m+1-k;m+1-k)}_{i+k,i}     &: \ \ & \sum_{k=1}^{m-i}(-1)^{k+1}\binom{m+1}{k} = 1 - (-1)^{m-i}\binom{m}{m-i} \\
                \bar{\Gamma}^{(m-k-1;m-k)}_{i+k,i} &: \ \ &\sum_{k=0}^{m-i-1}(-1)^{k+1}\binom{m+1}{k+1} = -1 +(-1)^{m-i}\binom{m}{m-i} \\
                \Gamma^{(k+1;k+1)}_{i-k,i} &: \ \ & \sum_{k=0}^{i-1}(-1)^{m+1-k}\binom{m+1}{k+1} = -(-1)^{m} - (-1)^{m-i}\binom{m}{i}
            \end{array}
         \eeq 
         Summing these contributions, at node $i$ we have
          \begin{align}
             a_{i,{\rm{arrows}}} &= 1 + (-1)^{m} ~ .
          \end{align}
We conclude that the anomaly cancellation condition is satisfied for all nodes in the quiver.

\paragraph{Anomaly-free fractional branes.}
Interestingly, there are more general solutions to the rank assignments that satisfy the anomaly cancellation conditions. A thorough study of this issue is beyond the scope of this paper, and it will be investigated elsewhere. Here, we just quote the result and consider some of its implications. The space of anomaly-free rank assignments for $Y^{1,0}(\mathbb{P}^m)$ is 2-dimensional and can be parametrized as follows:
\beq
(N_0,\ldots,N_m)=N(1,\ldots,1) + M (0,1,2,\ldots,m) ~ ,
\eeq
with $N$ and $M$ integers. Borrowing the nomenclature from $m\leq 3$, we will say that the $(1,\ldots,1)$ vector corresponds to {\it regular branes}, while more general ranks correspond to the inclusion of (anomaly-free) {\it fractional branes}.~\footnote{This is a standard nomenclature. While it is closely related to our other use of the term fractional brane, which is a bound state of wrapped branes associated to a single node in the quiver, we are confident that the distinction between the two meanings will be clear from the context.} Interestingly, all members of the $Y^{1,0}(\mathbb{P}^{m})$ admit a single type of anomaly-free fractional brane. This behavior generalizes the well-known example of $Y^{1,0}(\mathbb{P}^{1})$, {\it i.e.} the conifold. It is also reminiscent of what happens for the infinite family of $Y^{p,q}$ theories in 4d \cite{Benvenuti:2004dy}, all of which have a single type of anomaly-free fractional brane.

\subsubsection{Kontsevich bracket}

With the convention introduced in the previous section, we can write any $SU(m+1)$ invariant term in the superpotential as $A_{i,j}^{(m-1-c;m+1-k)}\Psi_{j,i}^{(c;k)}$, with $A_{i,j}^{(m-1-c;m+1-k)}$ a monomial and $\Psi_{j,i}^{(c;k)}$ an individual 
$SU(m+1)$ multiplet of arrows. We then have
            \begin{align}
                \pdv{}{\Psi_{j,i}^{(c;k)}} (A_{i,j}^{(m-1-c;m+1-k)}\Psi_{j,i}^{(c;k)}) = A_{i,j}^{(m-1-c;m+1-k)} ~ .
            \end{align}
As in \eref{commutation_relation_orbifolds}, for a cycle $A_{i,j}^{(c;k)}B^{(d;l)}_{j,i}$, with $_{i,j}^{(c;k)}$ and $B^{(d;l)}_{j,i}$ monomial of arrows, the commutation relation is
            \begin{align}
                A_{i,j}^{(c;k)}B^{(d;l)}_{j,i} &= (-1)^{cd + kl}B^{(d;l)}_{j,i}A^{(c;d)}_{i,j} ~ .
            \end{align}

             Since every term in the superpotential has degree $m-1$ and a total of $m+1$ $SU(m+1)$ indices, for the superpotential term we wrote above this commutation relation simplifies to 
             \begin{align}
                 A_{i,j}\Psi_{j,i}^{(c;k)} &= (-1)^{m(c+k)}\Psi_{j,i}^{(c;k)}A_{i,j} ~ .
             \end{align}
             
The various derivatives we need are given by
{\small
\begin{align}
                \pdv{W}{\Lambda_{i,i+k}^{(k-1;k)}} &= \delta_{0,i}s_{6}(k)X_{k,k-1}\Lambda_{k-1,m}^{m-k,m-k+1}X_{m,0} + \delta_{i+k,m}s_{6}(i+1)X_{m,0}\Lambda_{0,i+1}^{(i;i)}X_{i+1,k} \nonumber \\
                &+\delta_{0,i}\sum_{j=0}^{m-1-k}s_{7}(j,k)\Gamma_{k,k+j}^{(j+1;j+1)}\Lambda_{k+j,m}^{m-k-j-1;m-k-j}X_{m,0}\nonumber \\
                &+\delta_{i+k,m}\sum_{j=1}^{i}s_{7}(i-j,j)X_{m,0}\Lambda_{0,j}^{(j-1;j)}\Gamma_{j,i}^{(i-j;i-j)}\nonumber \\
                & +s_{1}(i+k,i-1)X_{i+k,i+k-1}\bar{\Gamma}_{i+k-1,i}^{(m-k;m+1-k)} + (-1)^{m}s_{2}(i-1,k)\bar{\Gamma}_{i+k,i-1}^{(m-k;m+1-k)}X_{i-1,i} \nonumber\\
                & + (-1)^{m}\sum_{j=k}^{i+k-1}s_{3}(i+k,j,k)\bar{\Gamma}_{i+k,i+k-j}^{(m-j-1;m-j)}\Gamma_{i+k-j,i}^{(j-k+1,j-k+1)} \nonumber\\
                & + (-1)^{m}\sum_{j=k}^{m-i-1}s_{4}(i+j,j-k,k)\Gamma_{i+k,i+j}^{(j-k+1;j-k+1)}\bar{\Gamma}_{i+j,i}^{(m-j-1;m-j)} \nonumber\\
                & + (-1)^{m}\sum_{j=k+1}^{m-i}s_{5}(i+k,j-k,k)\Lambda_{i+k,i+j}^{(j-k-1;j-k)}\bar{\Lambda}_{i+j,i}^{(m+1-j;m+1-j)} \nonumber\\
                & +  \sum_{j=k+1}^{m-i-k}s_{5}(i,k,j-k)\bar{\Lambda}^{(m+1-j;m+1-j)}_{i+k;i+k-j}\Lambda_{i+k-j;i}^{(j-k-1;j-k)} ~ ,\nonumber                 
\end{align} 
\begin{align}
                \pdv{W}{\bar{\Lambda}_{i+k,i}^{(m+1-k;m+1-k)}} &= \sum_{j=1}^{k-1}s_{5}(i+j,k-j,j)\Lambda_{i,i+j}^{(j-1;j-1)}\Lambda^{(k-j-1;k-j-1)}_{i+j,i+k} ~ , \nonumber\\[.23cm] 
                \pdv{W}{\bar{\Gamma}_{i+k,i}^{(m-k-1;m-k)}} &= (-1)^{m}s_{1}(i+k+1,k)\Lambda_{i,i+k+1}^{(k;k+1)}X_{i+k+1,i+k} + s_{2}(i,k-1)X_{i,i-1}\Lambda_{i-1,i+k}^{(k;k+1)} \nonumber\\
                &+\sum_{j=0}^{k-1}s_{3}(i+k,k,k-j)\Gamma_{i,i+j}^{(j+1,j+1)}\Lambda_{i+j,i+k}^{(k-j-1;k-j)} \nonumber\\
                &+ \sum_{j=1}^{k}s_{4}(i+j,j-k,j)\Lambda_{i,i+j}^{(j-1;j)}\Gamma_{i+j,i+k}^{(k-j+1;k-j+1)}  ~ , \nonumber\\[.23cm] 
                    \pdv{W}{\Gamma_{i,i+k}^{(k+1;k+1)}} &= \sum_{j=1}^{m-1-i}s_{3}(i+j,j,j-k)\Lambda_{i+k,i+j}^{(j-k-1;j-k)}\bar{\Gamma}_{i+j,i}^{(m-j-1;m-j)} \nonumber\\
                &+(-1)^{m}\sum_{j=0}^{m-1-i}s_{4}(i,k,j-k)\bar{\Gamma}_{i+k,i+k-j}^{(m-j-1;m-j)}\Lambda^{(j-k-1;j-k)}_{i+k-j,i}  \nonumber\\
                &+(-1)^{m} s_{7}(k,i)\Lambda_{i+k,m}^{(m-i-k-1;n-i-k)}X_{m,0}\Lambda_{0,i}^{(i-1;i)} ~ .
\end{align}
 }    
Since every term in the expansion of $\{W,W\}$ has degree $(m-2)$, the commutation rule for terms in this expansion is
             \begin{align}
                 \tilde{A}_{i,j}\Psi_{j,i}^{(c;k)} = (-1)^{m(k+c) + c}\Psi_{j,i}^{(c;k)}\tilde{A}_{i,j} ~.
             \end{align}
To determine $s_{1}, \cdots, s_{7}$ we first note that many of them can be made trivial by field redefinitions. We can fix $s_{6}(k)=1$ by redefining $X_{k,k-1} \to \pm X_{k,k-1}$ and fix $s_{7}(j,k)=1$ by redefining $\Gamma_{i,i+k} \to \pm \Gamma_{i,i+k}$. Lastly $s_{1}(i,k)$ can be chosen to be $1$ by redefining $\Lambda_{i,i+k} \to \pm\Lambda_{i,i+k}$. After eliminating these we find that Kontsevich bracket is satisfied for the following choice of signs:
\beq
\begin{array}{rcl}
                 s_{2}(i,k)   &= & (-1)^{k+1}~, \\[.075cm]
                 s_{3}(i,j,k) &= & (-1)^{j+1}~, \\[.075cm]
                 s_{4}(i,j,k) &= & (-1)^{m}~, \\[.075cm]
                 s_{5}(i,j,k) &= & (-1)^{j+m}~.
\end{array}
\eeq

\subsection{Moduli space}

Now, we verify that the proposed graded quivers give rise to the desired moduli spaces using perfect matchings. We will leave a detailed exposition to \cite{toappear1} and just present the main results here. First we consider the two points which are invariant under the $SU(m+1)$ global symmetry. The field content of the corresponding perfect matchings must involve complete representations of $SU(m+1)$. They are given by:
        \begin{align}
            \renewcommand{\arraystretch}{1.3}
            \begin{array}{|c|c|c|}
                \hline
                \mbox{      Point      }  & \mbox{      Chirals      } & \mbox{      Additional fields      } \\
                \hline
                v_0 & X_{m,0}        & \bar{\Lambda}^{(m+1-k;m+1-k)}_{i+k,i} \\
                              &                & \bar{\Gamma}_{i+k,i}^{(m-k-1;m-k)} \\
                \hline
                v_{m+2}  & X_{i+1,i}      & \bar{\Lambda}^{(m+1-k;m+1-k)}_{i+k,i} \\
                              &                & \Gamma_{i,i+k}^{(k+1;k+1)} \\
                \hline
            \end{array}
            \label{pms_Y10_1}     
        \end{align} 
 
Next, let us consider the perfect matchings for $v_{\mu}$, $\mu=1,\ldots,m+1$. As explained in \sref{section_Y10_geometries}, these points are permuted by the $SU(m+1)$ symmetry.  Picking one of them breaks $SU(m+1) \to SU(m)\times U(1)$. Under this breaking, a $k$-index antisymmetric representation $\Psi^{(c;k)}$ of $SU(m+1)$ decomposes into two representations of $SU(m)$, which we will denote $\Psi^{(c;k;\mu)}$ and $\Psi^{(c;k;\cancel{\mu})}$. Both of them are also antisymmetric and carry $k-1$ and $k$ $SU(m)$ indices, respectively. Explicitly:
        \begin{align}
            \Psi^{(c;k;\mu)}_{\nu_{1}\cdots\nu_{k-1}} &= \Psi^{(c;k)}_{\mu\nu_{1}\cdots\nu_{k-1}} \nonumber \\
            \Psi^{(c;k;\cancel{\mu})}_{\nu_{1}\cdots\nu_{k}} &= \Psi^{(c;k)}_{\nu_{1}\cdots\nu_{k}} \ \ \ \ \ \ \ \ \nu_{j} \ne \mu 
        \end{align}
Note that the conjugate of $\Psi^{(c;k,\mu)}$ is $\bar{\Psi}^{(m-c;m+1-k,\cancel{\mu})}$ and vice versa. The perfect matchings for $v_{\mu}$, $\mu=1\ldots m+1$, are given by:
        \begin{align}
            \renewcommand{\arraystretch}{1.3}
            \begin{array}{|c|c|c|}
                \hline
                \mbox{      Point      }  & \mbox{      Chirals      } & \mbox{      Additional fields      } \\
                \hline
               v_{\mu}, \mu=1\ldots m+1       & \Lambda^{(0;1;\mu)}_{i,i+1}     & \Lambda^{(k-1;k;\mu)}_{i,i+k} ~ , \, \bar{\Lambda}^{(m+1-k;m+1-k;\mu)}_{i+k,i} \\
                              &                                 & \bar{\Gamma}_{i+k,i}^{(m-k-1;m-k;\mu)}~ , \, \Gamma_{i,i+k}^{(k+1;k+1;\mu)} \\
                \hline
            \end{array}
             \label{pms_Y10_2}  
        \end{align}

\subsection{Partial resolution $C(Y^{1,0}(\mathbb{P}^{m}))\to \mathbb{C}^{m+2}$}

\label{section_Y10_partial_resolution}

\begin{figure}[ht]
    \centering
    \includegraphics[width=10cm]{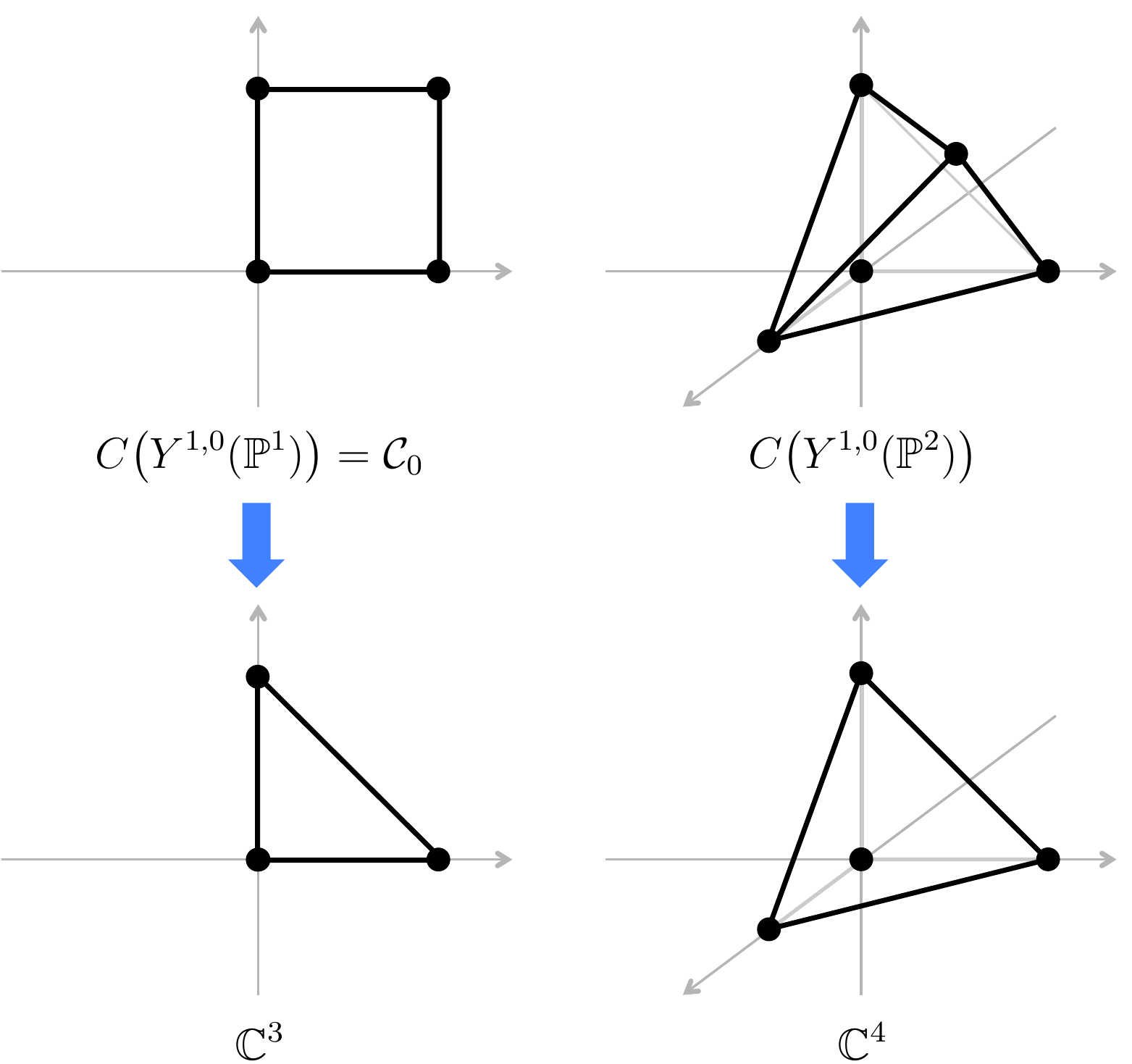}
\caption{$C(Y^{1,0}(\mathbb{P}^{m}))\to \mathbb{C}^{m+2}$ partial resolution for $m=1,2$.}
    \label{Y10m_to_Cmp2_Higgsing}
\end{figure}

Let us now consider yet another check of the proposed quiver theories. Removing the point $v_{m+2}$ in the toric diagram corresponds to the partial resolution:
\beq
C(Y^{1,0}(\mathbb{P}^{m}))\to \mathbb{C}^{m+2} ~ .
\eeq
\fref{Y10m_to_Cmp2_Higgsing} illustrates this resolution for $m=1,2$. This implies that the graded quivers associated to these geometries should be connected by higgsing, as we now explain. 

Let us determine that chiral fields that acquire a non-zero VEV in the corresponding higgsing. Denoting $P_{m+2}$ the chiral field content of the perfect matching associated to the removed point $v_{m+2}$, from \eref{pms_Y10_1}
we have:
\beq
P_{m+2}=\left\{X_{i+1,i},1\leq i \leq m \right\} ~ .
\label{chirals_pm+2}
\eeq
From \eref{pms_Y10_1} and \eref{pms_Y10_2}, we see that these chiral fields only appear in this perfect matching. This implies that given VEVs to all the chiral fields in \eref{chirals_pm+2} produces the desired partial resolution. 

We now consider how this higgsing gives rise to the quivers for $\mathbb{C}^{m+2}$, which were introduced in \sref{subsec: Cn quiver}. First, the VEVs for the $m$ bifundamental chiral fields in \eqref{chirals_pm+2} higgs the $m+1$ nodes in the quiver for $C(Y^{1,0}(\mathbb{P}^{m}))$ down to a single node, as expected.

Since the isometries of $C(Y^{1,0}(\mathbb{P}^{m}))$ and $ \mathbb{C}^{m+2}$ are $SU(m+1)$ and $SU(m+2)$, respectively, the global symmetry of the quiver theory must be enhanced from $SU(m+1)$ to $SU(m+2)$ by the higgsing. We note that all the chiral fields in \eref{chirals_pm+2} are singlets of $SU(m+1)$, which implies that the global symmetry would, at the very least, remain unbroken.

It is instructive to consider how the remaining fields form $SU(m+2)$ representations. It is straightforward, albeit tedious, to verify that the massless matter fields that survive the higgsing are all the arrows that were initially charged under node $0$, except for $X_{0,m}$. They are
        \begin{align}
             X_{m,0}                         & : &1 &\xrightarrow[(0)]{\phantom{aabbccd}1\phantom{aabbcc}}0 \nonumber \\
             \bar{\Lambda}_{m+1-k,0}^{(k;k)}        & : &k &\xrightarrow[(k)]{\phantom{aabb}\binom{m+1}{k}\phantom{aabb}} 0   && 1 \le k \le m                     
        \end{align} 
We thus have a multiplet of degree $k$ in the $k$-index totally antisymmetric representation of $SU(m+1)$ for every $k=0,\ldots,m$. The multiplet of degree $k$ and the conjugate of the multiplet of degree $m-k$ combine to form a degree $k$ field in the $(k+1)$-index totally antisymmetric representation of $SU(m+2)$ for $k=0,\ldots {m\ov 2}$.~\footnote{When $m$ is even, the field of degree $k={m\ov 2}$ coincides with the one of degree $m-k$. We thus obtain only half a multiplet for $k={m\ov 2}$, or the full multiplet by combining it with its conjugate.} This is precisely the field content for $\mathbb{C}^{m+2}$, as discussed in \sref{subsec: Cn quiver}.

\subsection{A simple duality cascade}

A beautiful property of the $Y^{1,0}(\mathbb{P}^{m})$ theories is that they have a single toric phase and that they enjoy a remarkably simple {\it duality cascade}, generalizing the well-known cascade for the conifold \cite{Klebanov:2000hb}. There is a single toric node, {\it i.e.} a node with two incoming chiral fields, which is node $0$. Similarly, node $m$ is a toric node under inverse duality. A duality on node $0$ results in the same theory, up to a cyclic permutation of the node labels. We will now explain how this comes about.

Let us first consider the ``flavors", namely the arrows charged under node $0$. Upon mutating node 0, they transform as follows
        \begin{align}
        \begin{array}{ccccc}
            X_{m,0} && \xrightarrow{\phantom{abcde}} && \tilde{X}_{0,m} \\
            X_{1,0} && \xrightarrow{\phantom{abcde}} && \tilde{X}_{0,1} \\
            \bar{\Lambda}_{m+1-k,0}^{(k;k)} && \xrightarrow{\phantom{abcde}} && \tilde{\Lambda}_{m+1-k,0}^{(k-1;k)} 
   \end{array}
   \label{flavors_dual}
        \end{align}
We will use a tilde to indicate the arrows of the mutated quiver. The fields on the right hand side of the last two rows reproduce the fields charged under node $m$ of the original theory if we relabel nodes as $i\to i-1 \mbox{ mod } (m+1)$. This is the first indication that effect of the mutation is a cyclic permutation of nodes.

        Next, let us consider the mesons generated by the mutation. There are two sets of them, coming from compositions with either $X_{m,0}$ or $X_{1,0}$. They are given by
        \begin{align}
        \begin{array}{ccccc}
            X_{1,0} \, \Lambda_{0,k}^{(k-1;k)}  && \xrightarrow{\phantom{abcde}} && \tilde{\Psi}_{1,k}^{(k-1;k)} \\
            X_{m,0} \, \Lambda_{0,k+1}^{(k;k+1)}  && \xrightarrow{\phantom{abcde}} && \tilde{\bar{\Gamma}}_{m,k+1}^{(k;k+1)} 
 \end{array}
        \end{align}
        All the arrows in the first set becomes massive while $\tilde{\bar{\Gamma}}_{m,1}^{(0;1)}$ also gets a mass. The relevant mass terms in the mutated superpotential and the terms in the original superpotential that give rise to them are:
\beq
\renewcommand{\arraystretch}{1.3}
            \begin{array}{|c|c|}            
                \hline
                   \mbox{   Term in the original superpotential  } & \mbox{Mass Term}  \\
                   \hline
                    X_{1,0}\Lambda_{0,m}^{(m-1;m)}X_{m,0}\Lambda_{0,1}^{(0;1)} & \tilde{\Psi}_{1,m}^{(m-1;m)}\tilde{\bar{\Gamma}}_{m,1}^{(0;1)} \\
                    X_{1,0}\Lambda_{0,k}^{(k-1;k)}\bar{\Gamma}_{k,1}^{(k-1;k)} & \ \ \tilde{\Psi}_{1,k}^{(m-k;m+1-k)}\tilde{\bar{\Gamma}}_{k,1}^{(k-1;k)} \ \ \\
                \hline
            \end{array}     
\eeq
After integrating out the massive fields, the ones that remain and are charged under node $1$ are $\tilde{X}_{0,1}$, $\tilde{X}_{2,1}$ and $\tilde{\Lambda}_{1,k}^{(k-1;k)}$. They correspond exactly to the set of arrows at toric node $0$ in the original theory. The mesons $\tilde{\bar{\Gamma}}_{m,k+1}^{(k-1;k)}$ for $k \ne m-1$ remain massless and are what is required to turn node $m$ of the mutated quiver into node $m-1$ of the original one.

Both the degree and representation under $SU(m+1)$ global symmetry of the arrows not charged under nodes 0, 1 or $m$ depend uniformly on the distance between the two nodes the arrow connects. None of these arrows are affected by mutation and relabeling $i\to i-1$ preserves distances. 

In summary, dualizing node $0$, we obtain the original quiver, up to an $i\to i-1$ cyclic relabeling of the nodes. When the nodes are cyclically ordered as in the examples in \fref{quivers_Y10m}, the net effect of the mutation is a clockwise rotation of the quiver. While we have focused on the quiver, it is straightforward to verify that we also obtain the original superpotential. 

After performing $m+1$ consecutive dualizations on the toric node at each step, we return to the initial quiver. This sequence of mutations therefore generalizes the notion of duality cascade to $m$-graded quivers.

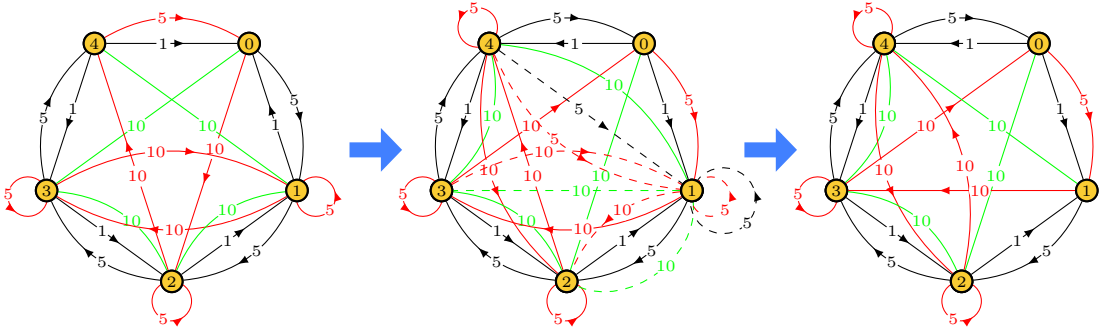
\begin{figure}[ht]
    \centering
    \newcommand{\pos}{0.43}
    \newcommand{\arrowHeadPosition}{0.6}
    \begin{tikzpicture}[scale=1.75 , decoration={markings,mark=at position \arrowHeadPosition with {\arrow{latex}}}] 
        \tikzstyle{every node}=[circle,thick,fill=yellow2,draw,inner sep=1pt,font=\tiny]
            \begin{scope}
                \draw(0.5877852522924731,0.8090169943749475) node ("A") {$0$};
                \draw(-0.587785252292473,0.8090169943749475) node ("B") {$1$};
                \draw[postaction={decorate}, red](-0.9510565162951536,-0.3090169943749473) arc(198.0:558.0:-0.15) node[pos = \pos, draw = none , fill = white , inner sep = 0]{$5$};
                \draw(-0.9510565162951536,-0.3090169943749473) node ("C") {$2$};
                \draw[postaction={decorate}, red](-1.8369701987210297e-16,-1.0) arc(270.0:630.0:-0.15) node[pos = \pos, draw = none , fill = white , inner sep = 0]{$5$};
                \draw(-1.8369701987210297e-16,-1.0) node ("D") {$3$};
                \draw[postaction={decorate}, red](0.9510565162951535,-0.3090169943749476) arc(342.0:702.0:-0.15) node[pos = \pos, draw = none , fill = white , inner sep = 0]{$5$};
                \draw(0.9510565162951535,-0.3090169943749476) node ("E") {$4$};
                \draw [postaction={decorate}, black]("B")to[bend right = 0] node[pos = \pos, draw = none , fill = white , inner sep = 0]{1}("A");
                \draw [postaction={decorate}, red]("B")to[bend left = 30] node[pos = \pos, draw = none , fill = white , inner sep = 0]{5}("A");
                \draw [green]("C")to[bend right= 0] node[pos = \pos, draw = none , fill = white , inner sep = 0]{10}("A");
                \draw [postaction={decorate}, red]("A")to[bend right = 0] node[pos = \pos, draw = none , fill = white , inner sep = 0]{10}("D");
                \draw [postaction={decorate}, black]("E")to[bend right = 0] node[pos = \pos, draw = none , fill = white , inner sep = 0]{1}("A");
                \draw [postaction={decorate}, black]("A")to[bend left = 30] node[pos = \pos, draw = none , fill = white , inner sep = 0]{5}("E");
                \draw [postaction={decorate}, black]("B")to[bend right = 0] node[pos = \pos, draw = none , fill = white , inner sep = 0]{1}("C");
                \draw [postaction={decorate}, black]("C")to[bend left = 30] node[pos = \pos, draw = none , fill = white , inner sep = 0]{5}("B");
                \draw [postaction={decorate}, red]("D")to[bend right = 0] node[pos = \pos, draw = none , fill = white , inner sep = 0]{10}("B");
                \draw [green]("E")to[bend right= 0] node[pos = \pos, draw = none , fill = white , inner sep = 0]{10}("B");
                \draw [postaction={decorate}, black]("C")to[bend right = 0] node[pos = \pos, draw = none , fill = white , inner sep = 0]{1}("D");
                \draw [postaction={decorate}, black]("D")to[bend left = 30] node[pos = \pos, draw = none , fill = white , inner sep = 0]{5}("C");
                \draw [green]("D")to[bend right= 30] node[pos = \pos, draw = none , fill = white , inner sep = 0]{10}("C");
                \draw [postaction={decorate}, red]("E")to[bend left = 30] node[pos = 0.49, draw = none , fill = white , inner sep = 0]{10}("C");
                \draw [postaction={decorate}, red]("C")to[bend left = 30] node[pos = \pos, draw = none , fill = white , inner sep = 0]{10}("E");
                \draw [postaction={decorate}, black]("D")to[bend right = 0] node[pos = \pos, draw = none , fill = white , inner sep = 0]{1}("E");
                \draw [postaction={decorate}, black]("E")to[bend left = 30] node[pos = \pos, draw = none , fill = white , inner sep = 0]{5}("D");
                \draw [green]("E")to[bend right= 30] node[pos = \pos, draw = none , fill = white , inner sep = 0]{10}("D");
                \draw(0.5877852522924731,0.8090169943749475) node ("A") {$0$};
                    \draw(-0.587785252292473,0.8090169943749475) node ("B") {$4$};
                    \draw(-0.9510565162951536,-0.3090169943749473) node ("C") {$3$};
                    \draw(-1.8369701987210297e-16,-1.0) node ("D") {$2$};
                    \draw(0.9510565162951535,-0.3090169943749476) node ("E") {$1$};
            \end{scope}
            \begin{scope}[shift = {(3,0)}]
                \tikzstyle{every node}=[circle,thick,fill=yellow2,draw,inner sep=1pt,font=\tiny]
                    \draw(0.5877852522924731,0.8090169943749475) node ("A") {$0$};
                    \draw(-0.587785252292473,0.8090169943749475) node ("B") {$1$};
                    \draw(-0.9510565162951536,-0.3090169943749473) node ("C") {$2$};
                    \draw(-1.8369701987210297e-16,-1.0) node ("D") {$3$};
                    \draw(0.9510565162951535,-0.3090169943749476) node ("E") {$4$};
                    \draw [postaction={decorate}, black]("A")to[bend right = 0] node[pos = \pos, draw = none , fill = white , inner sep = 0]{1}("B");
                    \draw [postaction={decorate}, black]("B")to[bend left = 30] node[pos = \pos, draw = none , fill = white , inner sep = 0]{5}("A");
                    \draw [postaction={decorate}, red]("C")to[bend right = 0] node[pos = \pos, draw = none , fill = white , inner sep = 0]{10}("A");
                    \draw [green]("D")to[bend right= 0] node[pos = \pos, draw = none , fill = white , inner sep = 0]{10}("A");
                    \draw [postaction={decorate}, black]("A")to[bend left = 0] node[pos = \pos, draw = none , fill = white , inner sep = 0]{1}("E");
                    \draw [postaction={decorate}, red]("A")to[bend left = 30] node[pos = \pos, draw = none , fill = white , inner sep = 0]{5}("E");
                    \draw[postaction={decorate}, red, ](-0.587785252292473,0.8090169943749475) arc(126.0:486.0:-0.15) node[pos = \pos, draw = none , fill = white , inner sep = 0]{$5$};
                    \draw(-0.587785252292473,0.8090169943749475) node ("B") {$1$};
                    \draw [postaction={decorate}, black]("B")to[bend right = 0] node[pos = \pos, draw = none , fill = white , inner sep = 0]{1}("C");
                    \draw [postaction={decorate}, black]("C")to[bend left = 30] node[pos = \pos, draw = none , fill = white , inner sep = 0]{5}("B");
                    \draw [green]("B")to[bend left= 30] node[pos = \pos, draw = none , fill = white , inner sep = 0]{10}("C");
                    \draw [postaction={decorate}, red]("D")to[bend right = 0] node[pos = \pos, draw = none , fill = white , inner sep = 0]{10}("B");
                    \draw [postaction={decorate}, red]("B")to[bend right = 30] node[pos = 0.48, draw = none , fill = white , inner sep = 0]{10}("D");
                    \draw [postaction={decorate}, dashed , black]("B")to[bend right = 0] node[pos = \pos, draw = none , fill = white , inner sep = 0]{5}("E");
                    \draw [postaction={decorate}, dashed , red]("B")to[bend right = 30] node[pos = \pos, draw = none , fill = white , inner sep = 0]{5}("E");
                    \draw [green]("E")to[bend right= 30] node[pos = 0.5, draw = none , fill = white , inner sep = 0]{10}("B");
                    \draw[postaction={decorate}, red, ](-0.9510565162951536,-0.3090169943749473) arc(198.0:558.0:-0.15) node[pos = \pos, draw = none , fill = white , inner sep = 0]{$5$};
                    \draw(-0.9510565162951536,-0.3090169943749473) node ("C") {$2$};
                    \draw [postaction={decorate}, black]("C")to[bend right = 0] node[pos = \pos, draw = none , fill = white , inner sep = 0]{1}("D");
                    \draw [postaction={decorate}, black]("D")to[bend left = 30] node[pos = \pos, draw = none , fill = white , inner sep = 0]{5}("C");
                    \draw [green]("D")to[bend right= 30] node[pos = \pos, draw = none , fill = white , inner sep = 0]{10}("C");
                    \draw [postaction={decorate}, red]("E")to[bend left = 30] node[pos = \pos, draw = none , fill = white , inner sep = 0]{10}("C");
                    \draw [postaction={decorate}, dashed , red]("C")to[bend left = 30] node[pos = 0.4, draw = none , fill = white , inner sep = 0]{10}("E");
                    \draw [dashed , green]("E")to[bend left= 0] node[pos = \pos, draw = none , fill = white , inner sep = 0]{10}("C");
                    \draw[postaction={decorate}, red, ](-1.8369701987210297e-16,-1.0) arc(270.0:630.0:-0.15) node[pos = \pos, draw = none , fill = white , inner sep = 0]{$5$};
                    \draw(-1.8369701987210297e-16,-1.0) node ("D") {$3$};
                    \draw [postaction={decorate}, black]("D")to[bend right = 0] node[pos = \pos, draw = none , fill = white , inner sep = 0]{1}("E");
                    \draw [postaction={decorate}, black]("E")to[bend left = 30] node[pos = \pos, draw = none , fill = white , inner sep = 0]{5}("D");
                    \draw [postaction={decorate}, dashed , red]("E")to[bend right = 30] node[pos = \pos, draw = none , fill = white , inner sep = 0]{10}("D");
                    \draw [dashed , green]("E")to[bend left= 60] node[pos = \pos, draw = none , fill = white , inner sep = 0]{10}("D");
                    \draw[postaction={decorate}, black, dashed , ](0.9510565162951535,-0.3090169943749476) arc(342.0:702.0:-0.25) node[pos = \pos, draw = none , fill = white , inner sep = 0]{$5$};
                    \draw(0.9510565162951535,-0.3090169943749476) node ("E") {$4$};
                    \draw[postaction={decorate}, red, dashed , ](0.9510565162951535,-0.3090169943749476) arc(342.0:702.0:-0.15) node[pos = \pos, draw = none , fill = white , inner sep = 0]{$5$};
                    \draw(0.9510565162951535,-0.3090169943749476) node ("E") {$4$};
                    \draw(0.5877852522924731,0.8090169943749475) node ("A") {$0$};
                    \draw(-0.587785252292473,0.8090169943749475) node ("B") {$4$};
                    \draw(-0.9510565162951536,-0.3090169943749473) node ("C") {$3$};
                    \draw(-1.8369701987210297e-16,-1.0) node ("D") {$2$};
                    \draw(0.9510565162951535,-0.3090169943749476) node ("E") {$1$};
                    \end{scope}
            \begin{scope}[shift = {(6,0)}]
               \tikzstyle{every node}=[circle,thick,fill=yellow2,draw,inner sep=1pt,font=\tiny]
                    \draw(0.5877852522924731,0.8090169943749475) node ("A") {$0$};
                    \draw(-0.587785252292473,0.8090169943749475) node ("B") {$1$};
                    \draw(-0.9510565162951536,-0.3090169943749473) node ("C") {$2$};
                    \draw(-1.8369701987210297e-16,-1.0) node ("D") {$3$};
                    \draw(0.9510565162951535,-0.3090169943749476) node ("E") {$4$};
                    \draw [postaction={decorate}, black]("A")to[bend right = 0] node[pos = \pos, draw = none , fill = white , inner sep = 0]{1}("B");
                    \draw [postaction={decorate}, black]("B")to[bend left = 30] node[pos = \pos, draw = none , fill = white , inner sep = 0]{5}("A");
                    \draw [postaction={decorate}, red]("C")to[bend right = 0] node[pos = \pos, draw = none , fill = white , inner sep = 0]{10}("A");
                    \draw [green]("D")to[bend right= 0] node[pos = \pos, draw = none , fill = white , inner sep = 0]{10}("A");
                    \draw [postaction={decorate}, black]("A")to[bend left = 0] node[pos = \pos, draw = none , fill = white , inner sep = 0]{1}("E");
                    \draw [postaction={decorate}, red]("A")to[bend left = 30] node[pos = \pos, draw = none , fill = white , inner sep = 0]{5}("E");
                    \draw[postaction={decorate}, red, ](-0.587785252292473,0.8090169943749475) arc(126.0:486.0:-0.15) node[pos = \pos, draw = none , fill = white , inner sep = 0]{$5$};
                    \draw(-0.587785252292473,0.8090169943749475) node ("B") {$1$};
                    \draw [postaction={decorate}, black]("B")to[bend right = 0] node[pos = \pos, draw = none , fill = white , inner sep = 0]{1}("C");
                    \draw [postaction={decorate}, black]("C")to[bend left = 30] node[pos = \pos, draw = none , fill = white , inner sep = 0]{5}("B");
                    \draw [green]("B")to[bend left= 30] node[pos = \pos, draw = none , fill = white , inner sep = 0]{10}("C");
                    \draw [postaction={decorate}, red]("D")to[bend right = 30] node[pos = \pos, draw = none , fill = white , inner sep = 0]{10}("B");
                    \draw [postaction={decorate}, red]("B")to[bend right = 30] node[pos = .48, draw = none , fill = white , inner sep = 0]{10}("D");
                    \draw [green]("E")to[bend right= 0] node[pos = \pos, draw = none , fill = white , inner sep = 0]{10}("B");
                    \draw[postaction={decorate}, red, ](-0.9510565162951536,-0.3090169943749473) arc(198.0:558.0:-0.15) node[pos = \pos, draw = none , fill = white , inner sep = 0]{$5$};
                    \draw(-0.9510565162951536,-0.3090169943749473) node ("C") {$2$};
                    \draw [postaction={decorate}, black]("C")to[bend right = 0] node[pos = \pos, draw = none , fill = white , inner sep = 0]{1}("D");
                    \draw [postaction={decorate}, black]("D")to[bend left = 30] node[pos = \pos, draw = none , fill = white , inner sep = 0]{5}("C");
                    \draw [green]("D")to[bend right= 30] node[pos = \pos, draw = none , fill = white , inner sep = 0]{10}("C");
                    \draw [postaction={decorate}, red]("E")to[bend left = 0] node[pos = 0.4125, draw = none , fill = white , inner sep = 0]{10}("C");
                    \draw[postaction={decorate}, red, ](-1.8369701987210297e-16,-1.0) arc(270.0:630.0:-0.15) node[pos = \pos, draw = none , fill = white , inner sep = 0]{$5$};
                    \draw(-1.8369701987210297e-16,-1.0) node ("D") {$3$};
                    \draw [postaction={decorate}, black]("D")to[bend right = 0] node[pos = \pos, draw = none , fill = white , inner sep = 0]{1}("E");
                    \draw [postaction={decorate}, black]("E")to[bend left = 30] node[pos = \pos, draw = none , fill = white , inner sep = 0]{5}("D");
                    \draw(0.5877852522924731,0.8090169943749475) node ("A") {$0$};
                    \draw(-0.587785252292473,0.8090169943749475) node ("B") {$4$};
                    \draw(-0.9510565162951536,-0.3090169943749473) node ("C") {$3$};
                    \draw(-1.8369701987210297e-16,-1.0) node ("D") {$2$};
                    \draw(0.9510565162951535,-0.3090169943749476) node ("E") {$1$};
            \end{scope}
                \draw [-{Triangle[angle = 90:4mm]},line width = 2mm,blue2](1.35,0) -- (1.75,0);
                \draw [-{Triangle[angle = 90:4mm]},line width = 2mm,blue2](4.35,0) -- (4.75,0);
    \end{tikzpicture}
\caption{a) Quiver diagram for $Y^{1,0}(\mathbb{P}^{4})$. b) Result of the mutation on node 0. Massive fields are represented by dashed arrows. c) After integrating out massive fields, we obtain the original quiver, up to an $i\to i-1$ cyclic relabeling of the nodes. This translates into a clockwise rotation of the quiver.}
    \label{mutation_m_4}
\end{figure}

\begin{figure}[ht]
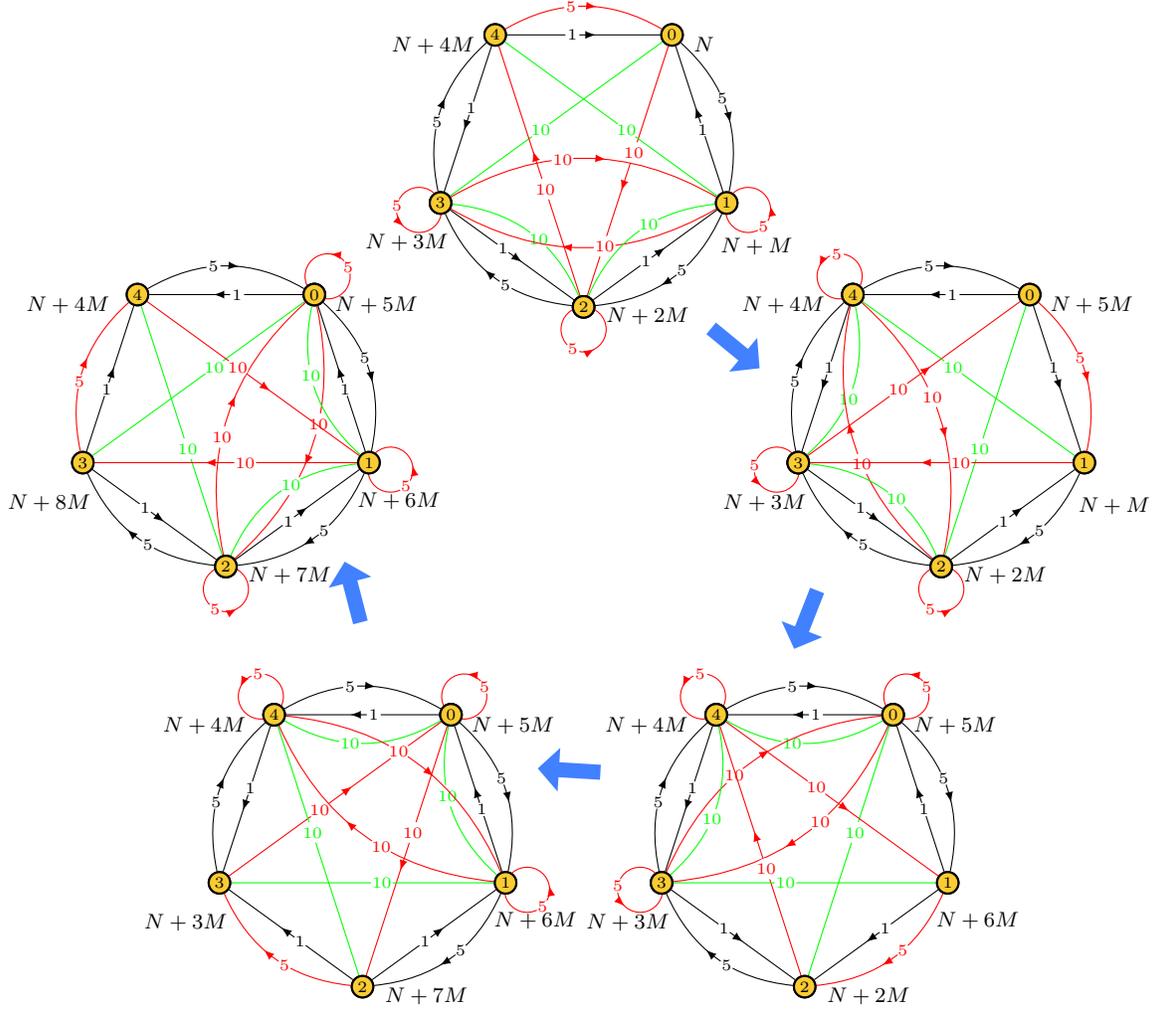

    \centering
    \newcommand{\pos}{0.42}
    \newcommand{\arrowHeadPosition}{0.58}
 
\caption{A period in the duality cascade for $Y^{1,0}(\mathbb{P}^{4})$, starting with $N$ regular and $M$ fractional branes. After each dualization, $M$ remains fixed and $N\to N+M$.}
    \label{cascade_m4}
\end{figure}

\fref{mutation_m_4} shows the transformation of the quiver for $m=4$. The intermediate step includes the massive fields, which are represented by dashed arrows. 
\fref{cascade_m4} shows a period in the cascade for $m=4$. We have included the ranks of the gauge groups associated to the nodes, in the presence of fractional branes, to follow their evolution. Interestingly, as it occurs in the well-known conifold cascade, the number of regular branes increases by 1 with every dualization while the number of fractional branes remains fixed. A full period hence returns to the original quiver with the regular branes increased by $(m+1) M$. For $m=1$, duality cascades admit a renormalization group interpretation. In that context, our choice of dualities corresponds to flowing towards the UV. The flow towards the IR, and the consequent decrease in the number of regular branes, is instead obtained by acting with inverse duality on the node that is toric under it.

\subsection{B-model computation}\label{sec: Bbranes Y10}

The B-model calculation of the graded quivers with superpotentials for the $Y^{1,0}(\mathbb{P}^{m})$ family is similar to the one in section~\ref{susbec: Bmod Cm+2/Zm+2}, with the notation of Appendix~\ref{App: alg geom}. The resolved local Calabi-Yau for this family is:
        \begin{align}
           \t {\bf X}_{m+2} = \mbox{Tot}(\Ol(-m)\oplus \Ol(-1) \to \mathbb{P}^{m})~.
        \end{align}
        Fractional branes are constructed from the exceptional collection on $\Pn{m}$, given by \eqref{pn_exceptional_collection} (with $m+1$ replaced by $m$), by using the embedding $i:\mathbb{P}^{m} \to \t {\bf X}_{m+2}$. They are:
        \begin{align}
             \big\{\shf_{j} \equiv i_{*}\Omega^{j}(j)[j] \;\big|\; 0 \le j \le m\big\}~.
        \end{align}
        To compute the generators of the $\Ext$ groups, we need the Koszul resolution for the fractional branes. It is given by:
        \begin{align}
            0 \rTo \mathcal{F}(m+1) \rTo^{\begin{pmatrix} -u_{\mu}e_{\mu} \nonumber \\ v_{\mu}e_{\mu}^{m}  \end{pmatrix}} \mathcal{F}(m) \oplus \mathcal{F}(1) \rTo^{\begin{pmatrix} v_{\mu}e_{\mu}^{m} &\; u_{\mu}e_{\mu} \end{pmatrix}} \mathcal{F} \rTo i_{*}\mathcal{F} ~ .
        \end{align}
        Here, $v_{\mu}$ is the coordinate of $\Ol(-m)$ fiber and $u_{\mu}$ is the coordinate of $\Ol(-1)$ fiber.

\subsubsection{Quiver}
        
The $\Ext$ group generators for these fractional branes naturally split into three groups and an additional generator, in obvious correspondence with the field content independently derived in \eref{field_content_Y10m}.

\smallskip

\paragraph{First group.}
The first group has a description very similar to the generators in the case of $\mathbb{C}^{m+2}/\mathbb{Z}_{m+2}$. They can be written as the antisymmetric composition of certain basic $\E^{1}$ generators. These are
            \begin{align}
                \lambda^{\mu_{1}\mu_{2}\cdots\mu_{k}}_{i,i+k} & \in  \cc^{0}(\Hom^{k}(\Sf_{i+k},\Sf_{i})) ~ , \nonumber \\
                \lambda^{\mu_{1}\mu_{2}\cdots\mu_{k}}_{i,i+k} & =   \frac{1}{k!}\lambda_{i,i+1}^{[\mu_{1}}\circ\lambda_{i+1,i+2}^{\mu_{2}}\circ \cdots \circ \lambda_{i+k-1,i+k}^{\mu_{k}]} ~ .
            \end{align} 
            $\lambda^{\mu_{1}\mu_{2}\cdots\mu_{k}}_{i,i+k}$ transform in the $k$-index antisymmetric representation of $SU(m+1)$. The basic generators $\lambda_{i,i+1}^{\mu}$, which transform in the fundamental representation of the global $SU(m+1)$ symmetry, are given by the chain map
            \begin{align}
                {\small\begin{diagram}
                    \om{i+1}(m+i+2) & \rTo & \om{i+1}(m+i+1) \oplus \om{i+1}(i+2)   & \rTo & \om{i+1}(i+1) \nonumber \\
                    \dTo^{\varphi^{\mu}}  &      & \dTo^{\tm{-\varphi^{\mu}}{0}{0}{-\varphi^{\mu}}} &      & \dTo_{\varphi^{\mu}} \nonumber \\
                    \om{i}(m+i+1)   & \rTo & \om{i}(m+i) \oplus \om{i}(i+1)       & \rTo & \om{i}(i)
                \end{diagram}}
            \end{align}
            Again, $\varphi^{\mu}$ are the global sections of $\Omega^{*}(-1)$ from \eqref{global_sections}. The Serre duals of these generators are determined along the familiar lines. They are
            \begin{align}
                 \bar{\lambda}_{i+k,i}^{\mu_{1},\mu_{2},\cdots,\mu_{m+1-k}} &\in \check{C}^{m}(\Hom^{2-k}(\Sf_{i+k},\Sf_{i})) ~ , \nonumber \\
                \bar{\lambda}_{i+k,i}^{\mu_{1},\mu_{2},\cdots,\mu_{m+1-k}} &= \frac{(m-i-k)!i!}{(m+1-k)!}\lambda_{i+k,m}^{[\mu_{1}\mu_{2}\cdots\mu_{m-i-k}}\circ \bar{\lambda}_{m,0}^{\mu_{m+1-i-k}} \circ \lambda_{0,i}^{\mu_{m+2-i-k}\cdots\mu_{m+1-k}]} ~ .
            \end{align} 
            With $\bar{\lambda}_{m,0}^{\mu_{m+1-i-k}}$ given by the chain map
            \begin{align}
                    {\small\begin{diagram}
                       &&&& \Ol(m+1) & \rTo & \Ol(m) \oplus \Ol(1)   & \rTo & \Ol \nonumber \\
                       &&&& \dTo^{\bar{\varphi}^{\mu}}  &       \nonumber \\
                    \om{m}(2m+1)   & \rTo & \om{m}(2m) \oplus \om{m}(m+1)       & \rTo & \om{m}(m)
                    \end{diagram}} 
            \end{align}
            
\smallskip

\paragraph{Second group.} 
The second group corresponds to the generators of $\cc^{1}(\Hom^{0}(\Sf_{i},\Sf_{i+1}))$. There is a set of generators ${x}_{i+1,i}$. They are singlets under $SU(m+1)$ defined by the chain maps
            \begin{align}
                {\small\begin{diagram}
                   && \om{i}(m+i+1)   & \rTo & \om{i}(m+i) \oplus \om{i}(i+1)       & \rTo & \om{i}(i) \nonumber \\
                   && \dTo^{\tiny\begin{pmatrix}x^{\prime} \nonumber \\ 0\end{pmatrix}}  &      & \dTo^{\tiny\begin{pmatrix}0 \nonumber \\ -x^{\prime}\end{pmatrix}}  \nonumber \\
                  \om{i+1}(m+i+2) & \rTo & \om{i+1}(m+i+1) \oplus \om{i+1}(i+2)   & \rTo & \om{i+1}(i+1) 
                \end{diagram}}
            \end{align}
            where $x \in \cc^{1}(\Omega)$. This means that locally for each $U_{\mu}\cap U_{\nu}$ there is one form $x_{\mu \nu}$ and this collection satisfies that for any $\mu,\nu$ and $\rho$
            \begin{align}
                x^{\prime}_{\mu\nu} + x^{\prime}_{\nu\rho} + x^{\prime}_{\rho \mu} = 0 ~.
            \end{align}
            Using \eqref{diagonal_transformation}, it can be verified that an explicit representative of this cohomology class is
            \begin{align}
                x^{\prime}_{0,i} &= w_{0,i}^{-1}\dd w_{0,i}  ~ , \nonumber \\
                x^{\prime}_{i,j} &= w_{i,i}^{-1}\dd w_{i,i} - w_{j,j}^{-1}\dd w_{j,j} ~ .
            \end{align}

\smallskip

\paragraph{Third group.}
With this in hand, the third set of $\Ext$ generators is
            \begin{align}
               \gamma^{\mu_{1}\mu_{2}\cdots\mu_{k+1}}_{i,i+k} & \in \cc^{1}(\Hom^{k+1}(\Sf_{i+k},\Sf_{i})) ~ , \nonumber \\ 
               \gamma^{\mu_{1}\mu_{2}\cdots\mu_{k+1}}_{i,i+k} & =  x_{i,i-1} \circ \lambda_{i-1,i+k}^{\mu_{1}\mu_{2}\cdots \mu_{k+1}} ~ . \label{gamma_definition_ext}
            \end{align}
            Motivated by the computation of $\bar{\lambda}$ presented above, in order to calculate the Serre duals of these arrows we start with the generators of $\E^{2}(\shf_{1},\shf_{m-1})$. These generators are $\bar{\gamma}_{m-1,1}^{\mu\nu} \in \check{C}^{m-1}(\Hom^{3-m}(\Sf_{1} ,\Sf_{m-1}))$ and are described by the chain map 
            \begin{align}
                {\small\begin{diagram}
                   && \om{}(m+2) & \rTo & \om{}(m+1) \oplus \om{}(2)   & \rTo & \om{}(1)  \nonumber \\
                   && \dTo^{\tiny\begin{pmatrix}0 \nonumber \\\bar{r}^{\mu\nu} \end{pmatrix}}  &      & \dTo^{\tiny\begin{pmatrix} \bar{r}^{\mu\nu} \nonumber \\ 0\end{pmatrix}}  \nonumber \\
                  \om{m-1}(2m)   & \rTo & \om{m-1}(2m-1) \oplus \om{m-1}(m)       & \rTo & \om{m-1}(m-1) 
                \end{diagram}}
            \end{align}
            where $\bar{r}^{\mu\nu}$ is an element of $\check{C}^{m-1}(\Omega^{m-2}\otimes \Ol(-2))$. Let us consider that $\bar{r}$ is given by the ansatz
            \begin{align}
                \bar{r}^{\mu\nu} = \frac{1}{2}\left(\phi^{\mu} \circ \kappa \circ \phi^{\nu} - \phi^{\nu} \circ \kappa \circ \phi^{\mu}\right) ~.
            \end{align}
            We observe that $\bar{r}^{\mu\nu} \in \check{C}^{m-1}(\Omega^{m-2}\otimes \Ol(-2))$ iff $\kappa \in \check{C}^{m-1}(\Omega^{m})$. Such a $\kappa$ corresponds to a local section of $\om{m}$ for every collection of $m$ patches satisfying that for $\cap_{\mu}U_{\mu}$
            \begin{align}
                \sum_{\mu}(-1)^{\mu}\kappa_{\hat{\mu}} = 0~,
            \end{align}
            where $\kappa_{\hat{\mu}}$ corresponds to collection with every patch except $U_{\mu}$. An explicit representative is
            \begin{align}
                \kappa_{\hat{i}} &= w_{0,i}\wedge_{j}w_{0,j}^{-1}\dd w_{0,j} ~ , \nonumber \\
                \kappa_{\hat{0}} &= \sum_{i}(-1)^{i}w_{i,i}^{-1}\wedge_{j}w_{j,j}^{-1}\dd w_{j,j} ~ . 
            \end{align}
            $\bar{\gamma}_{m-1,1}^{\mu\nu}$ allows us to determine the duals for all $\gamma^{\mu_{1}\mu_{2}\cdots\mu_{k+1}}_{i,i+k}$. These are
            \begin{align}
                \bar{\gamma}_{i+k,i}^{\mu_{1},\mu_{2},\cdots,\mu_{m-k}} &\in \check{C}^{m-1}(\Hom^{1-k}(\Sf_{i+k},\Sf_{i})) ~ , \nonumber \\
                \bar{\gamma}_{i+k,i}^{\mu_{1},\mu_{2},\cdots,\mu_{m-k}} &= c(i,k)\lambda_{i+k,m-1}^{[\mu_{1}\mu_{2}\cdots\mu_{m-i-k-1}}\circ \bar{\gamma}_{m-1,1}^{\mu_{m-i-k}\mu_{m+1-i-k}} \circ \lambda_{1,i}^{\mu_{m+2-i-k}\cdots\mu_{m-k}]} ~ .
            \end{align} 
             Where:
            \begin{align}
                c(i,k) = \frac{2(m-i-k-1)!(i-1)!}{(m-k)!} ~ . 
            \end{align}
            is just a conventional combinatorial factor.
 
 \smallskip

\paragraph{A lone generator.}
In addition to these three groups, there is another generator ${x}_{m,0}$.  It consists of the following map in $\check{C}^{m}(\Hom^{1-m}(\Sf_{m},\Sf_{0}))$:
            \begin{align}
                {\small\begin{diagram}
                   && \Ol(m+1) & \rTo & \Ol(m) \oplus \Ol(1)   & \rTo & \Ol(1)  \nonumber \\
                   && \dTo^{\tiny\begin{pmatrix}0 \nonumber \\\tilde{x} \end{pmatrix}}  &      & \dTo^{\tiny\begin{pmatrix} -\tilde{x} \nonumber \\ 0\end{pmatrix}}  \nonumber \\
                  \om{m}(2m+1)   & \rTo & \om{m}(2m) \oplus \om{m}(m+1)       & \rTo & \om{m}(m) 
                \end{diagram}}
            \end{align}
Proceeding along lines similar to the ones that result in \eqref{serre_duals}, we see that an explicit representative for $\tilde{x}$ is:
            \begin{align} 
                \tilde{x} = \wedge_{i}w_{0,i}^{-1}\dd w_{0,i} ~ .
            \end{align}
            
\smallskip

\noindent In summary, the $x$, $\lambda$ and $\gamma$ generators correspond precisely to the $X$, $\Lambda$ and $\Gamma$ fields in \eref{field_content_Y10m}. We have thus recovered the quivers for the entire $Y^{1,0}(\Pn{m})$ from the B-model.

\subsubsection{Superpotential}

\paragraph{Cubic terms.}
Since we have defined $\Ext$ generators as composition of simpler ones, it is straightforward to determine most of the $m_{2}$ products. For these pairs of generators, the $f_{2}$ vanish. We will mention a few of them here:
        \begin{align}
             m_{2}(\lambda^{\mu_{1}\mu_{2}\cdots\mu_{k}}_{i,i+k},\lambda^{\mu_{k+1}\mu_{2}\cdots\mu_{k+l}}_{j,j+l}) &= \delta_{i+k,j}\lambda^{\mu_{1}\cdots\mu_{k+l}}_{i,i+k+l}  ~ , \nonumber \\
             m_{2}({x_{n}}_{i,i-1} , \lambda^{\mu_{1}\cdots\mu_{k}}_{j,j+k}) &= \delta_{i-1,j}\gamma^{\mu_{1}\cdots\mu_{k}}_{i,i+k-1} ~ , \nonumber \\
             m_{2}(\gamma^{\mu_{1}\cdots\mu_{k+1}}_{i,i+k},\lambda^{\mu_{k+2}\cdots\mu_{k+l+1}}_{j,j+l}) &= \delta_{i+k,j}\gamma^{\mu_{1}\cdots\mu_{k+l+1}}_{i,i+k+l} ~ .
        \end{align} 
        Evaluation of $m_{2}(\lambda^{\mu_{1}\cdots\mu_{k}}_{i,i+k} , x_{j,j-1})$ is slightly more involved. We begin by pointing out a commutation relation:
        \begin{align}
            \varphi^{\mu} \circ x^{\prime} + x^{\prime}\circ \varphi^{\mu} = \delta \tilde{\pi}^{\mu} ~ ,      \label{phi_x_commutation} 
        \end{align}
      where the sheaf $\pi^{\mu}$ is defined to be the element of $\check{C}^0(\mathcal{O}(-1))$ such that:
        \begin{align}
            (\tilde{\pi}^\mu)_\nu = \delta^{\mu}_{\nu} e_\nu ~.
        \end{align}
        At the level of $\Ext$ generators, this commutation relation gives rise to the relation:
        \begin{align}
             \lambda^{\mu}_{i,i+1}\circ x_{i+1,i} = \delta \pi^{\mu}_{i,i} + X_{i,i-1} \circ \lambda^{\mu}_{i-1,i} , \label{lambda_x_commutation} ~,
        \end{align}
        where $\pi^{\mu}$ is defined by the chain map:
        \begin{align}
            {\small\begin{diagram}
                   && \om{i}(m+i+1)   & \rTo & \om{i}(m+i) \oplus \om{i}(i+1)       & \rTo & \om{i}(i) \nonumber \\
                   && \dTo^{\tiny\begin{pmatrix}\tilde{\pi}^{\mu} \nonumber \\ 0\end{pmatrix}}  &      & \dTo^{\tiny\begin{pmatrix}0 \nonumber \\ -\tilde{\pi}^{\mu}\end{pmatrix}}  \nonumber \\
                  \om{i}(m+i+1) & \rTo & \om{i}(m+i) \oplus \om{i}(i+1)   & \rTo & \om{i}(i) 
                \end{diagram}}
            \end{align}
        The first term in \eqref{lambda_x_commutation} is exact in \v{C}ech cohomology and contributes to $f_{2}$ while the second term is another generator and hence corresponds to $m_{2}$.

        Composing the above relation with more $\lambda$'s give us:
        \begin{align}
            \lambda^{\mu_{1}\cdots\mu_{k}}_{i,i+k}\circ x_{i,i+k-1} = x_{i,i-1}\circ\lambda^{\mu_{1}\cdots\mu_{k}}_{i-1,i+k-1} + \frac{1}{(k-1)!}\delta(\pi_{i,i}^{[\mu_{1}}\circ\lambda^{\mu_{2}\cdots\mu_{k}]}_{i,i+k-1}) ~. \label{lambda_x_composition}
        \end{align}
        The right hand side is again in a form that allows us to read off $m_{2}$ and $f_{2}$. We obtain:
        \begin{align}
            m_{2}(\lambda^{\mu_{1}\cdots\mu_{k}}_{i,i+k},x_{i+k,i+k-1}) &= \gamma^{\mu_{1}\cdots\mu_{k}}_{i,i+k-1} ~ , \nonumber \\
            f_{2}(\lambda^{\mu_{1}\cdots\mu_{k}}_{i,i+k},x_{i+k,i+k-1}) &= -\frac{1}{(k-1)!}\pi^{[\mu_{1}}_{i,i}\circ\lambda^{\mu_{2}\cdots\mu_{k}]}_{i,i+k-1} ~ . \label{f2_lambda_x}
        \end{align}
        Using $\gamma$'s definition composition in \eqref{gamma_definition_ext} and composing \eqref{lambda_x_composition} with $\lambda^{\mu}$'s on the right results in:
        \begin{align}
            m_{2}(\lambda^{\mu_{1}\cdots\mu_{k}}_{i,i+k},\gamma^{\mu_{k+1}\cdots\mu_{k+j+1}}_{i+k,i+k+j}) &= \gamma^{\mu_{1}\cdots\mu_{k+j+1}}_{i,i+k+j} ~ , \nonumber \\
                f_{2}(\lambda^{\mu_{1}\cdots\mu_{k}}_{i,i+k},\gamma^{\mu_{k+1}\cdots\mu_{k+j+1}}_{i+k,i+k+j}) &= -\frac{1}{(k+j)!}\pi^{[\mu_{1}}_{i,i}\circ\lambda^{\mu_{2}\cdots\mu_{k+j+1}]}_{i,i+k+j} ~ .  \label{f2_lambda_gamma} 
        \end{align}  
        This completes the reproduction of the cubic terms for this family, which were previously given in \eref{Y10_W_cubic}.
        
\paragraph{Quartic terms.}
To compute the quartic terms we need another set of non-vanishing $f_{2}$. These result from the composition of $x_{m,0}$ with $\lambda_{0,k}$. We start with:
         \begin{align}
             x_{m,0} \circ \lambda^{\mu}_{0,1} = \delta \sigma^{\mu}_{m,0} ~ , 
             \label{xm_lambda_composition}
         \end{align}
         where $\sigma^{\mu}$ is defined by the chain map:
         \begin{align}
            {\small\begin{diagram}
                   && \Omega(m+2)   & \rTo & \Omega(m+1) \oplus \Omega(2)       & \rTo & \Omega(1) \nonumber \\
                   && \dTo^{\tiny\begin{pmatrix}\tilde{\sigma}^{\mu} \nonumber \\ 0\end{pmatrix}}  &      & \dTo^{\tiny\begin{pmatrix}0 \nonumber \\ -\tilde{\sigma}^{\mu}\end{pmatrix}}  \nonumber \\
                  \om{m}(2m+1) & \rTo & \om{m}(2m) \oplus \om{m}(m+1)   & \rTo & \om{m}(m) 
                \end{diagram}}
        \end{align}
        $\tilde{\sigma}$ is an element of $\check{C}^{m-1}(\Omega^{m-1})$ given by:
        \begin{align}
            (\sigma^0)_{\hat{0}} &= 0 ~ , \nonumber \\
            (\sigma^0)_{\hat{j}} &= \wedge_{i \neq j} w_{0i}^{-1} \mathrm{d} w_{0i} \otimes e_0 ~ , \nonumber \\
            (\sigma^i)_{\hat{0}} &= w_{0i}^{-1} \wedge_{j\neq i} w_{0j}^{-1} \mathrm{d} w_{0j} \otimes e_0 ~ , \nonumber \\
            (\sigma^i)_{\hat{j}} &= 0 ~ .
        \end{align}
        Composing $\lambda^{\mu_{2}\cdots\mu_{k}}_{1,k}$ with \eqref{xm_lambda_composition} and doing a bit of algebra gives:
        \begin{align}
             m_{2}(x_{m,0},\lambda_{0,k}) &= 0 ~ , \nonumber \\
            f_{2}(x_{m,0},\lambda_{0,k}) &= -\frac{1}{k!}\gamma^{[\mu_{1}}_{m,1}\circ \lambda^{\mu_{2}\cdots\mu_{k}]}_{1,k} ~ .
        \end{align}
        Combining this with the earlier results for $f_{2}$ in \eqref{f2_lambda_x} we can compute that:
        \begin{multline}
            x_{m,0}\circ f_{2}(\lambda_{0,k}^{\mu_{1}\cdots\mu_{k}},x_{k,k-1}) - f_{2}(X_{m,0},\lambda^{\mu_{1}\cdots\mu_{k}}_{0,k})\circ x_{k,k-1}\\
                = \bar{\lambda}_{m,k-1}^{\mu_{1}\cdots\mu_{k}} + \frac{k-1}{k!}\delta (\gamma^{[\mu_{1}}_{m,1}\circ \pi^{\mu_{2}}_{1,1}\circ \lambda^{\mu_{3} \cdots \mu_{k}]}_{1,k-1}) ~ .
        \end{multline}
        Using this, we conclude that:
        \begin{align}
            m_{3}(x_{m,0},\lambda^{\mu_{1}\cdots\mu_{k}}_{0,k},x_{k,k-1}) &= \bar{\lambda}^{\mu_{1}\cdots\mu_{k}}_{m,k-1} ~ .          
        \end{align} 
        Similarly combining \eqref{xm_lambda_composition} and \eqref{f2_lambda_gamma} results in:
        \begin{align}
            m_{3}(X_{m,0},\lambda^{\mu_{1}\cdots\mu_{k}}_{0,k},\gamma^{\mu_{k+1}\cdots \mu_{k+j+1}}_{k,k+j}) = \bar{\lambda}^{\mu_{1}\cdots\mu_{k+j+1}}_{m,k+j}~.
        \end{align}
        This gives us all the quartic terms in the superpotential. At this point we note that although $f_{3}$ is nontrivial, using consideration of global symmetry and the degree constraint mentioned earlier it can be shown that it cannot result in any additional terms in the superpotential. Hence the quartic terms agree with the ones we wrote for graded quiver.

\paragraph{Absence of higher order terms.}
In principle, we should continue the computations to determine whether the superpotential contains higher order terms. These terms would correspond to gauge invariants of order $m-1$. It is a relatively straightforward exercise to verify that the $SU(m+1) \times U(1)^{m+1}$ global symmetry, whose existence follows from the underlying CY geometry and which is already fixed by the previously computed cubic and quartic terms in the superpotential, rules out any higher order term.

\smallskip

Summarizing the results in this section, we have recovered the superpotential for the entire $Y^{1,0}(\mathbb{P}^m)$ family, which was given in \eref{Y10_W_cubic} and \eref{Y10_W_quartic}.

\section{The $\mathbb{F}_0^{(m)}$ family}\label{sec: F0m}

Our last class of examples is a family of geometries that we denote $\mathbb{F}_0^{(m)}$, which correspond to the affine cones over the $(\mathbb{P}^1)^{m+1}$, a direct product of $m+1$ $\mathbb{P}^1$'s.

\subsection{The toric geometries}

The toric diagram for $\mathbb{F}_0^{(m)}$ is the $(m+1)$-dimensional polytope consisting of the following points.
\beq
\begin{array}{c}
(0,\ldots,0) \\
(\pm 1,0,\ldots, 0) \\
\vdots \\
(0,\ldots,0,\pm1)
 \end{array}
 \label{toric_F0m_general}
\eeq 
These geometries have an $SU(2)^{m+1}$ isometry, which translates into a global symmetry of the corresponding quiver theories. The Newton polynomials contain $2m+3$ terms, of which $m+2$ can be scaled to 1. The remaining $m+1$ coefficients encode the sizes of the $\mathbb{P}^1$'s. The behavior of the mirror geometry as a function of these coefficients was studied in detail for $m=1,2$ in \cite{Franco:2016qxh}. 

\begin{figure}[ht]
	\centering
	\includegraphics[width=13cm]{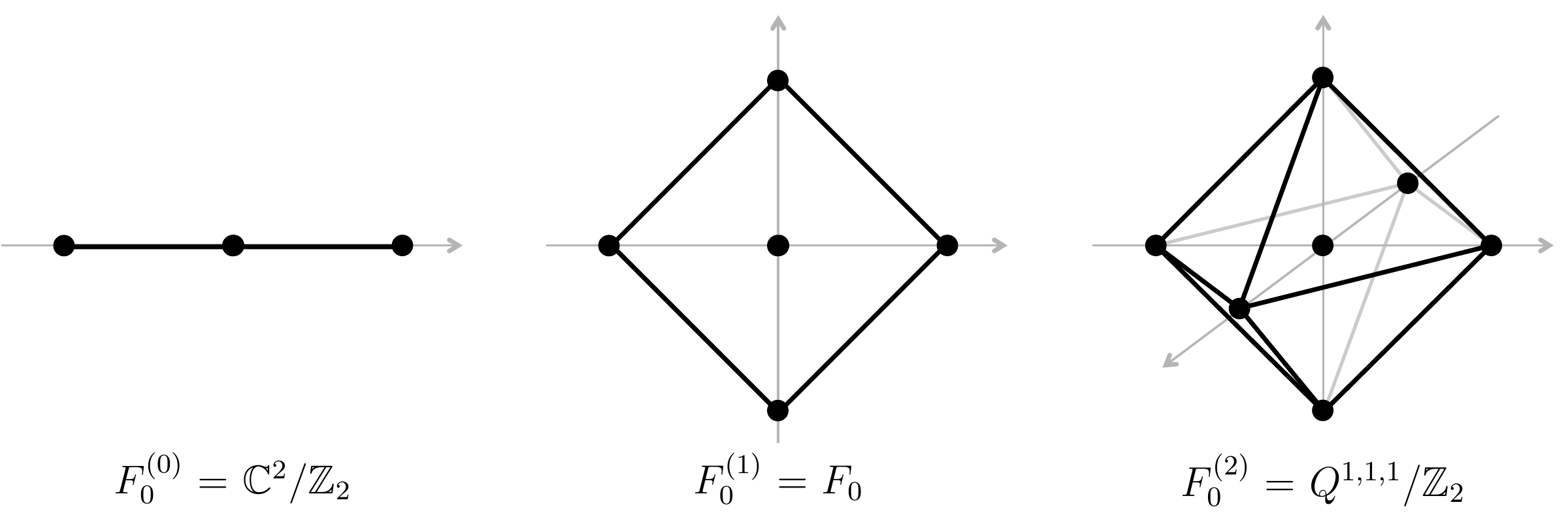}
\caption{Toric diagrams for $\mathbb{F}_0^{(m)}$ with $m=0,1,2$.}
	\label{toric_diagrams_first_F0ms}
\end{figure}

This family contains and naturally generalizes some interesting geometries. In particular, its first members are:
\beq
\begin{array}{ccl}
\mathbb{F}_0^{(0)} & = & \mathbb{C}^2/\mathbb{Z}_2~,\\
\mathbb{F}_0^{(1)} & = & \mathbb{F}_0~, \\
\mathbb{F}_0^{(2)} & = & C(Q^{1,1,1}/\mathbb{Z}_2)~,
\end{array}
\eeq
whose toric diagrams are shown in \fref{toric_diagrams_first_F0ms}.

This is an extremely interesting family of geometries because, contrary to the previous classes of theories, for $m > 0$ they give rise to multiple toric phases related by the corresponding order $m+1$ dualities. 
The $m=1$ \cite{Feng:2002zw} and $2$ \cite{Franco:2016nwv,Franco:2016qxh,Franco:2018qsc} cases have been extensively studied in the literature. In particular, $\mathbb{F}_0^{(1)}$ has 2 toric phases and $\mathbb{F}_0^{(2)}$ has 14 toric phases.

\subsection{The graded quivers}

\label{section_quivers_F0m}

A simple way of constructing a toric phase for each of these geometries is by iterative orbifold reduction \cite{Franco:2016fxm}. The quiver for $\mathbb{F}_0^{(m)}$ has $2^{m+1}$ nodes. This is also clear from the toric diagram, which doubles its normalized volume every time $m$ is increased by 1, as well as from the fact that $\chi((\mathbb{P}^1)^{m+1})= 2^{m+1}$. For later use, it is convenient to label the nodes using $(m+1)$-dimensional vectors with $0$ or $1$ entries, {\it i.e.} in binary.

\paragraph{Quiver.}
The quiver is constructed as follows. Consider two nodes $\alpha$ and $\beta$ labeled by vectors $\vec{\alpha}$ and $\vec{\beta}$. Let us define
\beq
d_{\alpha\beta}=\sum_{i=1}^{m+1} (\beta_i-\alpha_i) ~.
\label{d_ab}
\eeq

Then: 
\begin{itemize}
\item There is an arrow from $\alpha$ to $\beta$ iff $d_{\alpha\beta}>0$, i.e. iff $\beta_i\geq \alpha_i$ for all $1\leq i \leq m+1$.
\item The degree of the arrow is 
\be
c=d_{\alpha\beta}-1 ~.
\label{degrees_F0^m}
\eeq
\item The multiplicity of the arrow is $2^{c+1}$. More specifically, the arrow represents $2^{c+1}$ fields that transform in the 
\be
{\bf 2}_1^{\beta_1-\alpha_1}\times {\bf 2}_2^{\beta_2-\alpha_2}\times \ldots \times {\bf 2}_{m+1}^{\beta_{m+1}-\alpha_{m+1}}
\eeq
representation of the $SU(2)^{m+1}$ global symmetry, where the subindices run over the different $SU(2)$ factors.
\end{itemize}
As usual, we can restrict to fields with $c\leq {m\ov 2}$ by conjugating the arrows with $c>{m\ov 2}$.

\paragraph{Superpotential.}
As for the $\mathbb{C}^{m+2}/\mathbb{Z}_{m+2}$ family, it is possible to show the construction of these models via iterative orbifold reduction implies that all the terms in the superpotential are cubic. The superpotential terms are given by cubic terms of degree $m-1$ combined into $SU(2)^{m+1}$ invariants. Once again, it is possible to show that terms for all possible integer partitions of $m-1$ into three integers are present. In fact we can regard the purely cubic superpotential as the characteristic property of the specific toric phases of $\mathbb{F}_0^{(m)}$ that we construct.

Let us be more explicit about the superpotential for the $\mathbb{F}_0^{(m)}$ family.  From our previous discussion of the field content, there is an arrow connecting nodes $i$ and $j$ whenever $d_{ij}\neq 0$. We will consider the arrow $X_{ij}$ which has $d_{ij}>0$ as the field while we will write $\overline{X}_{ji}$ for its conjugate.~\footnote{Note the convention we use for this argument is not the usual one in which we restrict to degrees $c\leq {m\ov 2}$. For example the arrow directed from $(1,1,\cdots,1)$ to $(0,0,\cdots,0)$ is a chiral but in this notation it will be written as the conjugate of $X_{(0,0,\cdots,0),(1,1,\cdots,1)}$.} It is also useful to define a partial ordering relation $\succ$ between two nodes by $j \succ i$ iff $d_{ij}>0$. 

The superpotential can then be written as  
\beq
W = \sum_{i}\sum_{j \succ i}\sum_{k \succ j}s(i,j,k)X_{ij}X_{jk}\bar{X}_{ki} ~ ,
\label{W_F0m_general}
\eeq
where we omit $SU(2)^{m+1}$ indices and their contractions, and the $s(i,j,k)$ are signs that are necessary for the vanishing of $\{W,W\}$. According to \eref{degrees_F0^m}, $X_{ij}$ has degree $d_{ij}-1$, $X_{jk}$ has degree $d_{jk}-1$ and $\bar{X}_{ki}$ has degree $m+1-d_{ik}$. Gauge invariance implies that $d_{ik} = d_{ij} + d_{jk}$, which in turn implies that the degree of any such term is equal to $m-1$ and it is hence present in the superpotential.

\subsection*{Periodic quivers}

Arguably the simplest representation of theories in the $\mathbb{F}_0^{(m)}$ family is in terms of periodic quivers on $\mathbb{T}^{m+1}$. We can imagine the unit cell has length 2 in every direction and the vector labels we just discussed give the positions of the nodes. Pairs of chiral fields aligned with the $i^{th}$ direction are the $SU(2)_i$ doublets connecting these nodes. These hypercubic structure is completed with additional arrows that form degree $m-1$ triangles representing the cubic terms in the superpotential.

\subsubsection{Generalized anomaly cancellation}

Let us restrict to the case in which all gauge groups have rank $N$. Let $i$ be a node having $k$ entries which are zero, in the binary notation. Then, normalizing by $N$, the contribution of the arrows to the anomaly at node $i$ is:
\beq
\begin{array}{cl}
          a_{\rm{arrows}} & = \sum_{l=1}^{k}\binom{k}{l}(-1)^{l-1}2^{l} + \sum_{l=1}^{m+1-k}\binom{m+1-k}{l}(-1)^{m+1-l}2^{l} \\[.25cm]
             & = -\sum_{l=1}^{k}\binom{k}{l}(-2)^{l} + (-1)^{m+1} \sum_{l=1}^{m+1-k}(-2)^{l} \\[.25cm]
             &= -[(-1)^{l}-1] + (-1)^{m+1}[(-1)^{m+1-l} - 1] \\[.25cm]
             &= 1 + (-1)^m~.
\end{array}
\eeq 
Thus, the anomaly-free condition is satisfied.

\subsubsection{Kontsevich bracket} 

Now we will show that $\{W,W\}=0$ when the coefficients in the superpotential are chosen to be $s(i,j,k) = (-1)^{d(i,j) + m \, d(i,k)}$. First we make a preliminary comment about the way indices are contracted in the superpotential using the $SU(2)$ invariant tensor $\epsilon^{\mu\nu}$. Note that for any term in the superpotential one of these indices will always be contracted with a barred field and the other one with an unbarred field. Even though we do not show these indices in the interest of a clean notation, we will stick to a convention in which the first index contracts with the unbarred field and the second one with barred field. Tiptoeing this convention, in the expressions below the first index of the implicit $\epsilon^{\mu\nu}$ is free for the derivatives with respect to unbarred fields, while the second index is free for the derivatives with respect to barred fields.

With this in mind, the derivatives we need are
        \begin{align}
      \pdv{W}{\bar{X}_{ki}} &= \sum_{j|k \succ j \succ i}(-1)^{d(i,j)+m\, d(i,k)}X_{ij}X_{jk} ~ , \nonumber \\
      \pdv{W}{X_{ik}}       &= \sum_{l|k \succ i \succ l}(-1)^{d(l,i)+m \, d(l,k)}(-1)^{(m-c_{lk})(c_{li} + c_{ik})}\bar{X}_{kl}X_{li} \nonumber \\
       &+ \sum_{l|l \succ k \succ i}(-1)^{d(i,k)+m \, d(i,l)}(-1)^{c_{ik}(c_{kl} + m - c_{il})}X_{kl}\bar{X}_{li} ~ .
    \end{align}
    Here $c_{ij}$ is the degree of $X_{i,j}$ i.e $c_{ij} = d(i,j) - 1$. Working $\mbox{mod } 2$ for any $k \succ j \succ i$ we have
   \beq
     d(i,j) + d(j,k) + d(i,k) = 0 \ \ \  \Rightarrow \ \ \ c_{ij} + c_{jk} + c_{ik} = 1 ~ .
   \eeq
   
Using the fact that $c_{ij}(c_{ij}+1) = 0 \mbox{ mod } 2$ for any $i,j$ we get
\beq
       (-1)^{c_{lk}(c_{li} + c_{ik})} = (-1)^{c_{ik}(c_{kl} + c_{il})} = 1 ~ .
 \eeq 
    With these relations $\{W,W\}$ becomes
    \begin{align}
 \sum_{i}\sum_{k\succ i}\pdv{W}{\bar{X}_{ki}}\pdv{W}{X_{ik}} &= \sum_{i,j,k,l|l\succ k \succ j \succ i}X_{ij}X_{jk}X_{kl}\bar{X}_{li}\Big[(-1)^{mc_{ik}}(-1)^{d(i,k) + d(i,j) +m d(i,k)+m \, d(i,l)} \nonumber \\
                   &+ (-1)^{(m+1)c_{il} + c_{ij}(c_{jk} + c_{kl} + m - c_{li})}(-1)^{d(j,k)+d(i,j)+m \, d(i,l)+m \, d(j,l)}\Big]  ~ .
    \end{align}
      Simplifying this expression using the $\mbox{mod } 2$ relations above, we conclude that  $\{W,W\}=0$.

\subsection{Moduli space} 

\label{section_pms_F0}

Now we explain how the perfect matchings indeed give rise to $\mathbb{F}_{0}^{(m)}$ as the moduli space. First we turn to the central point of the toric diagram \eref{toric_F0m_general}. Since the origin is invariant under the global $SU(2)^{m+1}$ symmetry, the perfect matchings associated to this point contain full representations of it. There is one such perfect matching which is immediately evident from the way we have written the superpotential. It consists of all arrows
        \begin{align}
             \{\bar{X}_{i,j}|i \succ j\} ~ .
        \end{align}
 Writing it in terms of barred fields, makes it manifest that this is a perfect matching due to the form of the superpotential \eref{W_F0m_general}. The chiral fields in this perfect matching are in $\bar{X}_{(1,\cdots,1),(0,\cdots,0)}$ which has dimension $2^{m+1}$ and transforms as $\mathbf{2}_{1} \times \cdots \times \mathbf{2}_{m+1}$.

The central point contains additional perfect matchings. Indeed we know that for $F_{0}^{(1)}$ there are $5$ perfect matchings corresponding to the central point \cite{Feng:2002zw} while $F_{0}^{(2)}$ has $19$ \cite{Franco:2018qsc}. It is straightforward to determine these extra perfect matchings and they will be presented in a forthcoming work \cite{toappear1}. Their explicit field content is rather involved and not illuminating for our current discussion.
        
Next let us consider the corners of the toric diagram, for which $x_{\mu} = \pm 1$ and all the other coordinates are zero. $SU(2)_{\mu}$ transforms these two points into one another so picking one of them breaks $SU(2)_{\mu} \to U(1)\times U(1)$. We need to consider how a representation $X_{i,j}$ of $SU(2)^{m+1}$ splits under this reduced symmetry. There are two cases:
        \begin{itemize}
            \item
                $i_{\mu} = j_{\mu}$. In this case the original multiplet transforms trivially under $SU(2)_{\mu}$ and remains intact. Its conjugate also remains intact.
            \item
                $j_{\mu} - i_{\mu} = 1$. In this case $X_{i,j}$ splits into two multiplets: $X_{i,j}^{+}$  and $X_{i,j}^{-}$ both of which transform as
                \begin{align}
                    \mathbf{2}_{1}^{j_{1}-i_{1}}\times \cdots \times\mathbf{2}_{\mu-1}^{j_{\mu-1}-i_{\mu-1}}\times \mathbf{2}_{\mu+1}^{j_{\mu+1}-i_{\mu+1}}\times \cdots \times \mathbf{2}_{m+1}^{j_{m+1}-i_{m+1}}
                \end{align}
                under the remaining $SU(2)^{m}$.

We will again choose to make all the quantum numbers explicit so that the conjugate of $X_{i,j}^{+}$ is $\bar{X}_{j,i}^{-}$.
        \end{itemize}
        The superpotential also splits into two parts
        \begin{align}
            W = W_{0} + W_{+-} ~ .
        \end{align}
        $W_{0}$ consists of terms which contain no fields charged under $SU(2)_{\mu}$. $W_{+-}$ consists of terms with two arrows charged under $SU(2)_{\mu}$; one unbarred and one barred. Under the reduced symmetry, such a term splits as
        \begin{align}
            X_{i,j}X_{j,k}\bar{X}_{k,i} \to X_{i,j}^{+}X_{j,k}\bar{X}^{-}_{k,i} - X_{i,j}^{-}X_{j,k}\bar{X}^{+}_{k,i} && j_{\mu} - i_{\mu} = 1 ~ , \nonumber \\
            X_{i,j}X_{j,k}\bar{X}_{k,i} \to X_{i,j}X_{j,k}^{+}\bar{X}_{k,i}^{-} - X_{i,j}X_{j,k}^{-}\bar{X}_{k,i}^{+} && k_{\mu} - j_{\mu} = 1 ~ .
        \end{align}
        With this, it is straightforward to verify that the following collection $P_{\mu}^{+}$ of fields is a perfect matching
        \begin{itemize}
            \item
                If $j_{\mu} - i_{\mu} = 1$, then $P_{\mu}^{+}$ contains $X_{i,j}^{+}$ and the conjugate of $X_{i,j}^{-}$ i.e $\bar{X}_{j,i}^{+}$. These arrows cover each term in $W_{+-}$ exactly once and do not cover any term in $W_{0}$.
            \item
                If $j_{\mu} - i_{\mu} = 0$, then $p_{\mu}^{-}$ contains $\bar{X}_{j,i}$. These arrows cover each term of $W_{0}$ exactly once and do not cover any term in $W_{+-}$.
        \end{itemize}
        Above we have assumed $j \succ i$, which is the condition for the existence of an arrow between $i$ and $j$. 

        The perfect matching $P_{\mu}^{+}$ corresponds to $x_{\mu} = 1$. Its chiral field content, which we will denote by $p^{+}_{\mu}$ is
        \begin{align}
            p_{\mu}^{+} = \left\{X^{+}_{(a_{1},\cdots , a_{\mu-1} 0 , a_{\mu+1} , \cdots a_{m+1}),(a_{1},\cdots , a_{\mu-1} , 1 , a_{\mu+1} , \cdots a_{m+1})}\right\}\cup \left\{\bar{X}^{+}_{(1,\cdots,1),(0,\cdots,0)}\right\} ~ .
            \label{P+_F0}
        \end{align}
Regarding the fields on the right brackets, note that since $X^{-}_{(0,\cdots,0),(1,\cdots,1)}$ have degree $m$, their conjugates $\bar{X}^{+}_{(1,\cdots,1),(0,\cdots,0)}$ are indeed chiral fields, i.e. they have degree 0. We can rewrite \eref{P+_F0} as
        \begin{align}
            p_{\mu}^{+} = \left\{X^{+}_{(a_{1},\cdots , a_{\mu-1} 0 , a_{\mu+1} , \cdots a_{m+1}),(a_{1},\cdots , a_{\mu-1} , 1 , a_{\mu+1} , \cdots a_{m+1})}\right\}\cup \left\{X^{-}_{(0,\cdots,0),(1,\cdots,1)}\right\} ~ .
\label{p_mu+F0}
        \end{align}

        Similarly the perfect matching corresponding to $x_{\mu} = -1$ is the collection $P_{\mu}^{-}$ of the following arrows:
        \begin{itemize}
            \item
                If $j_{\mu} - i_{\mu} = 1$, then $P_{\mu}^{-}$ contains $X_{i,j}^{-}$ and the conjugate of $X_{i,j}^{+}$ i.e $\bar{X}_{j,i}^{-}$.
            \item
                If $j_{\mu} - i_{\mu} = 0$, then $P_{\mu}^{-}$ contains $\bar{X}_{j,i}$. 
        \end{itemize}
        The chiral field content $p_{\mu}^{-}$ of this perfect matching is
        \begin{align}
            p_{\mu}^{-} = \left\{X^{-}_{(a_{1},\cdots , a_{\mu-1} 0 , a_{\mu+1} , \cdots a_{m+1}),(a_{1},\cdots , a_{\mu-1} , 1 , a_{\mu+1} , \cdots a_{m+1})}\right\}\cup \left\{\bar{X}^{-}_{(1,\cdots,1),(0,\cdots,0)}\right\} ~ ,
        \end{align}
which can be rewritten as
        \begin{align}
            p_{\mu}^{-} = \left\{X^{-}_{(a_{1},\cdots , a_{\mu-1} 0 , a_{\mu+1} , \cdots a_{m+1}),(a_{1},\cdots , a_{\mu-1} , 1 , a_{\mu+1} , \cdots a_{m+1})}\right\}\cup \left\{X^{+}_{(0,\cdots,0),(1,\cdots,1)}\right\} ~ .
\label{p_mu-F0}
        \end{align}

\subsection{Examples}

The periodic quivers for these theories are rather simple, but they become hard to visualize beyond $m=2$ due to their high dimensionality. The exponential growth of the number of gauge groups makes their ordinary quivers look rather complicated. However, we consider it is instructive to explicitly present the quivers for $m=1,2,3$. $F_0^{(0)}$ is $\mathbb{C}^2/\mathbb{Z}_2$, and its quiver was given in \fref{quivers_CnZn}.

\begin{figure}[H]
	\centering
	\includegraphics[width=6cm]{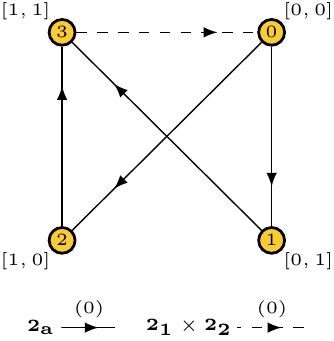}
\caption{Quiver diagram for $F_0^{(1)}$.}
	\label{quiver_F01}
\end{figure}

\fref{quiver_F01} shows the quiver diagram for $F_0^{(1)}$. This is the well-known phase 2 of $F_0$ (see e.g. \cite{Feng:2002zw}).

\begin{figure}[H]
	\centering
	\includegraphics[width=9cm]{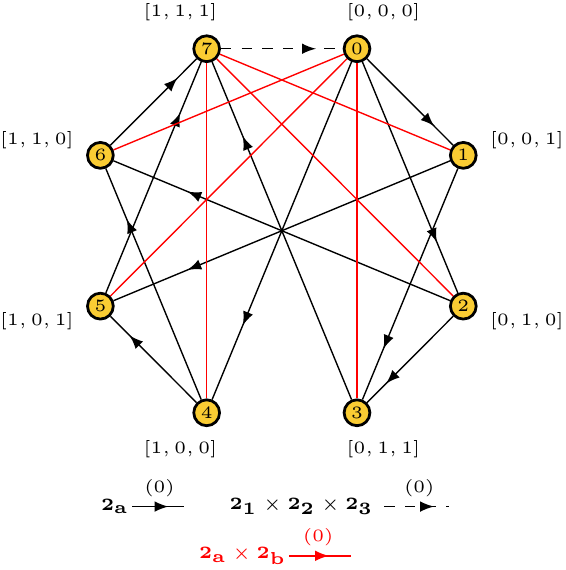}
\caption{Quiver diagram for $F_0^{(2)}$.}
	\label{quiver_F02}
\end{figure}

The quiver for $F_0^{(2)}$ is presented in \fref{quiver_F02}. This is phase $L$ of $Q^{1,1,1}/\mathbb{Z}_2$ in the classification of \cite{Franco:2018qsc}. The periodic quiver for this phase, which explicitly shows plaquettes for all the superpotential terms, can be found in the appendix of \cite{Franco:2018qsc}.

\begin{figure}[H]
	\centering
	\includegraphics[width=11cm]{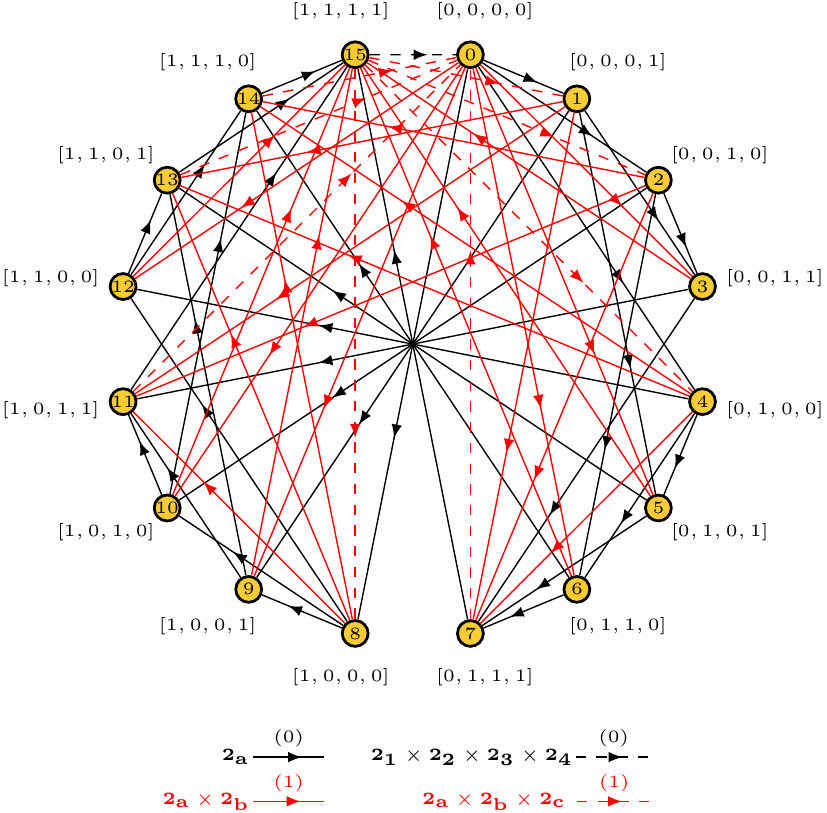}
\caption{Quiver diagram for $F_0^{(3)}$.}
	\label{quiver_F03}
\end{figure}

Finally, \fref{quiver_F03} shows the quiver for $F_0^{(3)}$.

The field content for this theory can be summarized in the following table:

{\footnotesize
\beq
\begin{array}{|c|c|}
            \hline
            \mbox{Field} &  SU(2)^{4} \mbox{ representation} \\
            \hline
                X_{(0,a,b,c),(1,a,b,c)}  & \mathbf{2}_{1} \\
                X_{(a,0,b,c),(a,1,b,c)}  & \mathbf{2}_{2} \\
                X_{(a,b,0,c),(a,b,1,c)}  & \mathbf{2}_{3} \\
                X_{(a,b,c,0),(a,b,c,1)}  & \mathbf{2}_{4} \\
            \hline
                \Lambda_{(1,1,a,b),(0,0,a,b)} & \mathbf{2}_{1} \times \mathbf{2}_{2} \\
                \Lambda_{(1,a,1,b),(0,a,0,b)} & \mathbf{2}_{1} \times \mathbf{2}_{3} \\
                \Lambda_{(1,a,b,1),(0,a,b,0)} & \mathbf{2}_{1} \times \mathbf{2}_{4} \\
                \Lambda_{(a,1,1,b),(a,0,0,b)} & \mathbf{2}_{2} \times \mathbf{2}_{3} \\
                \Lambda_{(a,1,1,b),(a,0,0,b)} & \mathbf{2}_{2} \times \mathbf{2}_{2} \\
                \Lambda_{(a,b,1,1),(a,b,0,0)} & \mathbf{2}_{3} \times \mathbf{2}_{4} \\ 
            \hline
                \Lambda_{(0,0,0,a),(1,1,1,a)} & \mathbf{2}_{1}\times \mathbf{2}_{2} \times \mathbf{2}_{3} \\
                \Lambda_{(0,0,a,0),(1,1,a,1)} & \mathbf{2}_{1}\times \mathbf{2}_{2} \times \mathbf{2}_{4} \\
                \Lambda_{(0,a,0,a),(1,a,1,1)} & \mathbf{2}_{1}\times \mathbf{2}_{3} \times \mathbf{2}_{4} \\
                \Lambda_{(a,0,0,0),(a,1,1,1)} & \mathbf{2}_{2}\times \mathbf{2}_{3} \times \mathbf{2}_{4} \\
            \hline 
                \ \ \ \ \ X_{(1,1,1,1),(0,0,0,0)} \ \ \ \ \ & \ \ \ \ \ \mathbf{2}_{1}\times \mathbf{2}_{2} \times \mathbf{2}_{3} \times \mathbf{2}_{4} \ \ \ \ \ \\
            \hline
 \end{array}    
\eeq
}

Its superpotential contains the following terms:
\beq
\begin{array}{ccr}
          W_{J} & =  & \sum_{a,b}\Lambda_{(1,1,a,b),(0,0,a,b)}X_{(0,0,a,b),(1,0,a,b)}X_{(1,0,a,b),(1,1,a,b)} \\[.15cm]
                & & +  \Lambda_{(0,0,0,0),(1,1,1,0)}X_{(1,1,1,0),(1,1,1,1)}X_{(1,1,1,1),(0,0,0,0)}\\[.15cm]
                & & +  \Lambda_{(0,0,0,1),(1,1,1,1)}X_{(1,1,1,1),(0,0,0,0)}X_{(0,0,0,0)(0,0,0,1)} ~ , \\[.25cm]
          W_{H} & = & \sum_{a}\bar{\Lambda}_{(1,1,1,a)(0,0,0,a)}\bar{\Lambda}_{(0,0,0,a),(0,1,1,a)}X_{(0,1,1,a),(1,1,1,a)}\\[.15cm]
                & & +  \bar{\Lambda}_{(0,0,0,0),(1,1,0,0)}\bar{\Lambda}_{(1,1,0,0),(1,1,1,1)}X_{(1,1,1,1),(0,0,0,0)} ~ .
\end{array}
\eeq
where the global $SU(2)^{4}$ indices and their contractions have been suppressed. The rest of terms can be obtained from these by permuting the entries in the vector labels of nodes. Here we have used the $J$- and $H$-term notation for superpotential terms in the case of $m=3$ \cite{Franco:2016tcm,Franco:2017lpa}.

\subsection{$\mathbb{F}_0^{(m)}\to \mathbb{F}_0^{(m-1)}\times \mathbb{C}$ partial resolution}

\label{section_F0_partial_resolution}

\begin{figure}[ht]
	\centering
	\includegraphics[width=11cm]{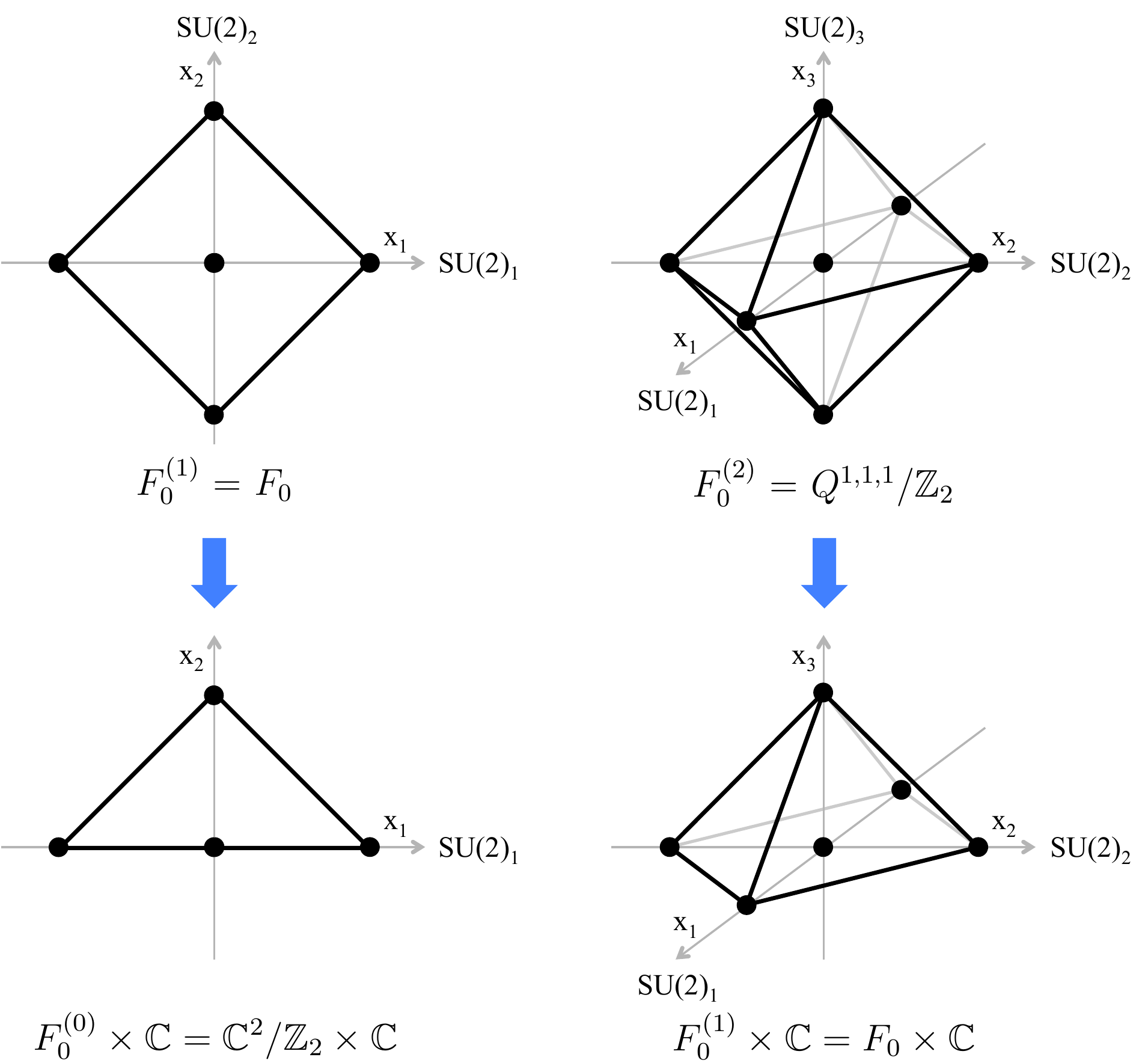}
\caption{$\mathbb{F}_0^{(m)}\to F_0^{(m-1)}\times \mathbb{C}$ partial resolution for $m=1,2$.}
	\label{F0mp1_to_F0mxC_Higgsing}
\end{figure}

The underlying geometry implies that there exists an interesting connection between consecutive members of this family of quiver theories. Removing any corner of the toric diagram for $\mathbb{F}_0^{(m)}$ results in the toric diagram for $\mathbb{F}_0^{(m-1)}\times \mathbb{C}$, namely the toric diagram for $F_0^{(m-1)}$ plus an additional point. This operation corresponds to the following partial resolution
\beq
\mathbb{F}_0^{(m)} \to F_0^{(m-1)}\times \mathbb{C} ~ . 
\eeq
\fref{F0mp1_to_F0mxC_Higgsing} illustrates this process in the cases of $F_0^{(1)}$ and $F_0^{(2)}$ as starting points. As we now explain, at the level of the quiver such a partial resolution translates into a higgsing from $\mathbb{F}_0^{(m)}$ to the dimensional reduction of the $F_0^{(m-1)}$ theory. 

It is convenient to recall the geometric origin of the $SU(2)^{m+1}$ global symmetry. The toric diagram for $\mathbb{F}_0^{(m)}$, which is given by \eref{toric_F0m_general}, is $(m+1)$-dimensional and contains $2^{m+1}$ corners. There is a pair of opposite corners for each direction $x_\mu$, $\mu=1,\ldots,m+1$, which in turn corresponds to the $SU(2)_\mu$ factor of the global symmetry.  In \fref{F0mp1_to_F0mxC_Higgsing}, we have indicated the correspondence between pairs of corners and global symmetry factors. 

Without loss of generality, let us consider removing $p_{m+1}^-$ (removing any of the other corners is equivalent by symmetry). Partial resolution maps to a higgsing of the quiver theory. Based on general considerations, it is natural to expect that deleting this corner corresponds to giving non-zero VEVs to the $2^m$ chiral fields $X^-_{(a_1,\ldots,a_m,0)(a_1,\ldots,a_m,1)}$. Below we discuss how this expectation turns out to be correct.

\paragraph{Global symmetry.} 
Since we give VEVs to fields that transform exclusively in the ${\bf 2}_{m+1}$ representation, we have the following pattern of global symmetry breaking 
\beq
SU(2)_1 \times \ldots \times SU(2)_{m} \times SU(2)_{m+1} \to SU(2)_1 \times \ldots \times SU(2)_{m} ~ ,
\eeq
namely the $SU(2)_{m+2}$ factor disappears. This is in precise agreement with the geometric expectation.

\paragraph{Quiver.} 
The $2^m$ VEVs for bifundamental chirals reduce the number of gauge groups to a half as follows. The VEV for $X^-_{(a_1,\ldots,a_m,0)(a_1,\ldots,a_m,1)}$ higgses the gauge symmetry associated to nodes $(a_1,\ldots,a_m,0)$ and $(a_1,\ldots,a_m,1)$ to the diagonal subgroup. The corresponding recombined nodes can be naturally identified by the remaining labels, i.e. by the vectors $(a_1,\ldots,a_m)$. We thus have
\beq
(a_1,\ldots,a_m,0) \times (a_1,\ldots,a_m,1) \to (a_1,\ldots,a_m) ~ .
\eeq
The change in the number of gauge groups is in agreement with the fact that the volume of the toric diagram is halved by this particular partial resolution.

Let us now study the matter content of the resulting quiver. All fields which are singlets of $SU(2)_{m+1}$ survive in the final theory. These fields, now connecting the recombined nodes, give exactly the matter content of $F_0^{(m-1)}$. 

Next, let us consider the fields that transform as doublets of $SU(2)_{m+1}$ (and maybe doublets of additional $SU(2)_\mu$ factors). First, the chiral fields $X^+_{(a_1,\ldots,a_m,0)(a_1,\ldots,a_m,1)}$, which form $SU(2)_{m+1}$ doublets with the chiral fields acquiring VEVs, survive in the final theory. Originally transforming in bifundamental representations, they turn into adjoints of the corresponding recombined nodes $(a_1,\ldots,a_m)$. We can interpret such adjoint chiral fields as the ones arising from the dimensional reduction of vector multiplets.

Finally, combining the cubic superpotential \eref{W_F0m_general} with the VEVs for the fields $X^+_{(a_1,\ldots,a_m,0)(a_1,\ldots,a_m,1)}$ gives rise to masses for all other $X^-$ fields, where the superindex refers to just the $SU(2)_{m+1}$ quantum number, so they can be integrated out. The associated $X^+$ fields remain massless and give rise to a copy of the matter content for $F_0^{(m-1)}$, but with the degrees of fields increased by 1. 

Summarizing the previous discussion, the final quiver corresponds to the dimensional reduction of $F_0^{(m-1)}$, as expected from the geometry. It is also straightforward to verify that this process generates the desired superpotential.

\paragraph{Perfect matchings.} 

From \sref{section_pms_F0}, we see that the only corner perfect matching that contains chiral fields acquiring a VEV is $p_{m+1}^-$. This implies that the proposed set of VEVs precisely remove the corner associated to $p_{m+1}^-$, while all the others remain. It is also possible to verify that some of the perfect matchings at the origin of the toric diagram are removed, while others survive. In summary, the proposed higgsing exactly produces the desired partial resolution.

\subsection{B-model computation}

The computations for this family follow the same pattern as in previous examples. We start with the resolution of these singularities as the total space of the canonical line bundle over $(\mathbb{P}^{1})^{m+1}$. It is given by:
        \begin{align}
           \t {\bf X}_{m+2} = \mbox{Tot}(\Ol(-2,-2, \cdots, -2) \to \mathbb{P}_{1}^{1}\times \mathbb{P}_{2}^{1} \times \cdots \times \mathbb{P}_{m+1}^{1}) ~ .
            \label{F_0_resolution}
        \end{align}
For $m=0$, this coincides with the resolution $\CO(-2) \rightarrow \mathbb{P}^1$ of $\mathbb{C}^{2}/\mathbb{Z}_{2}$, which we discussed in \sref{susbec: Bmod Cm+2/Zm+2}. Since for $\mathbb{P}^{1}$, $\Ol(-2) \cong \Omega$, the exceptional collection on $\mathbb{P}^{1}$ reads:
        \begin{align}
            \{\mathcal{E}_{1}\equiv \Ol(-1)[1]~,\,  \mathcal{E}_{0} \equiv \Ol \}~.
        \end{align}
        An exceptional collection on $(\mathbb{P}^{1})^{m+1}$ has $2^{m+1}$ elements, which are the line bundles:
        \begin{align}
            \big\{\mathcal{E}_{i} \equiv \mathcal{E}_{i_{1}}\otimes \mathcal{E}_{i_{2}}\otimes  \cdots \otimes \mathcal{E}_{i_{m+1}}\; \big|\; i_{\mu} \in \{0,1\}\big\}~.
        \end{align}
        Here, the index $i$ is a binary vector of length $m+1$. The sheaves in the exceptional collection on $\t {\bf X}_{m+2}$ are then of the form:
        \be
         \mathcal{F}_{i} \equiv i_{*}\mathcal{E}_{i}~,
         \ee
 with the embedding $i:(\mathbb{P}^{1})^{m+1} \to \t {\bf X}_{m+2}$. 
        
The next step is to find the Koszul resolution of these sheaves. The Koszul resolution for $m =0$ is the same as Koszul resolution for $m=0$ in \eqref{Koszul_resolution_CnZn}. For general $m$, the Koszul resolution is given by:~\footnote{The notation $\mathcal{E}(p_{1},\cdots,p_{k})$ denotes the sheaf $\mathcal{E}$ tensored with $\CO(p_1, \cdots, p_k)$.}
        \begin{align}
            \begin{diagram}
                0 & \rTo & \mathcal{E}(2,2,\cdots,2) & \rTo^{\;\;\omega} & \mathcal{E} & \rTo & i_{*}\mathcal{E} & \rTo & 0~,
            \end{diagram} 
        \end{align}
        where the map $\omega$ is an $m+1$ fold product of the map $v_{\mu}e_{\mu}^{2}$ we found earlier for $\mathbb{C}^{2}/\mathbb{Z}_{2}$---see Appendix~\ref{App: alg geom}.

\subsubsection{Quiver fields}
        
\paragraph{Basic case: $m=0$.} 
To compute the generator of $\Ext$ groups, it is useful to start from $m=0$. We call $y_{0,1}^{s}$, with $s=\pm$, the generators of $\cc^{0}(\Hom^{1}(\mathcal{F}_{1},\mathcal{F}_{0}))$. They are defined by:
            \begin{diagram}
                 \mathcal{O}(1)& \rTo & \mathcal{O}(-1) \nonumber \\
                \dTo^{z^{s}} & & \dTo^{-z^{s}} \nonumber \\
                \mathcal{O}(2) & \rTo & \mathcal{O} 
            \end{diagram}
Here,  $z^{s}$ correspond to the global sections of $\mathcal{O}(1)$ and, as explained earlier, the global sections of $\Ol(p)$ are determined by homogeneous polynomials of degree $p$ in the homogeneous coordinate. Labeling the homogeneous coordinates of $\mathbb{P}^{1}$ by $z^{\pm}$, we see that each of them gives rise to a generator $y_{0,1}^{\pm}$, which together transform in the fundamental representation of the $SU(2)$ global symmetry.

The Serre duals $\bar{y}_{1,0}^{s}$ are in $\cc^{1}(\Hom^{0}(\mathcal{F}_{1},\mathcal{F}_{0}))$. They correspond to the chain maps:
            \begin{diagram}
                &&\mathcal{O}(2) & \rTo & \mathcal{O} \nonumber \\
                &&\dTo^{\bar{z}^{s}}\nonumber \\
                \mathcal{O}(1) & \rTo & \mathcal{O}(-1)
            \end{diagram}
            Here the $\bar{z}^{s}$ are generators of $\cc^{1}(\Ol(-3))$. Locally, in the patch where $z^{+}\ne 0$, they are: 
            \begin{align}
                \bar{z}^{+} &= w_{+}^{-2}e_{+}^{3}~, \\
                \bar{z}^{-} &= -w_{+}^{-1}e_{+}^{3}~. 
            \end{align}
            $w_{+}$ is the local coordinate of this patch and, as before, $e_{+}$ is the basis of $\mathcal{O}(-1)$ in this patch.
            Composing $y_{0,1}^{t}$ and $\bar{y}_{1,0}^{s}$ results in:
            \begin{align}
                \bar{y}_{1,0}^{s}\circ y_{0,1}^{t} &= \epsilon^{st}y_{1,1}~, \nonumber \\
                y_{0,1}^{s}\circ\bar{y}_{1,0}^{t}  &=-\epsilon^{st}y_{0,0}~, \label{F_0_barz_pm}
            \end{align}
            with $y_{i,i}$ being the generators of $\E^{2}(\mathcal{F}_{i},\mathcal{F}_{i})$. They are defined by the chain map:
            \begin{diagram}
                &&\mathcal{O}(-i+2) & \rTo & \mathcal{O}(-i) \nonumber \\
                &&\dTo^{\bar{z}^{0}}\nonumber \\
                \mathcal{O}(-i+2) & \rTo & \mathcal{O}(-i)
            \end{diagram}
            where $\bar{z}^{0}$ is the sole generator of $\cc^{1}(\mathcal{O}(-2))$, given locally by:
            \begin{align}
                \bar{z}^{0} = w_{+}^{-1}e_{+}^{2} \label{F_0_barz_0}~.
            \end{align}

\paragraph{General $m$.}
It is straightforward to determine the quiver for general $m$, using the information we gained for the $m=0$ case. Given a pair of fractional branes $\mathcal{F}_{i}$ and $\mathcal{F}_{j}$, we consider the following chain maps $x^{s}_{i,j}$ 
                \begin{diagram}
                 \mathcal{O}(-j_{1}+2,\cdots,-j_{m+1}+2)& \rTo & \mathcal{O}(-j_{1},\cdots,-j_{m+1}) \nonumber \\
                \dTo^{\prod_{\mu=1}^{m+1}\xi_{\mu}^{s_{\mu}}} & & \dTo^{\prod_{\mu=1}^{m}(-1)^{j_{\mu}-i_{\mu}}\xi_{\mu}^{s_{\mu}}} \nonumber \\
                \mathcal{O}(-i_{1}+2,\cdots,-i_{m+1}+2) & \rTo & \mathcal{O}(-i_{1},\cdots,-i_{m+1}) 
                \end{diagram}
                where $\xi_{\mu}^{s_{\mu}}$ is a global section of $\mathcal{O}(j_{\mu}-i_{\mu})$. Hence, we can divide the $(\mathcal{F}_{i},\mathcal{F}_{j})$ pairs into two cases:
                \begin{enumerate}
                    \item There exists a $\mu$ such that $j_{\mu} = 0$ and $i_{\mu} = 1$. In this case, $\xi_{\mu}^{s_{\mu}}$ must be a global section of $\mathcal{O}(-1)$ over the $\mu^{th}$ $\mathbb{P}^{1}$. Since $\mathcal{O}(-1)$ has no global sections, $\E^{c}(\mathcal{F}_{j},\mathcal{F}_{i})$ is empty for all $c$. 
                    \item
                        $j_{\mu} \ge i_{\mu}$ for all $\mu$. In this case, the $\xi_{\mu}^{s_{\mu}}$ fall into two classes:
                        \begin{enumerate}
                            \item[(2.a)] 
                                If $j_{\mu} = i_{\mu}$, then $\xi_{\mu}^{s_{\mu}}$ is a local section of $\Ol$, so there is only one possibility for it i.e. $1$.
                            \item[(2.b)] 
                                If $j_{\mu} = 1$ and $i_{\mu} = 0$, then $\xi_{\mu}^{s_{\mu}}$ is a global section of $\mathcal{O}(1)$. In this case, there are two possibilities for it: $z^{\pm}_{\mu}$, {\it i.e.} the two homogeneous coordinates of $\mathbb{P}^{1}_{\mu}$. This also means that $x^{s}_{i,j}$ transforms in the fundamental representation of the $SU(2)_{\mu}$ factor of the global symmetry. 
                        \end{enumerate}
                        Combining (2.a) and (2.b), we conclude that $x^{s} \in \cc^{0}(\Hom^{k}(\mathcal{F}_{i} , \mathcal{F}_{j}))$ with $k = \sum_{\mu} (j_{\mu} - i_{\mu})$. There are $2^{k+1}$ of these generators.
                \end{enumerate}
                
                This completes our derivation of the quiver, which is in perfect agreement with the one found in \sref{section_quivers_F0m} using generalized orbifold reduction.
                                
Finally, let us compute the Serre duals $\bar{x}^{t}_{j,i}$ of these arrows. They are given by the chain maps:
                \begin{diagram}
                    &&\mathcal{O}(-i_{1}+2,\cdots,-i_{m+1}+2) & \rTo & \mathcal{O}(-i_{1},\cdots,-i_{m+1}) \nonumber \\
                    &&\dTo^{\prod_{\mu=1}^{m+1}\bar{\xi}_{\mu}^{t_{\mu}}}\nonumber \\
                    \mathcal{O}(-j_{1}+2,\cdots,-j_{m+1}+2) & \rTo & \mathcal{O}(-j_{1},\cdots,-j_{m+1}) 
                \end{diagram}
                As is occurs for $\xi_{\mu}^{s_{\mu}}$, $\bar{\xi}_{\mu}^{t_{\mu}}$ only exist for $j_{\mu} \ge i_{\mu}$ and we will need to deal with the corresponding two cases separately:
                {\renewcommand{\labelenumi}{(\alph{enumi})}
                \begin{enumerate}
                    \item
                        If $j_{\mu} = i_{\mu}$ then $\bar{\xi}_{\mu}^{t_{\mu}} \in \cc^{1}(\mathcal{O}(-2))$, so the only possibility is $\bar{z}^{0}_{\mu}$. The $\bar{z}^{0}$ is given in \eqref{F_0_barz_0} and the subscript indicates that the base is $\mathbb{P}^{1}_{\mu}$.
                    \item
                        If $j_{\mu} = 1$ and $i_{\mu} = 0$, then $\bar{\xi}_{\mu}^{t_{\mu}} \in \cc^{1}(\mathcal{O}(-3))$ and there are two possibilities, namely $\bar{\xi}_{\mu}^{\pm} = \bar{z}_{\mu}^{\pm}$. Again the subscript indicates that the base is $\mathbb{P}^{1}_{\mu}$ with $\bar{z}^{\pm}$ defined in \eqref{F_0_barz_pm}.
                \end{enumerate}
                Hence $\bar{x}^{t}_{j,i} \in \cc^{m+1}(\Hom^{1-k}(\mathcal{F}_{i},\mathcal{F}_{j}))$ and they are indeed the Serre duals of $\bar{x}^{s}_{i,j}$.
                }

\subsubsection{Superpotential}

The cubic superpotential terms follow straightforwardly from the composition. Following our definition of $x_{i,j}^{s}$ and $x_{j,k}^{t}$ and composing them results in:
            \begin{align}
                m_{2}(x_{i,j}^{s},x_{j,k}^{t}) = x_{i,k}^{s\, t} ~ .
            \end{align}
            Here the $s\,t$ in the superscript means that the fundamental $SU(2)$ indices of $x_{i,j}$ and $x_{j,k}$ are concatenated. Since the $f_{2}$'s are all trivially zero, there are no higher products. We then reproduce the superpotential \eref{W_F0m_general}.

\section{Conclusions} \label{label_section_conclusions}

It was recently shown that $m$-graded quivers with superpotentials provide a mathematical framework that elegantly unifies the description of minimally SUSY gauge theories in even dimension \cite{Franco:2017lpa}. The cases of $m=0,1,2,3$ correspond to 6d $\mathcal{N}=(0,1)$, 4d $\mathcal{N}=1$, 2d $\mathcal{N}=(0,2)$ and 0d $\mathcal{N}=1$ field theories, respectively. A rich class of such theories can be engineered in terms of Type IIB D$(5-2m)$-branes probing CY $(m+2)$-folds. One of the primary motivations for this paper was to establish the physical significance of $m$-graded quivers for $m>3$. Naively, it may seem that it is physically impossible to go beyond $m=3$, since it would require the gauge theory to live below 0d and the CY$_{m+2}$ to go beyond the critical dimension of Type IIB string theory. In this work we have shown that $m$-graded quivers describe the open string sector of the topological B-model on CY $(m+2)$-folds, for any $m$.

To illustrate this correspondence, we constructed toric quivers associated to three infinite families of toric singularities indexed by $m$.~\footnote{As usual, due to the dualities discussed in Appendix \ref{app: mutations}, the map between geometry and quivers is not one-to-one. For a given CY singularity, it is possible to start from the quiver we presented and construct all the corresponding duals, by quiver mutations.} We first derived these families using a variety of powerful tools that are available in the toric case, which include: algebraic dimensional reduction (sometimes combined with orbifolding), orbifold reduction, 3d printing and partial resolution. We independently derived all these quiver theories via B-model computations.

Our results provide the first explicit examples of $m$-graded quivers with superpotentials for CY $(m+2)$-folds with $m > 4$. Previously, only a few orbifold examples had been presented for $m=4$ \cite{Franco:2017lpa} and $m=3$ \cite{Diaconescu:2000ec,Douglas:2002fr,Franco:2016tcm,Closset:2017yte}. Quivers for more general geometries were studied only up to $m=2$, both in physics and mathematics.  

In this work, we considerably expanded the exploration of quiver theories associated to CY $(m+2)$-folds. Until now, quiver gauge theories were typically studied at fixed $m$. For each $m$ (and only for $m \leq 2$, so far), one could then consider various infinite families of geometries and construct their dual quiver gauge theories. In the toric case, this approach was significantly accelerated by the study of brane tilings ($m=1$) and brane brick models ($m=2$). In this work, we have included a new ``theory space'' direction to the problem, considering all possible CY dimensions at once. New tools for studying toric quivers, for any $m$, will be discussed in \cite{toappear1}.

Various interesting aspects of SUSY gauge theories extend to the more general context of $m$-graded quivers. For instance, we have shown that some of these theories admit periodic duality cascades. Generalizing the well-known behavior of the conifold, we presented explicit examples based on the $C(Y^{1,0}(\mathbb{P}^m))$ family, in which the number of fractional branes remains constant while the number of regular branes depends linearly on the step of the cascade. It would be interesting to investigate the significance of such formal cascades for arbitrary $m$. Interestingly, gravity duals with a running number of regular branes exist for systems of branes at CY 4-folds, namely for $m=2$~\cite{Herzog:2000rz}. It would be interesting to elucidate whether those solutions have a field theoretic interpretation in terms of cascades of trialities.

It is natural to expect that order $m+1$ dualities correspond to mutations of exceptional collections of B-branes. This expectation is supported by the known $m=1$ \cite{Herzog:2003zc,Herzog:2004qw,Aspinwall:2004vm} and $m=2$ \cite{Closset:2017yte} cases, mirror symmetry \cite{Franco:2016qxh,Franco:2016tcm} and the general discussion in \cite{Franco:2017lpa}. We plan to elaborate on this correspondence in the near future.

\acknowledgments

We would like to thank Eric Sharpe for collaboration at the initial stages of this project. SF is also grateful to Gregg Musiker for previous collaborations on related topics and to Cumrun Vafa for suggesting some of the ideas developed in this paper. The work of CC is supported by a Royal Society University Research Fellowship. The work of AH and SF was supported by the U.S. National Science Foundation grants PHY-1518967 and PHY-1820721. AH and SF would like to thank the Simons Center for Geometry and Physics for hospitality during the completion of this work. SF is also grateful to the Kavli Institute for Theoretical Physics, where part of this worked was performed. The KITP is supported in part by the National Science Foundation under Grant No. NSF PHY-1748958.

\appendix

\section{B-model computation of quivers and superpotentials}\label{App: alg geom}

In this appendix, we provide a brief review of the sheaf computation of quivers and superpotentials in the B-model. In the main body of the paper, we use the methods outlined here to derive the quivers and superpotentials for several infinite families of theories. For more details, the interested reader can consult \cite{Aspinwall:2004bs, Katz:2002gh}. For detailed reviews of B-branes, we refer to \cite{Sharpe:2003dr, Aspinwall:2004jr}.
        
 The D-branes compatible with the B-twist of Type II string theory are called B-branes. Mathematically, these branes, denoted by $\CE$, are objects of the derived category $D^b({{\bf X}_{m+2}})$ of the $(m+2)$-complex-dimensional target space ${{\bf X}_{m+2}}$. The B-model open string states with boundary conditions on the two objects $\mathcal{E}$ and $\mathcal{F}$ are counted by the $\Ext$ groups:
        \beq\label{app Ext}
        \bigoplus_{d=1}^{m+2} \mathrm{Ext}^d_{{{\bf X}_{m+2}}} (\mathcal{E},\mathcal{F})~.
        \eeq
 Each element of the group \eqref{app Ext} is interpreted as an open string state ``stretched from the brane $\CE$ to the brane $\CF$.''
        The OPE relations between open string vertex operators are encoded in the $A_\infty$ structure of the derived category. Thus, the $A_\infty$ structure controls the terms appearing in the ``spacetime'' superpotential; see \cite{Herbst:2004jp} and references therein.

\subsection{Ext groups}

The B-branes we consider are complex submanifolds of some local Calabi-Yau $\t{\bf X}_{m+2}$, a smooth resolution of the CY singularity ${\bf X}_{m+2}$. Assume $S$ is a complex submanifold of $\t{\bf X}_{m+2}$, and $\mathcal{E}_1$ and $\mathcal{E}_2$ are holomorphic vector bundles over $S$. If we denote the embedding of $S$ in $\t{\bf X}_{m+2}$ by $i$, then the objects in $D^b(\t{\bf X}_{m+2})$ corresponding to $\mathcal{E}_1$ and $\mathcal{E}_2$ are $i_{*}\mathcal{E}_1$ and $i_{*}\mathcal{E}_2$, respectively. The B-model spectrum of open strings between two D-branes on $S$, with gauge bundles $\mathcal{E}_1$ and $\mathcal{E}_2$, is given by:
        \begin{equation}
        \bigoplus_{d=0}^{m+2} \mathrm{Ext}^d_{\t{\bf X}_{m+2}} (i_*\mathcal{E}_1,i_*\mathcal{E}_2)~.
        \end{equation}
        The $\Ext$ groups above are determined by the following spectral sequence \cite{Katz:2002gh}:
        \begin{equation}\label{ss}
        E^{p,q}_2: H^p(S, \mathcal{E}_1^\vee \otimes \mathcal{E}_2 \otimes \wedge^q \mathcal{N}_S) \, \Rightarrow \, \mathrm{Ext}^{p+q}_{\t{\bf X}_{m+2}}(i_*\mathcal{E}_1,i_*\mathcal{E}_2)~,
        \end{equation}
        where $\mathcal{N}_S$ is the normal bundle of $S$ in $\t{\bf X}_{m+2}$.
        In many cases, the spectral sequence \eqref{ss} trivializes---that is:
        \beq
        \mathrm{Ext}^{d}_{\t{\bf X}_{m+2}}(i_*\mathcal{E}_1,i_*\mathcal{E}_2) \cong \bigoplus_{p+q=d} H^p(S, \mathcal{E}_1^\vee \otimes \mathcal{E}_2 \otimes \wedge^q \mathcal{N}_S)~.
        \eeq
        In such cases, we can determine the $\Ext$ groups by computing cohomology groups. If $S$ is a direct product of projective spaces, the cohomology groups can be calculated by the Borel-Weil-Bott theorem \cite{10.2307/1969996, 10.2307/1970237}, which expresses the $\Ext$ groups as representations of the global symmetry.

\subsection{$A_\infty$ structure}

The derived category $D^b(\t{\bf X}_{m+2})$ is an $A_\infty$-category. By definition, an $A_\infty$-category $\mathcal{C}$ consists of a collection of objects, Obj$(\mathcal{C})$, a $\mathbb{Z}$-graded vector space of morphisms Hom$_{\mathcal{C}}(E,F)$ for any $E,F \in \mathrm{Obj}(\mathcal{C})$ and, for every $k \geq 1$, $k$-linear maps:
        \beq
        m_k:\mathrm{Hom}_{\mathcal{C}}(E_{k-1},E_k) \otimes \cdots \otimes \mathrm{Hom}_{\mathcal{C}}(E_0,E_1) \rightarrow \mathrm{Hom}_{\mathcal{C}}(E_0,E_k)~,
        \eeq
        of degree $2-k$, satisfying the $A_\infty$ relations:
        \beq
        \sum_{p,q}(-1)^{k-p-q+pq} m_{k-p+1}(a_k,\cdots,a_{p+q+1},m_p(a_{p+q},\cdots,a_{q+1}),a_q,\cdots,a_1)=0~,
        \eeq
        for every $k>0$.
        We will follow the method proposed in \cite{Aspinwall:2004bs} to compute the composition maps $m_k$ of $D^b(\t{\bf X}_{m+2})$.

        Any object in $D^b(\t{\bf X}_{m+2})$ can be represented by a cochain complex $\mathcal{E}^\bullet$ 
        of locally-free sheaves over ${\t{\bf X}_{m+2}}$. For any pair of complexes, the $\Ext$ groups $\mathrm{Ext}^d_{\t{\bf X}_{m+2}}(\mathcal{E}^\bullet,\mathcal{F}^\bullet)$ can be viewed as the cohomology of the single complex associated with the double complex $(K^{\bullet,\bullet},d,\delta)$ with:
        \beq
        K^{p,q}(\mathcal{E}^\bullet,\mathcal{F}^\bullet) = \check{C}^p (\mathcal{U}, \mathcal{H}om^q(\mathcal{E}^\bullet,\mathcal{F}^\bullet))~,
        \eeq
        where $\check{C}^p(\mathcal{U},\cdot)$ denotes the \u{C}ech cochains of degree $p$ associated with some acyclic covering $\mathcal{U}$, and $\mathcal{H}om^q$ denotes the maps of degree $q$ between complexes, {\it i.e.}:
        \beq
        \mathcal{H}om^q(\mathcal{E}^\bullet,\mathcal{F}^\bullet) = \bigoplus_i \mathcal{H}om(\mathcal{E}^i,\mathcal{F}^{i+q}) ~ .
        \eeq
        In the double complex $(K^{\bullet,\bullet},d,\delta)$, $d$ is the differential of \u{C}ech cochains and $\delta$ is defined as follows. Let $\partial_j$ and $\partial'_k$ be differentials of $\mathcal{E}^j$ and $\mathcal{F}^k$ respectively, then for any $\sum_i \phi_{q,i} \in \mathcal{H}om^q(\mathcal{E}^\bullet,\mathcal{F}^\bullet)$ with $\phi_{q,i} \in \mathcal{H}om(\mathcal{E}^i,\mathcal{F}^{i+q})$, we have:
        \beq
        \delta_q \phi_{q,i} = \partial'_{q+i} \circ \phi_{q,i} - (-1)^q \phi_{i+1,q} \circ \partial_i~.
        \eeq
        For any $\mathcal{E}^\bullet$ and $\mathcal{F}^\bullet$, we associate to every $a \in \mathrm{Ext}^d_{\t{\bf X}_{m+2}}(\mathcal{E}^\bullet,\mathcal{F}^\bullet)$ an element $\iota(a) \in \oplus_{p+q=d}K^{p,q}(\mathcal{E}^\bullet,\mathcal{F}^\bullet)$, such that the cohomology class of $\iota(a)$ is $a$. Then, there exist maps:
        \beq
        f_k: \mathrm{Ext}^\bullet_{\t{\bf X}_{m+2}}(\mathcal{E}_{k-1}^\bullet,\mathcal{E}_k^\bullet) \otimes \cdots \otimes \mathrm{Ext}^\bullet_{\t{\bf X}_{m+2}}(\mathcal{E}_0^\bullet,\mathcal{E}_1^\bullet) \rightarrow \oplus_{p,q}K^{p,q}(\mathcal{E}_0^\bullet,\mathcal{E}_k^\bullet)~,
        \eeq
        of degree $1-k$ for any $k\geq 1$, such that:
        \beq
        f_1 = \iota~,
        \eeq
        and
        \begin{equation}\label{fk}
        \sum_{r+s+t=k} (-1)^{r+st} f_{n+1-s}(\mathrm{id}^{\otimes r} \otimes m_s \otimes \mathrm{id}^{\otimes t}) = \sum_{2 \leq r \leq n \atop i_1+\cdots +i_r = k} (-1)^w f_{i_1} \circ f_{i_2} \circ \cdots \circ f_{i_r} + d f_k~,
        \end{equation}
        where $w=(r-1)(i_1-1)+(r-2)(i_2-1)+\cdots +(i_{r-1}-1)$ and $\circ$ denotes the composition of maps in $\oplus_{p,q} K^{p,q}(\bullet,\bullet)$.
           For example, we have:
        \begin{equation}\label{f2}
        \iota m_2 = \iota \circ \iota + d f_2~,
        \end{equation}
        and
        \begin{equation}\label{f3}
        \iota m_3 = f_2(\mathrm{id} \otimes m_2) - f_2(m_2 \otimes \mathrm{id}) + (\iota \circ f_2) - (f_2 \circ \iota) + d f_3~.
        \end{equation}
To compute the $A_\infty$ structure, the first step is to find representatives for a basis of the $\Ext$ groups, which in turn defines $\iota$. Then, we can employ \eqref{fk} to compute the composition maps $m_k$. Specifically, we can use \eqref{f2} to determine $m_2$ and $f_2$, then use \eqref{f3} to determine $m_3$ and $f_3$ and so forth.

        In the theories we consider, the B-branes of interest are of the form:
        \be 
        i_* \mathcal{E}~,
        \ee
         with $i$ the embedding of a complex submanifold $S$ in $\t{\bf X}_{m+2}$, and $\mathcal{E}$ a holomorphic vector bundle over $S$. Suppose that $\mathcal{E}^\bullet_l$ is the Koszul resolution of $i_* \mathcal{E}_l$:
        \beq
        \cdots \rightarrow \mathcal{E}^{-i}_l \rightarrow \mathcal{E}^{-i+1}_l \rightarrow \cdots \rightarrow \mathcal{E}^{0}_l \rightarrow i_* \mathcal{E}_l \rightarrow 0 ~ .
        \eeq
        Then, $\mathrm{Ext}^d_{\t{\bf X}_{m+2}} (i_*\mathcal{E}_1,i_*\mathcal{E}_2)$ is the same as $\mathrm{Ext}^d_{\t{\bf X}_{m+2}} (\mathcal{E}_1^\bullet,\mathcal{E}_2^\bullet)$, so that we can use the method discussed above to compute the composition maps $m_k$.

\subsection{Superpotential}

Given a graded quiver with nodes corresponding to coherent sheaves $i_* \mathcal{E}_j, j=1,\cdots,n$, where $n$ is the number of nodes, we can read off the superpotential from the composition maps $m_k$. To that end, we fix a basis $\phi^{(d)\mu}_{j_2,j_1}$ for each $\mathrm{Ext}^d_{\t{\bf X}_{m+2}}(i_* \mathcal{E}_{j_1},i_* \mathcal{E}_{j_2})$. Following the convention described in the main text, we will label the corresponding quiver field by $(\Phi_{j_{1},j_{2}}^{(k-1)})_{\mu}$. Note that the $\Ext$ generator and the field have conjugate indices and differ in degree by $1$. The label $\mu$ runs over the generators. For the examples we are considering, it coincides with the flavor symmetry index.  For each $j$, $\mathrm{Ext}^{m+2}_{\t{\bf X}_{m+2}}(i_* \mathcal{E}_{j},i_* \mathcal{E}_{j})$ is 1-dimensional. If $\overline{\phi}^{(n - k)\bar{\mu}}_{j_1,j_2}$ is the generator corresponding to the Serre dual of $\phi^{(k)\mu}_{j_2,j_1}$, then:
        \beq
        m_2(\overline{\phi}^{(n - m)\bar{\mu}}_{j_1,j_2},\phi^{(m)\mu}_{j_2,j_1}) = \phi^{(n)}_{j_1,j_1}~,
        \eeq
        for any $j_1$ and $\mu$. By choosing a basis, we fix the normalization of the trace map $\gamma_j:\mathrm{Ext}^{m+2}_{\t{\bf X}_{m+2}}(i_* \mathcal{E}_{j},i_* \mathcal{E}_{j}) \rightarrow \mathbb{C}$ defined by:
        \beq
        \gamma_j(\phi^{n}_{j,j}) = 1~.
        \eeq
        For any generator $\phi^{(l)}_{j,i} \in \mathrm{Ext}^l_{\t{\bf X}_{m+2}}(i_* \mathcal{E}_{i},i_* \mathcal{E}_{j})$, we consider all the paths connecting nodes $i$ and $j$ in the quiver. If there exist fields $\Phi^{(n_1)}_{s_1,i}, \Phi^{(n_2)}_{s_2,s_1},\cdots,\Phi^{(n_k)}_{j,s_{k-1}}$ along some path with $k$ arrows such that:
        \begin{equation}\label{c+}
        \gamma_i(m_2(\overline{\phi}^{(n - l)}_{i,j}, m_k(\phi^{(n_k)}_{j,s_{k-1}},\cdots,\phi^{(n_1)}_{s_1,i})))
        \end{equation}
        is nonzero, then there is a term proportional to:
                \beq
     \Phi^{(n_1-1)}_{i,s_1}\cdots\Phi^{(n_k-1)}_{s_{k-1},j}\overline{\Phi}^{(n - m-1)}_{j,i}~,
        \eeq
        with the coefficient equal to \eqref{c+} in the superpotential. Similarly, if there exist fields $\Phi^{(n_1)}_{s_1,j}, \Phi^{(n_2)}_{s_2,s_1},\cdots,\Phi^{(n_k)}_{i,s_{k-1}}$ along some path in the opposite direction such that:
        \begin{equation}\label{c-}
        \gamma_j(m_2(\phi^{(l)}_{j,i}, m_k(\phi^{(n_k)}_{i,s_{k-1}},\cdots,\phi^{(n_1)}_{s_1,j})))
        \end{equation}
        is nonzero, then there is a term proportional to
        \begin{equation}
\Phi^{(n_1-1)}_{j,s_1,} \cdots \Phi^{(n_k-1)}_{s_{k-1},i} \Phi^{(l-1)}_{i,j}~,
\label{potential_term_quiver}
        \end{equation}
        with the coefficient equal to \eqref{c-} in the superpotential. Every term in the superpotential can be computed this way, thus the $A_\infty$ structure of the derived category completely determines the superpotential.

        Note that, since $\gamma$ is only non-zero on $\mathrm{Ext}^{m+2}$ generators, and since $m_{k}$ has degree $2-k$, the ``superpotential coupling'' \eqref{c-} is non-zero only if:
        \begin{align}
            l + \sum_{j=1}^{k}n_{j} = m + k~.
        \end{align}
        This is simply the ghost-number selection rule for disk correlators in the B-model.
It directly follows that the only terms that can appear in the superpotential have quiver degree:
\be
{\rm deg}\big(\Phi^{(n_1-1)}_{j,s_1,} \cdots \Phi^{(n_k-1)}_{s_{k-1},i} \Phi^{(l-1)}_{i,j} \big)= l-1 + \sum_{j=1}^k (n_j-1)= m-1~.
\ee
 Hence, the degree constraint for the superpotential of an $m$-graded quiver is automatically satisfied.

\subsection{Sheaves on $\Pn{n}$: a primer}
    
In order to derive the quivers and superpotentials for the geometries considered in this paper using the technology we have just discussed, it is useful to review some notions about sheaves on $\Pn{n}$. In the rest of this section, we present several elementary results about \v{C}ech cohomology with sections taking values in such sheaves. 

        Let us start with the presentation of $\Pn{n}$ in the homogeneous coordinates. Starting from $\mathbb{C}^{n+1}$, we obtain $\mathbb{P}^{n}$ by identifying:
        \begin{align}
            (z_{0},\cdots,z_{n}) \sim \lambda (z_{0},\cdots,z_{n})~.
        \end{align}
        From this presentation, we can pass on to standard charts on $\Pn{n}$. There are $n+1$ of these charts, denoted by $U_{\mu}$. $U_{\mu}$ covers the complex lines for which $z_{\mu} \ne 0$. We will denote the $i^{th}$ local coordinate on $U_{\mu}$ by $w_{\mu,i}$, with $1\le i \le n$. The explicit map between the two presentations is:
        \begin{align}
            w_{\mu,i} = \left\{\begin{array}{ccc}
                                    \frac{z_{0}}{z_{\mu}} & & i = \mu~, \nonumber \\
                                    \frac{z_{i}}{z_{\mu}} & & i \ne \mu~.
                               \end{array}
                              \right. \label{transition_functions}
        \end{align}

\subsubsection{Sheaves $\Ol(p)$}
        
The tautological line bundle, denoted by $\Ol(-1)$, is the sheaf on $\Pn{n}$ which assigns to each point in it the line it represents in $\mathbb{C}^{n+1}$. We denote the basis of this sheaf on the chart $U_{\mu}$ by $e_{\mu}$. The transition functions between different charts are then represented by the equation:
            \begin{align}
                e_{i} = w_{0,i}^{-1}e_{0}~.
            \end{align}
            
            The sheaf $\Ol(-p)$, for $p > 0$, is the sheaf which locally has as its basis the $p^{th}$ tensor power $e_{\mu}^{p}$ of $e_{\mu}$.
   The sheaf $\Ol(p)$, for $p>0$, is defined to be the dual sheaf of $\Ol(-p)$. In particular let $e^{*}_{\mu}$ be the basis of $\Ol(1)$ in the chart $\mu$ then the transition functions for it are determined by:
            \begin{align}
                e^{*}_{i} = w_{0,i}e_{0}^{*} ~ .     
            \end{align}
            $(e^{*}_{\mu})^{p}$ form a basis of $\Ol(p)$ in $U_{\mu}$. Finally, $\Ol(0)$, which is often denoted as $\Ol$, is the trivial sheaf.

\subsubsection{Tangent and cotangent bundles} 
        
One-forms $\dd w_{0,i}$ form a basis of the cotangent bundle in the $\mu^{th}$ chart. The transition matrix can be found using \eqref{transition_functions}. We will not reproduce all of them here, but will mention an identity that will be useful for our calculations, namely:
            \begin{align}
                w_{0,i}^{-i}\dd w_{0,i} = -w_{i,i}^{-1}\dd w_{i,i} ~ . \label{diagonal_transformation}
            \end{align}

            $\om{p}$ is the $p^{th}$ antisymmetric tensor power of $\Omega$. The transition functions again follow straightforwardly, albeit tediously, from \eqref{transition_functions}. The situation is simplest for the highest non-trivial power, {\it i.e.} $\om{n}$, also called the determinant bundle. Its basis is $\wedge_{i}\dd w_{\mu,i}$ and the transition function is the determinant of the transition matrix for $\Omega$:
            \begin{align}
                \wedge_{j} \dd w_{0,j} = w_{i,i}^{-n-1}\wedge_{j}\dd w_{i,j}~.
                \label{det_transition}
            \end{align}
            The tangent bundle $\Omega^{*}$ is the dual of the cotangent bundle. In the local coordinates of the chart $U_{\mu}$, its basis is given by the vector fields $\pdv{}{w_{\mu,i}}$. Locally, the action of vector fields on the differential form is given by contraction or interior derivation.

\subsubsection{\v{C}ech cohomology}
        
Next, we turn to the computation of some sheaf-valued \v{C}ech cohomology groups on $\Pn{n}$. We will also organize them into representations of $SU(n+1)$, with its action on $\Pn{n}$ induced from $\C^{n+1}$. The most basic of these are $\cc^{0}$, which correspond to the global section of the said sheaves.

$\Ol(-p)$ has no global sections for $p > 0$. The same is true for $\om{p}$. However their dual bundles do have global sections. For $\Ol(p)$ with $p \ge 0$, a local section is determined by a homogeneous polynomial of degree $p$ in the homogeneous coordinates $z_{\mu}$. These obviously transform in the symmetric $(p,0)$-index tensor representation of $SU(n+1)$, which has dimension $\binom{n+1+p}{p}$. 

            The tangent bundle $\Omega^{*}$ has $(n+1)^{2}-1$ global sections. In the homogeneous coordinates, these are given by:
            \begin{align}
                 z_{\mu}\pdv{}{z_{\nu}}~,
             \end{align} 
            with the linear relation $\sum_{\mu}z_{\mu}\pdv{}{z_{\mu}} = 0$.
They transform in the adjoint representation of $SU(n+1)$.

             More relevant for us will be the sheaf $\Omega^{*}(-1)$.~\footnote{For any sheaf $F$ we define $F(p)$ to be $F$ tensored with $\Ol(p)$.} 
            It has $(n+1)$ of global sections transforming in the $(0,1)$-index representation of $SU(n+1)$. Locally in $U_{0}$, they can be written as:
             \begin{align}
                 \varphi^{0} &= -\sum_{i}w_{0,i}\pdv{}{w_{0,i}}\otimes e_{0}~, \nonumber \nonumber \\
                 \varphi^{i} &= \pdv{}{w_{0,i}}\otimes e_{0}~. \label{global_sections}
             \end{align}
             The maps between two sheaves $E$ and $F$ form a sheaf denoted by $\Hom(E,F)$. The sections \eref{global_sections} can also be regarded as the global sections of $\Hom(\Omega,\Ol(-1))$. More generally, they can be regarded as global sections of $\Hom(\Omega^{i+1}(j+1),\Omega^{i}(j))$.

             We can also easily compute the global sections of $\Hom(\Omega^{i+k}(j+k),\Omega^{i}(j))$. These are given by antisymmetric compositions of $\lambda^{i}$ defined above and they transform in the antisymmetric $k$-index~\footnote{More formally $(0,k)$, but throughout the paper all the representations we mention are of this form and we will just write $k$ to simplify the notation.} representation of $SU(n+1)$. Concretely, a basis of them is given by:
             \begin{align}
                \varphi^{\mu_{1}\cdots \mu_{k}} = \frac{1}{k!}\varphi^{[\mu_{1}}\circ \varphi^{\mu_{2}} \circ  \cdots \circ \varphi^{\mu_{k}]}~. \label{section_composition}   
            \end{align} 
            The square brackets represent the antisymmetrization of the indices they enclose.

        \subsubsection{Serre duality}

Serre duality is one of the most important properties of these sheaf-valued cohomology groups. In the present case, it is the statement that there is an isomorphism between $\cc^{i}(E)$ and $\cc^{n-i}(E^{*}(-n-1))^{*}$. 

Let us see how this plays out in the case of $\Hom(\Omega^{n}(n+j),\mathcal{O}(j))$, which we computed in the last section. Its dual sheaf is $\Hom(\Ol(j),\Omega^{n}(n+j)) \cong \Omega(n)$. So, to exhibit Serre duality we need to find $\cc^{n}(\Omega^{n}(-1))$.

  An element of $\Omega^{n}(-1)$ is a top form with coefficients in $\Ol(-1)$. It being in the $n^{th}$ \v{C}ech cohomology means that it is holomorphic in $\cap_{\mu}U_{\mu}$, {\it i.e.} intersection of all $n+1$ charts, but not holomorphic in any intersection of $n$ charts. Let us consider the ansatz that a member $\bar{\varphi}$ of this cohomology group is given in the coordinates of $U_{0}$ by:
            \begin{align}
                \bar{\varphi} = \wedge_{i}w^{p_{i}}_{0,i}\dd w_{0,i} \otimes e_{0}~.
            \end{align}
            Using \eqref{transition_functions} and \eqref{det_transition}, we see that, in the local coordinates of patch $U_{i}$, we can write $\bar{\varphi}$ as:
            \begin{align}
                \bar{\varphi} = (-1)^{i}w_{i,i}^{-n-2-\sum_{j}p_{j}}\dd w_{i,i}\wedge_{j\ne i}^{m}w^{p_{j}}_{i,j}\dd w_{i,j} \otimes e_{i}~.
            \end{align}
            The holomorphy constraint described above means that:
            \begin{align}
                p_{i} < 0 \qquad \mbox{and} \qquad -\sum_{i}p_{i} < n+2~.
            \end{align}
            Hence, there are $n+1$ choices of $\bar{\varphi}$:
            \begin{align}
                \bar{\varphi}^{0} &= \wedge_{j}w^{-1}_{0,j}\dd w_{0,j} \otimes e_{0}~, \nonumber \nonumber \\
                \bar{\varphi}^{i} &= w_{0,i}^{-1}\wedge_{j}w^{-1}_{0,j}\dd w_{0,j} \otimes e_{0}~. \label{serre_duals}
            \end{align}
   The dimension $n+1$ is indeed the one we would have expected from Serre duality. Note that $\bar{\varphi}$ transforms in the $1$-index representation of $SU(n+1)$ which is conjugate to the representation in which elements of $\Hom(\Omega^{n}(n+j),\mathcal{O}(j))$, {\it i.e.} $\varphi^{i_{1}\cdots i_{n}}$, transform.

\section{Graded quiver mutations}\label{app: mutations}

Graded quivers with superpotentials enjoy order $m+1$ mutations, which reproduce the dualities of the corresponding gauge theories for $m\leq 3$ and generalize them for $m>3$.
In this appendix, we summarize the effect of a mutation on a node, which we identify as $\star$ \cite{Franco:2017lpa}.

\begin{figure}[H]
	\centering
	\includegraphics[width=12.5cm]{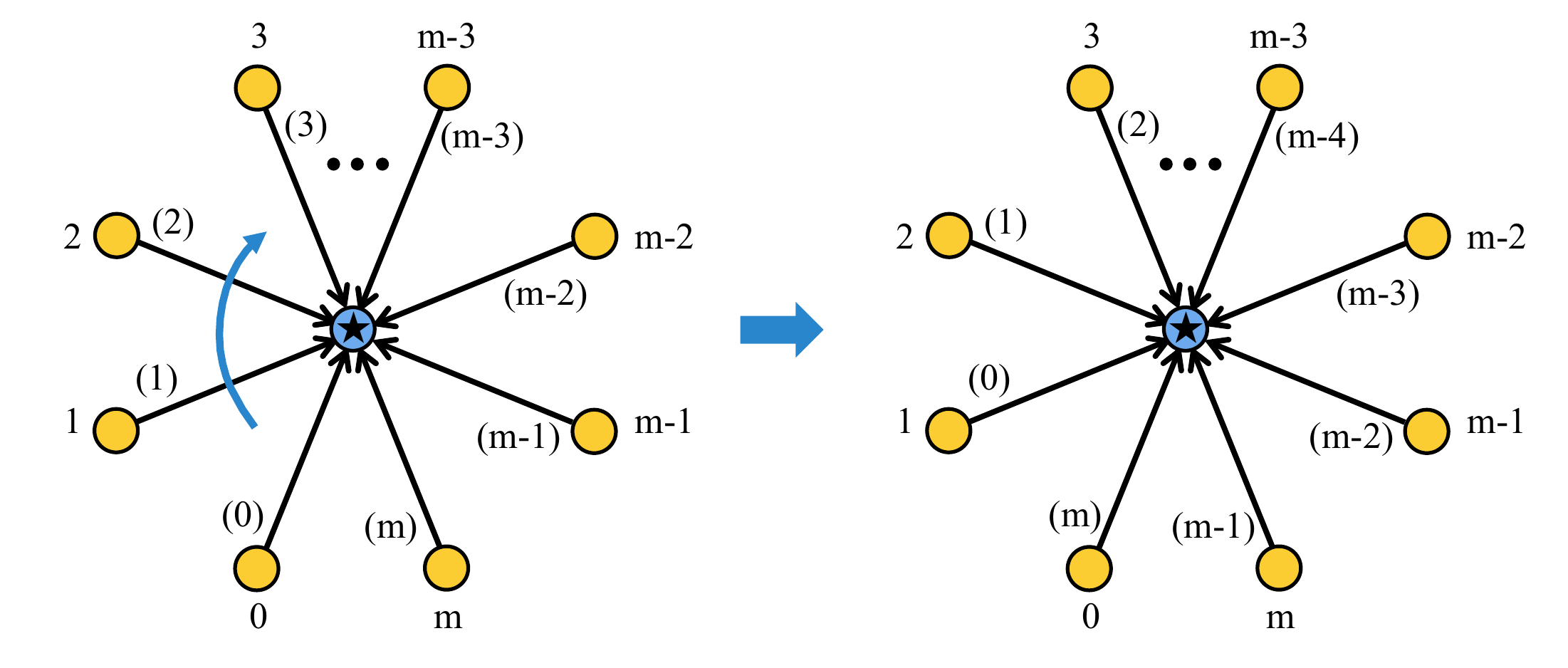}
\caption{The transformation of flavors upon a mutation on node $\star$ can be implemented as a rotation of the degrees of the arrows.}
	\label{mutation_flavors}
\end{figure}

\paragraph{1. Flavors.}
As it is standard, we refer to the arrows connected to the mutated node as {\it flavors}. It is possible to take all flavors as incoming into the mutated note, simply by trading any arrow that is oriented outward for its conjugate. Once this is done, there is a natural cyclic order for flavors around the node, in which the degree of incoming arrows increases clockwise, as shown on the left of \fref{mutation_flavors}. There can be multiple or no arrows of a given degree.

Under the mutation, the flavors transform as follows:
\paragraph{2. Rotation of the degrees.} {Replace every incoming arrow} \xymatrix{ i \ar@{->}[r]^{(c)} & \star} with the arrow 
\xymatrix{ i \ar@{->}[r]^{(c-1)} & \star}.
In terms of the cyclic ordering of flavors previously introduced, this transformation is elegantly implemented as a clockwise rotation of the degrees of the flavors while keeping the spectator nodes fixed, as shown in \fref{mutation_flavors}.

\paragraph{2. Mesons.} 
The second step in the transformation of the quiver involves the addition of composite arrows, to which we refer as {\it mesons}. For every $2$-path \xymatrix{ i \ar@{->}[r]^{(0)} & \star \ar@{->}[r]^{(c)} & j} in $\overline{Q}$, where $c\neq m$, {add a new arrow} \xymatrix{ i  \ar@/^0.5pc/[rr]^{(c)} & \star & j}. In other words, we generate all possible mesons involving incoming chiral fields. Sometimes, we might chose to represent the field to be composed with a chiral field as an arrow that goes into the mutated node. The orientation of both arrows, both incoming, naively seems incompatible for composition. The general rule above is equivalent to saying that, in such cases, we use the conjugate of the incoming chiral field for the composition. This phenomenon, dubbed {\it anticomposition}, was first discussed in the physics literature in the context of quadrality of 0d $\mathcal{N}=1$ theories \cite{Franco:2016tcm}. 
\begin{figure}[H]
	\centering
	\includegraphics[width=12cm]{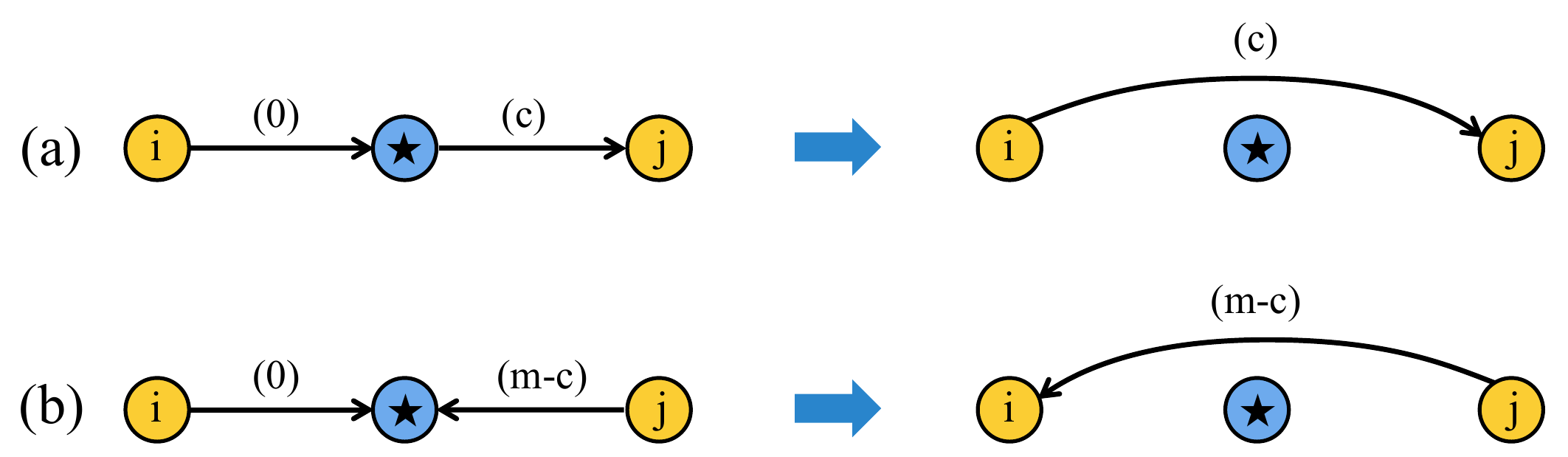}
\caption{a) Composition of arrows into a meson. b) The same process interpreted as anticomposition.}
	\label{composition_anticomposition}
\end{figure}

\paragraph{3. Superpotential.} 
Under mutation, the superpotential transforms according to the following rules:

\begin{itemize}

\item[{\bf 3.a)}] {\bf Cubic dual flavors-meson couplings.} For every $2$-path, \xymatrix{ i \ar@{->}[r]^{(0)} & \star \ar@{->}[r]^{(c)} & j} in $\overline{Q}$, with $c\neq m$, add the new arrow \xymatrix{ i \ar@{->}[r]^{(c)} & j} in $\overline{Q}$ and the new cubic term $\Phi_{ij}^{(c)}\Phi_{\star j}^{(c+1)}\Phi_{i\star}^{(m)} =
\Phi_{ij}^{(c)}\Phi_{j\star}^{(m-c-1)}\Phi_{\star i}^{(0)}$ to $W$. \fref{mutation_cubic_couplings} shows the general form of these terms, which are in one-to-one correspondence with the mesons.

\begin{figure}[H]
	\centering
	\includegraphics[width=11cm]{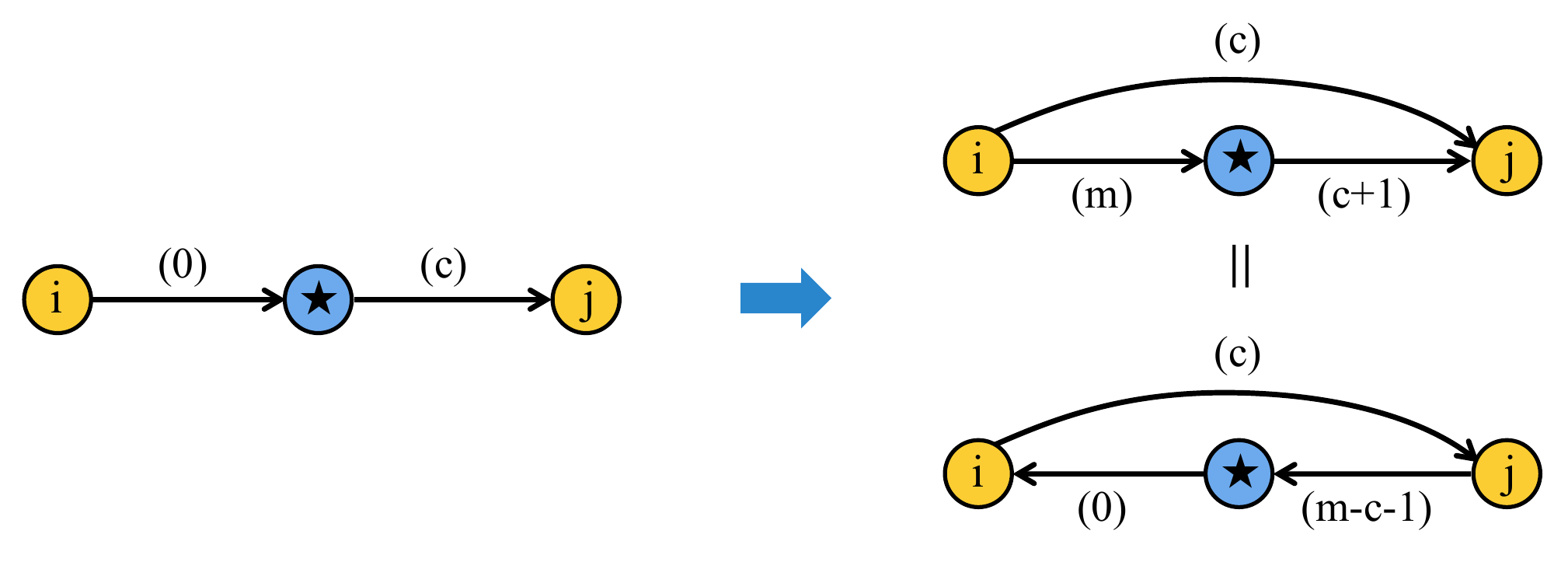}
\caption{New cubic terms coupling mesons to dual flavors.}
	\label{mutation_cubic_couplings}
\end{figure}

\end{itemize}

The remaining rules concern pre-existing terms in the superpotential. First of all, terms that do not go through the mutated noted are not modified. The transformation of terms that contain the mutated node depends on the degrees of the arrows that are connected to it in the corresponding cycle.

\newpage

\begin{itemize}

\item[{\bf 3.b)}] Replace instances of $\Phi_{i\star}^{(0)}\Phi_{\star j}^{(c)}$ in $W$ with the meson $\Phi_{ij}^{(c)}$ that results from composing the two arrows.

\begin{figure}[H]
	\centering
	\includegraphics[height=3.8cm]{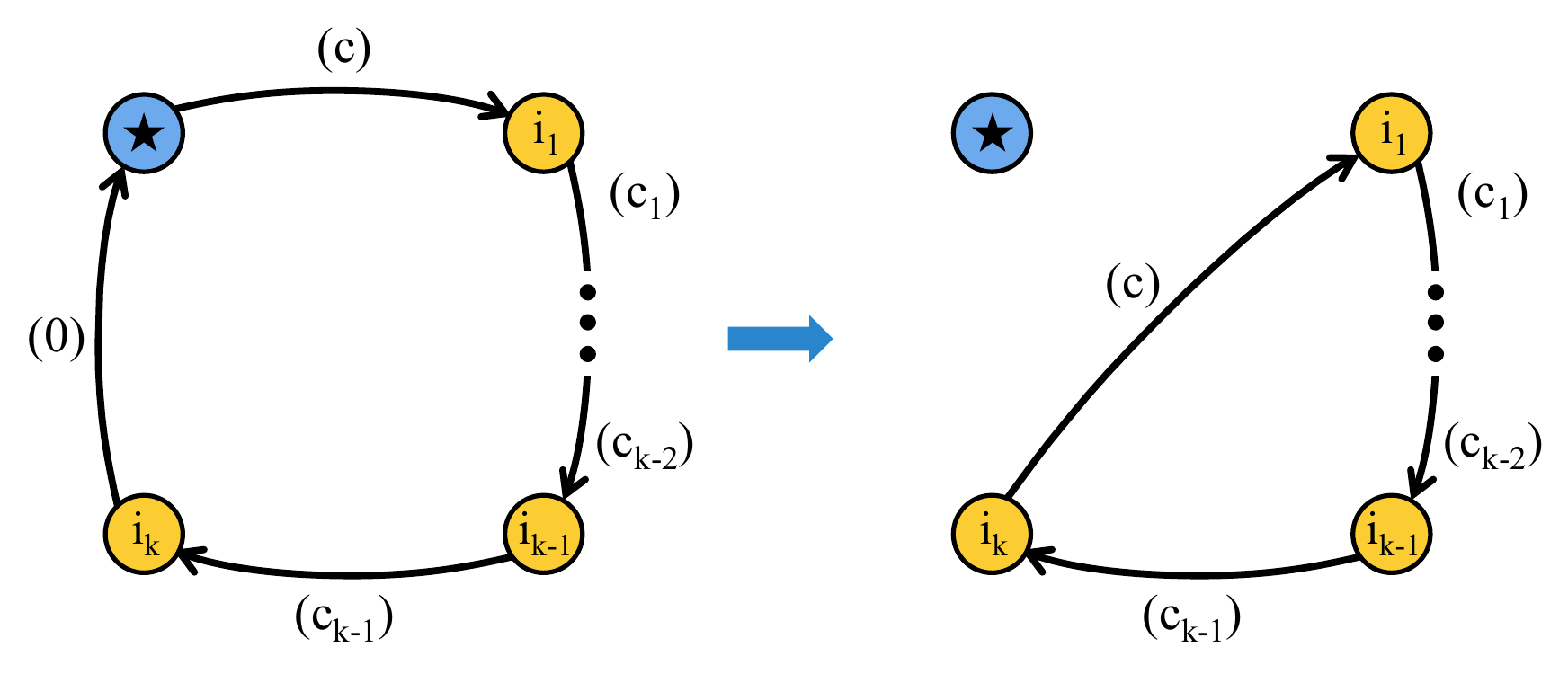}
\caption{Mutation of a superpotential term with a 2-path giving rise to a meson.}
	\label{W_1-mesons}
\end{figure}

\item[{\bf 3.c)}] Replace instances of $\Phi_{i\star}^{(c)}\Phi_{\star j}^{(d)}$ in $W$, where $c\neq 0$ and $d$ is arbitrary with the product $\Phi_{i \star}^{(c-1)}\Phi_{\star j}^{(d+1)}$---that is, we write each closed path in $W$ in terms of the new arrows. 

\begin{figure}[H]
	\centering
	\includegraphics[height=3.8cm]{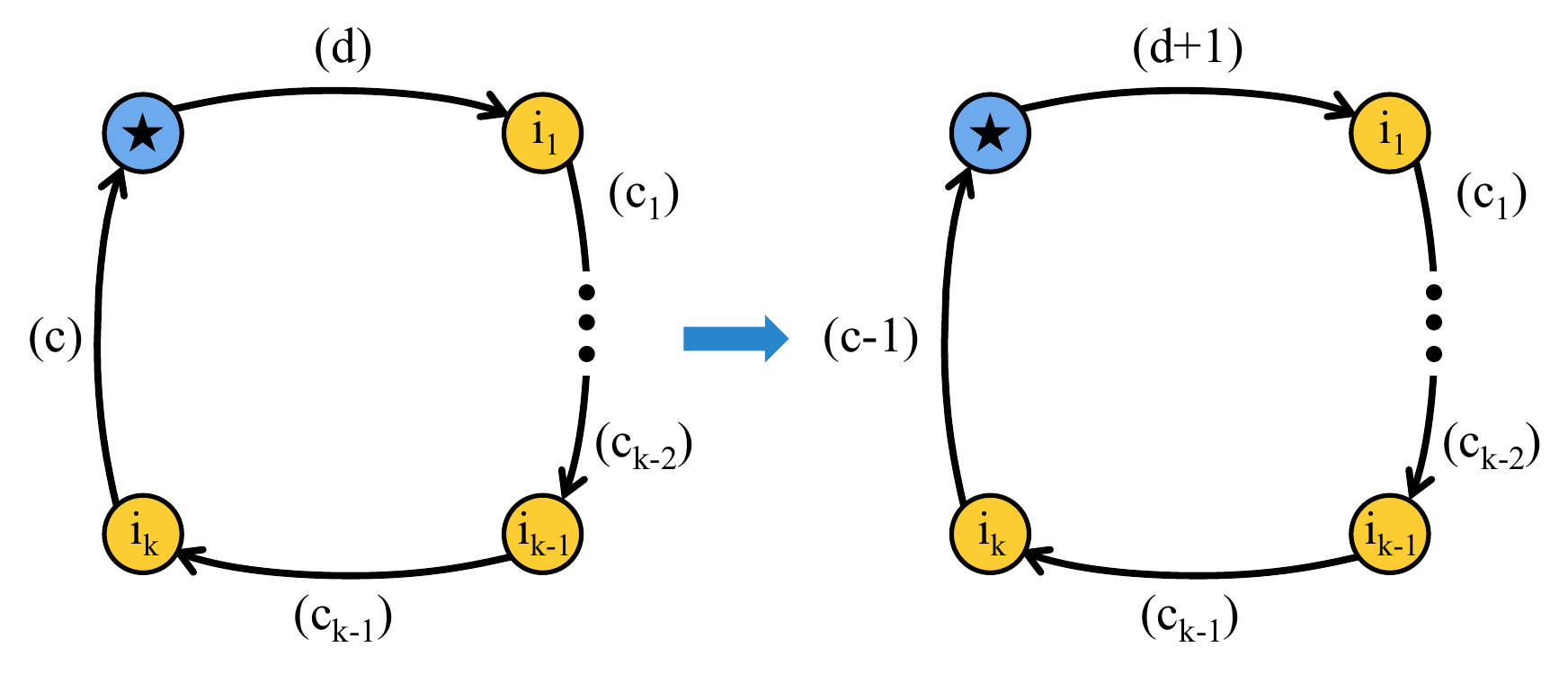}
\caption{Mutation of a superpotential term with a 2-path that goes through the mutated node but does not generate a meson.}
	\label{W_0-mesons}
\end{figure}

\item[{\bf 3.d)}] Additionally, if there is an incoming chiral arrow $\Phi_{i_0 \star}^{(0)}$ at the mutated node, an additional term in $W$ is generated by duplicating this cycle, replacing instances of $\Phi_{i\star}^{(c)}\Phi_{\star j}^{(d)}$ with the product of mesons $\Phi_{i i_0}^{(c)}\Phi_{i_0 j}^{(d)}$, which follow from (anti)composing $\Phi_{i\star}^{(c)}$ and $\Phi_{\star j}^{(d)}$ with $\Phi_{i_0 \star}^{(0)}$.

\begin{figure}[H]
	\centering
	\includegraphics[height=3.9cm]{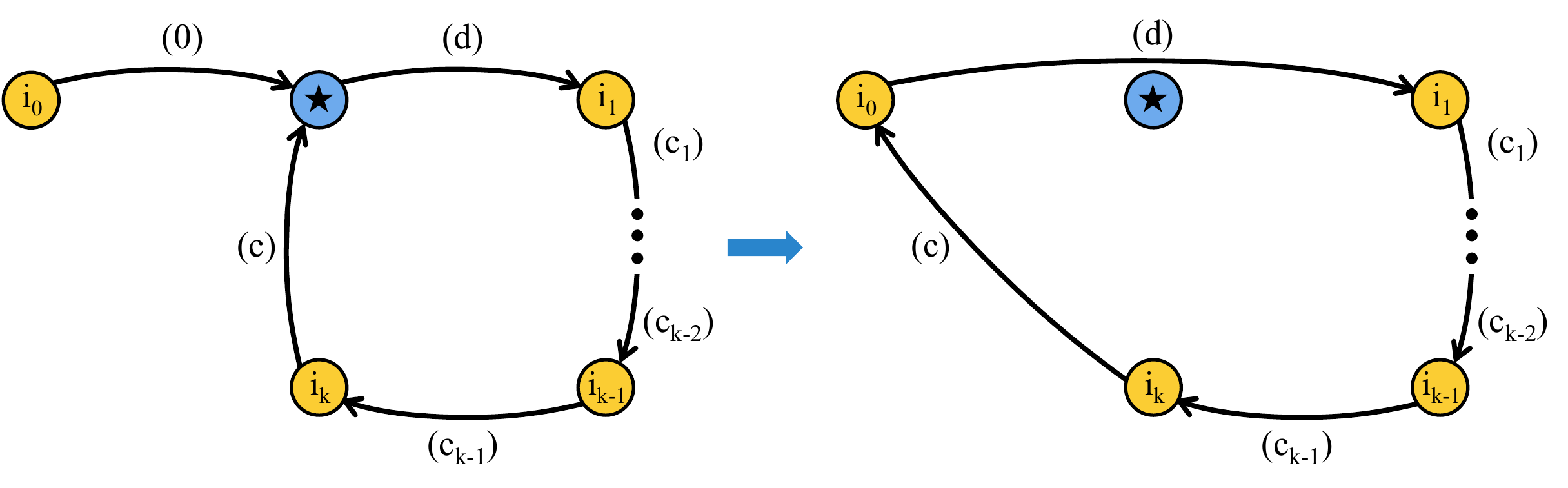}
\caption{Mutation of a superpotential term in the presence of an additional chiral field going intothe mutated node.}
	\label{W_2-mesons}
\end{figure}

\item[{\bf 3.e)}] Finally, we can ``integrate out'' massive arrows, which corresponds to removing all the 2-cycles that appear in the superpotential while imposing the ``equations of motion'' for the corresponding arrows \cite{Franco:2017lpa}.

\end{itemize}

\noindent Note that rules {\bf 3.c)} and {\bf 3.d)} become relevant for $m\geq 2$.

\paragraph{4. Ranks.} 
Finally, one can study how quiver representations transform under mutations. Let us assign the ranks $N_i$ to the quiver nodes. Then, the rank $N_\star$ of the mutated node transforms as:
\beq
N_\star'=N_0-N_\star~,
\label{mutation_ranks}
\eeq
where $N_0$ indicates the total number of incoming chiral fields.
Periodicity of the rank after $(m+1)$ consecutive mutations on the same node requires that, for every node:
\be
\begin{array}{rrl}
{\rm if}\; m\in 2\Z+1\;:\; \;\;& 0 = N_m - N_{m-1}+ \ldots - N_1 + N_0~, \\[.15 cm]
{\rm if}\;  m\in 2\Z\; :\; \;\;& 2 N_\star = N_m - N_{m-1}+ \ldots - N_1 + N_0~.
\end{array}
\label{anomalies_even_odd_m}
\eeq
This coincides with the generalized anomaly cancellation conditions discussed in \sref{subsec: anomaly free}.

\bibliographystyle{JHEP}
\bibliography{mybib}

\end{document}

%% file: pref.tex
\newcommand{\be}{\begin{equation}}
\newcommand{\ee}{\end{equation}}
\newcommand{\beq}{\begin{equation}}
\newcommand{\beql}[1]{\begin{equation}\label{#1}}
\newcommand{\eeq}{\end{equation}}
\newcommand{\ba}{\begin{array}}
\newcommand{\ea}{\end{array}}
\newcommand{\bea}{\begin{eqnarray}}
\newcommand{\beal}[1]{\begin{eqnarray}\label{#1}}
\newcommand{\eea}{\end{eqnarray}}
\newcommand{\ben}{\begin{enumerate}}
\newcommand{\een}{\end{enumerate}}
\newcommand{\bean}{\begin{eqnarray*}}
\newcommand{\eean}{\end{eqnarray*}}
\newcommand{\eref}[1]{(\ref{#1})}
\newcommand{\sref}[1]{\S\ref{#1}}

\newcommand{\fref}[1]{Figure \ref{#1}}
\newcommand{\btab}[1]{\begin{tabular}{#1}}
\newcommand{\etab}{\end{tabular}}

\def \Black {\pdfsetcolor{0 0 0 1}}

\newcommand{\comment}[1]{}
\newcommand{\CF}{{\cal F}}

\newcommand{\CN}{{\cal N}}

\newcommand{\qed}{\nobreak \ifvmode \relax \else
      \ifdim\lastskip<1.5em \hskip-\lastskip
      \hskip1.5em plus0em minus0.5em \fi \nobreak
      \vrule height0.75em width0.5em depth0.25em\fi}

%% file: paper.bbl
\providecommand{\href}[2]{#2}\begingroup\raggedright\begin{thebibliography}{10}

\bibitem{Douglas:1996sw}
M.~R. Douglas and G.~W. Moore, \emph{{D-branes, Quivers, and ALE Instantons}},
  \href{http://arxiv.org/abs/hep-th/9603167}{{\tt hep-th/9603167}}.

\bibitem{Morrison:1998cs}
D.~R. Morrison and M.~R. Plesser, \emph{{Nonspherical horizons. 1.}},
  {\emph{Adv.Theor.Math.Phys.} {\bf 3} (1999) 1--81},
  [\href{http://arxiv.org/abs/hep-th/9810201}{{\tt hep-th/9810201}}].

\bibitem{Beasley:1999uz}
C.~Beasley, B.~R. Greene, C.~Lazaroiu and M.~Plesser, \emph{{D3-branes on
  partial resolutions of Abelian quotient singularities of Calabi-Yau
  threefolds}},
  \href{http://dx.doi.org/10.1016/S0550-3213(99)00646-X}{\emph{Nucl.Phys.} {\bf
  B566} (2000) 599--640}, [\href{http://arxiv.org/abs/hep-th/9907186}{{\tt
  hep-th/9907186}}].

\bibitem{Feng:2000mi}
B.~Feng, A.~Hanany and Y.-H. He, \emph{{D-brane gauge theories from toric
  singularities and toric duality}},
  \href{http://dx.doi.org/10.1016/S0550-3213(00)00699-4}{\emph{Nucl. Phys.}
  {\bf B595} (2001) 165--200}, [\href{http://arxiv.org/abs/hep-th/0003085}{{\tt
  hep-th/0003085}}].

\bibitem{Beasley:2001zp}
C.~E. Beasley and M.~R. Plesser, \emph{{Toric duality is Seiberg duality}},
  \href{http://dx.doi.org/10.1088/1126-6708/2001/12/001}{\emph{JHEP} {\bf 0112}
  (2001) 001}, [\href{http://arxiv.org/abs/hep-th/0109053}{{\tt
  hep-th/0109053}}].

\bibitem{Feng:2001xr}
B.~Feng, A.~Hanany and Y.-H. He, \emph{{Phase structure of D-brane gauge
  theories and toric duality}},
  \href{http://dx.doi.org/10.1088/1126-6708/2001/08/040}{\emph{JHEP} {\bf 08}
  (2001) 040}, [\href{http://arxiv.org/abs/hep-th/0104259}{{\tt
  hep-th/0104259}}].

\bibitem{Feng:2001bn}
B.~Feng, A.~Hanany, Y.-H. He and A.~M. Uranga, \emph{{Toric duality as Seiberg
  duality and brane diamonds}},
  \href{http://dx.doi.org/10.1088/1126-6708/2001/12/035}{\emph{JHEP} {\bf 12}
  (2001) 035}, [\href{http://arxiv.org/abs/hep-th/0109063}{{\tt
  hep-th/0109063}}].

\bibitem{Feng:2002zw}
B.~Feng, S.~Franco, A.~Hanany and Y.-H. He, \emph{{Symmetries of toric
  duality}}, \href{http://dx.doi.org/10.1088/1126-6708/2002/12/076}{\emph{JHEP}
  {\bf 12} (2002) 076}, [\href{http://arxiv.org/abs/hep-th/0205144}{{\tt
  hep-th/0205144}}].

\bibitem{Wijnholt:2002qz}
M.~Wijnholt, \emph{{Large volume perspective on branes at singularities}},
  \href{http://dx.doi.org/10.4310/ATMP.2003.v7.n6.a6}{\emph{Adv. Theor. Math.
  Phys.} {\bf 7} (2003) 1117--1153},
  [\href{http://arxiv.org/abs/hep-th/0212021}{{\tt hep-th/0212021}}].

\bibitem{Benvenuti:2004dy}
S.~Benvenuti, S.~Franco, A.~Hanany, D.~Martelli and J.~Sparks, \emph{{An
  infinite family of superconformal quiver gauge theories with Sasaki-Einstein
  duals}}, \href{http://dx.doi.org/10.1088/1126-6708/2005/06/064}{\emph{JHEP}
  {\bf 06} (2005) 064}, [\href{http://arxiv.org/abs/hep-th/0411264}{{\tt
  hep-th/0411264}}].

\bibitem{Franco:2005rj}
S.~Franco, A.~Hanany, K.~D. Kennaway, D.~Vegh and B.~Wecht, \emph{{Brane Dimers
  and Quiver Gauge Theories}},
  \href{http://dx.doi.org/10.1088/1126-6708/2006/01/096}{\emph{JHEP} {\bf 01}
  (2006) 096}, [\href{http://arxiv.org/abs/hep-th/0504110}{{\tt
  hep-th/0504110}}].

\bibitem{Benvenuti:2005ja}
S.~Benvenuti and M.~Kruczenski, \emph{{From Sasaki-Einstein spaces to quivers
  via BPS geodesics: L**p,q|r}},
  \href{http://dx.doi.org/10.1088/1126-6708/2006/04/033}{\emph{JHEP} {\bf 04}
  (2006) 033}, [\href{http://arxiv.org/abs/hep-th/0505206}{{\tt
  hep-th/0505206}}].

\bibitem{Franco:2005sm}
S.~Franco et~al., \emph{{Gauge theories from toric geometry and brane
  tilings}}, \href{http://dx.doi.org/10.1088/1126-6708/2006/01/128}{\emph{JHEP}
  {\bf 01} (2006) 128}, [\href{http://arxiv.org/abs/hep-th/0505211}{{\tt
  hep-th/0505211}}].

\bibitem{Butti:2005sw}
A.~Butti, D.~Forcella and A.~Zaffaroni, \emph{{The Dual superconformal theory
  for L**pqr manifolds}},
  \href{http://dx.doi.org/10.1088/1126-6708/2005/09/018}{\emph{JHEP} {\bf 09}
  (2005) 018}, [\href{http://arxiv.org/abs/hep-th/0505220}{{\tt
  hep-th/0505220}}].

\bibitem{Franco:2017lpa}
S.~Franco and G.~Musiker, \emph{{Higher Cluster Categories and QFT Dualities}},
  \href{http://dx.doi.org/10.1103/PhysRevD.98.046021}{\emph{Phys. Rev.} {\bf
  D98} (2018) 046021}, [\href{http://arxiv.org/abs/1711.01270}{{\tt
  1711.01270}}].

\bibitem{Closset:2017yte}
C.~Closset, J.~Guo and E.~Sharpe, \emph{{B-branes and supersymmetric quivers in
  2d}}, \href{http://dx.doi.org/10.1007/JHEP02(2018)051}{\emph{JHEP} {\bf 02}
  (2018) 051}, [\href{http://arxiv.org/abs/1711.10195}{{\tt 1711.10195}}].

\bibitem{Closset:2017xsc}
C.~Closset, D.~Ghim and R.-K. Seong, \emph{{Supersymmetric gauged matrix models
  from dimensional reduction on a sphere}},
  \href{http://dx.doi.org/10.1007/JHEP05(2018)026}{\emph{JHEP} {\bf 05} (2018)
  026}, [\href{http://arxiv.org/abs/1712.10023}{{\tt 1712.10023}}].

\bibitem{Eager:2018oww}
R.~Eager and I.~Saberi, \emph{{Holomorphic field theories and Calabi--Yau
  algebras}},  \href{http://arxiv.org/abs/1805.02084}{{\tt 1805.02084}}.

\bibitem{GarciaCompean:1998kh}
H.~Garcia-Compean and A.~M. Uranga, \emph{{Brane box realization of chiral
  gauge theories in two-dimensions}},
  \href{http://dx.doi.org/10.1016/S0550-3213(98)00725-1}{\emph{Nucl.Phys.} {\bf
  B539} (1999) 329--366}, [\href{http://arxiv.org/abs/hep-th/9806177}{{\tt
  hep-th/9806177}}].

\bibitem{Franco:2015tna}
S.~Franco, D.~Ghim, S.~Lee, R.-K. Seong and D.~Yokoyama, \emph{{2d (0,2) Quiver
  Gauge Theories and D-Branes}},
  \href{http://dx.doi.org/10.1007/JHEP09(2015)072}{\emph{JHEP} {\bf 09} (2015)
  072}, [\href{http://arxiv.org/abs/1506.03818}{{\tt 1506.03818}}].

\bibitem{Franco:2015tya}
S.~Franco, S.~Lee and R.-K. Seong, \emph{{Brane Brick Models, Toric Calabi-Yau
  4-Folds and 2d (0,2) Quivers}},
  \href{http://dx.doi.org/10.1007/JHEP02(2016)047}{\emph{JHEP} {\bf 02} (2016)
  047}, [\href{http://arxiv.org/abs/1510.01744}{{\tt 1510.01744}}].

\bibitem{Franco:2016nwv}
S.~Franco, S.~Lee and R.-K. Seong, \emph{{Brane brick models and 2d (0, 2)
  triality}}, \href{http://dx.doi.org/10.1007/JHEP05(2016)020}{\emph{JHEP} {\bf
  05} (2016) 020}, [\href{http://arxiv.org/abs/1602.01834}{{\tt 1602.01834}}].

\bibitem{Franco:2016fxm}
S.~Franco, S.~Lee and R.-K. Seong, \emph{{Orbifold Reduction and 2d (0,2) Gauge
  Theories}}, \href{http://dx.doi.org/10.1007/JHEP03(2017)016}{\emph{JHEP} {\bf
  03} (2017) 016}, [\href{http://arxiv.org/abs/1609.07144}{{\tt 1609.07144}}].

\bibitem{Franco:2016tcm}
S.~Franco, S.~Lee, R.-K. Seong and C.~Vafa, \emph{{Quadrality for
  Supersymmetric Matrix Models}},
  \href{http://dx.doi.org/10.1007/JHEP07(2017)053}{\emph{JHEP} {\bf 07} (2017)
  053}, [\href{http://arxiv.org/abs/1612.06859}{{\tt 1612.06859}}].

\bibitem{Franco:2018qsc}
S.~Franco and A.~Hasan, \emph{{$3d$ Printing of $2d$ $\mathcal{N}=(0,2)$ Gauge
  Theories}}, \href{http://dx.doi.org/10.1007/JHEP05(2018)082}{\emph{JHEP} {\bf
  05} (2018) 082}, [\href{http://arxiv.org/abs/1801.00799}{{\tt 1801.00799}}].

\bibitem{Seiberg:1994bp}
N.~Seiberg, \emph{{The Power of holomorphy: Exact results in 4-D SUSY field
  theories}},  in \emph{{PASCOS '94: Proceedings, 4th International Symposium
  on Particles, Strings and Cosmology, Syracuse, New York, USA, May 19-24,
  1994}}, pp.~0357--369, 1994.
\newblock \href{http://arxiv.org/abs/hep-th/9408013}{{\tt hep-th/9408013}}.

\bibitem{Sharpe:1999qz}
E.~R. Sharpe, \emph{{D-branes, derived categories, and Grothendieck groups}},
  \href{http://dx.doi.org/10.1016/S0550-3213(99)00535-0}{\emph{Nucl. Phys.}
  {\bf B561} (1999) 433--450}, [\href{http://arxiv.org/abs/hep-th/9902116}{{\tt
  hep-th/9902116}}].

\bibitem{Douglas:2000gi}
M.~R. Douglas, \emph{{D-branes, categories and N=1 supersymmetry}},
  \href{http://dx.doi.org/10.1063/1.1374448}{\emph{J. Math. Phys.} {\bf 42}
  (2001) 2818--2843}, [\href{http://arxiv.org/abs/hep-th/0011017}{{\tt
  hep-th/0011017}}].

\bibitem{Katz:2002gh}
S.~H. Katz and E.~Sharpe, \emph{{D-branes, open string vertex operators, and
  Ext groups}}, \href{http://dx.doi.org/10.4310/ATMP.2002.v6.n6.a1}{\emph{Adv.
  Theor. Math. Phys.} {\bf 6} (2003) 979--1030},
  [\href{http://arxiv.org/abs/hep-th/0208104}{{\tt hep-th/0208104}}].

\bibitem{Katz:2002jh}
S.~H. Katz, T.~Pantev and E.~Sharpe, \emph{{D branes, orbifolds, and Ext
  groups}},
  \href{http://dx.doi.org/10.1016/j.nuclphysb.2003.09.022}{\emph{Nucl. Phys.}
  {\bf B673} (2003) 263--300}, [\href{http://arxiv.org/abs/hep-th/0212218}{{\tt
  hep-th/0212218}}].

\bibitem{1999math......8027B}
T.~{Bridgeland}, A.~{King} and M.~{Reid}, \emph{{Mukai implies McKay: the McKay
  correspondence as an equivalence of derived categories}},
  \href{http://arxiv.org/abs/math/9908027}{{\tt math/9908027}}.

\bibitem{Bridgeland:2005fr}
T.~Bridgeland, \emph{{T-structures on some local Calabi-Yau varieties}},
  \href{http://arxiv.org/abs/math/0502050}{{\tt math/0502050}}.

\bibitem{lam2014calabi}
Y.~T. Lam, \emph{Calabi-yau categories and quivers with superpotential}, .

\bibitem{Franco:2016qxh}
S.~Franco, S.~Lee, R.-K. Seong and C.~Vafa, \emph{{Brane Brick Models in the
  Mirror}}, \href{http://dx.doi.org/10.1007/JHEP02(2017)106}{\emph{JHEP} {\bf
  02} (2017) 106}, [\href{http://arxiv.org/abs/1609.01723}{{\tt 1609.01723}}].

\bibitem{Brunner:1997gf}
I.~Brunner and A.~Karch, \emph{{Branes at orbifolds versus Hanany Witten in
  six-dimensions}},
  \href{http://dx.doi.org/10.1088/1126-6708/1998/03/003}{\emph{JHEP} {\bf 03}
  (1998) 003}, [\href{http://arxiv.org/abs/hep-th/9712143}{{\tt
  hep-th/9712143}}].

\bibitem{toappear1}
S.~Franco and A.~Hasan, \emph{Graded Quivers, Generalized Dimer Models and
  Toric Geometry, Work in progress}.

\bibitem{Feng:2005gw}
B.~Feng, Y.-H. He, K.~D. Kennaway and C.~Vafa, \emph{{Dimer models from mirror
  symmetry and quivering amoebae}},
  \href{http://dx.doi.org/10.4310/ATMP.2008.v12.n3.a2}{\emph{Adv. Theor. Math.
  Phys.} {\bf 12} (2008) 489--545},
  [\href{http://arxiv.org/abs/hep-th/0511287}{{\tt hep-th/0511287}}].

\bibitem{Futaki:2014mpa}
M.~Futaki and K.~Ueda, \emph{{Tropical Coamoeba and Torus-Equivariant
  Homological Mirror Symmetry for the Projective Space}},
  \href{http://dx.doi.org/10.1007/s00220-014-2155-1}{\emph{Commun. Math. Phys.}
  {\bf 332} (2014) 53--87}.

\bibitem{MR3590528}
S.~Oppermann, \emph{Quivers for silting mutation}, {\emph{Adv. Math.} {\bf 307}
  (2017) 684--714}.

\bibitem{Buan08}
A.~{Bakke Buan} and H.~{Thomas}, \emph{{Coloured quiver mutation for higher
  cluster categories}}, {\emph{ArXiv e-prints} (Sept., 2008) },
  [\href{http://arxiv.org/abs/0809.0691}{{\tt 0809.0691}}].

\bibitem{Ladkani}
S.~Ladkani, \emph{{Finite-dimensional algebras are $(m>2)$-Calabi-Yau Tilted}},
  {\emph{ArXiv e-prints} (2016) }, [\href{http://arxiv.org/abs/1603.09709}{{\tt
  1603.09709}}].

\bibitem{Witten:1993yc}
E.~Witten, \emph{{Phases of N = 2 theories in two dimensions}},
  \href{http://dx.doi.org/10.1016/0550-3213(93)90033-L}{\emph{Nucl. Phys.} {\bf
  B403} (1993) 159--222}, [\href{http://arxiv.org/abs/hep-th/9301042}{{\tt
  hep-th/9301042}}].

\bibitem{Cachazo:2001sg}
F.~Cachazo, B.~Fiol, K.~A. Intriligator, S.~Katz and C.~Vafa, \emph{{A
  Geometric unification of dualities}},
  \href{http://dx.doi.org/10.1016/S0550-3213(02)00078-0}{\emph{Nucl. Phys.}
  {\bf B628} (2002) 3--78}, [\href{http://arxiv.org/abs/hep-th/0110028}{{\tt
  hep-th/0110028}}].

\bibitem{Herzog:2003zc}
C.~P. Herzog, \emph{{Exceptional collections and del Pezzo gauge theories}},
  \href{http://dx.doi.org/10.1088/1126-6708/2004/04/069}{\emph{JHEP} {\bf 04}
  (2004) 069}, [\href{http://arxiv.org/abs/hep-th/0310262}{{\tt
  hep-th/0310262}}].

\bibitem{Aspinwall:2004vm}
P.~S. Aspinwall and I.~V. Melnikov, \emph{{D-branes on vanishing del Pezzo
  surfaces}},
  \href{http://dx.doi.org/10.1088/1126-6708/2004/12/042}{\emph{JHEP} {\bf 12}
  (2004) 042}, [\href{http://arxiv.org/abs/hep-th/0405134}{{\tt
  hep-th/0405134}}].

\bibitem{Hanany:2006nm}
A.~Hanany, C.~P. Herzog and D.~Vegh, \emph{{Brane tilings and exceptional
  collections}},
  \href{http://dx.doi.org/10.1088/1126-6708/2006/07/001}{\emph{JHEP} {\bf 07}
  (2006) 001}, [\href{http://arxiv.org/abs/hep-th/0602041}{{\tt
  hep-th/0602041}}].

\bibitem{Herzog:2006bu}
C.~P. Herzog and R.~L. Karp, \emph{{On the geometry of quiver gauge theories
  (Stacking exceptional collections)}},
  \href{http://dx.doi.org/10.4310/ATMP.2009.v13.n3.a1}{\emph{Adv. Theor. Math.
  Phys.} {\bf 13} (2009) 599--636},
  [\href{http://arxiv.org/abs/hep-th/0605177}{{\tt hep-th/0605177}}].

\bibitem{Herbst:2004jp}
M.~Herbst, C.-I. Lazaroiu and W.~Lerche, \emph{{Superpotentials, A(infinity)
  relations and WDVV equations for open topological strings}},
  \href{http://dx.doi.org/10.1088/1126-6708/2005/02/071}{\emph{JHEP} {\bf 02}
  (2005) 071}, [\href{http://arxiv.org/abs/hep-th/0402110}{{\tt
  hep-th/0402110}}].

\bibitem{Aspinwall:2004bs}
P.~S. Aspinwall and S.~H. Katz, \emph{{Computation of superpotentials for
  D-branes}}, \href{http://dx.doi.org/10.1007/s00220-006-1527-6}{\emph{Commun.
  Math. Phys.} {\bf 264} (2006) 227--253},
  [\href{http://arxiv.org/abs/hep-th/0412209}{{\tt hep-th/0412209}}].

\bibitem{kad}
T.~V. Kadeishvili, \emph{The algebraic structure in the homology of an
  {$A_{\infty}$} algebra}, {\emph{Sobshch. Akad. Nauk. Gruzin. SSR} {\bf 108}
  (1982) 249--252}.

\bibitem{Aspinwall:2004jr}
P.~S. Aspinwall, \emph{{D-branes on Calabi-Yau manifolds}},  in \emph{{Progress
  in string theory. Proceedings, Summer School, TASI 2003, Boulder, USA, June
  2-27, 2003}}, pp.~1--152, 2004.
\newblock \href{http://arxiv.org/abs/hep-th/0403166}{{\tt hep-th/0403166}}.
\newblock \href{http://dx.doi.org/10.1142/9789812775108_0001}{DOI}.

\bibitem{Klebanov:1998hh}
I.~R. Klebanov and E.~Witten, \emph{{Superconformal field theory on
  three-branes at a Calabi-Yau singularity}},
  \href{http://dx.doi.org/10.1016/S0550-3213(98)00654-3}{\emph{Nucl.Phys.} {\bf
  B536} (1998) 199--218}, [\href{http://arxiv.org/abs/hep-th/9807080}{{\tt
  hep-th/9807080}}].

\bibitem{Gauntlett:2004hh}
J.~P. Gauntlett, D.~Martelli, J.~F. Sparks and D.~Waldram, \emph{{A New
  infinite class of Sasaki-Einstein manifolds}},
  \href{http://dx.doi.org/10.4310/ATMP.2004.v8.n6.a3}{\emph{Adv. Theor. Math.
  Phys.} {\bf 8} (2004) 987--1000},
  [\href{http://arxiv.org/abs/hep-th/0403038}{{\tt hep-th/0403038}}].

\bibitem{toappear2}
S.~Franco and A.~Hasan, \emph{Work in progress}.

\bibitem{Franco:2017cjj}
S.~Franco, D.~Ghim, S.~Lee and R.-K. Seong, \emph{{Elliptic Genera of 2d (0,2)
  Gauge Theories from Brane Brick Models}},
  \href{http://dx.doi.org/10.1007/JHEP06(2017)068}{\emph{JHEP} {\bf 06} (2017)
  068}, [\href{http://arxiv.org/abs/1702.02948}{{\tt 1702.02948}}].

\bibitem{Klebanov:2000hb}
I.~R. Klebanov and M.~J. Strassler, \emph{{Supergravity and a confining gauge
  theory: Duality cascades and chi SB resolution of naked singularities}},
  \href{http://dx.doi.org/10.1088/1126-6708/2000/08/052}{\emph{JHEP} {\bf 08}
  (2000) 052}, [\href{http://arxiv.org/abs/hep-th/0007191}{{\tt
  hep-th/0007191}}].

\bibitem{Diaconescu:2000ec}
D.-E. Diaconescu and M.~R. Douglas, \emph{{D-branes on stringy Calabi-Yau
  manifolds}},  \href{http://arxiv.org/abs/hep-th/0006224}{{\tt
  hep-th/0006224}}.

\bibitem{Douglas:2002fr}
M.~R. Douglas, S.~Govindarajan, T.~Jayaraman and A.~Tomasiello, \emph{{D branes
  on Calabi-Yau manifolds and superpotentials}},
  \href{http://dx.doi.org/10.1007/s00220-004-1091-x}{\emph{Commun. Math. Phys.}
  {\bf 248} (2004) 85--118}, [\href{http://arxiv.org/abs/hep-th/0203173}{{\tt
  hep-th/0203173}}].

\bibitem{Herzog:2000rz}
C.~P. Herzog and I.~R. Klebanov, \emph{{Gravity duals of fractional branes in
  various dimensions}},
  \href{http://dx.doi.org/10.1103/PhysRevD.63.126005}{\emph{Phys. Rev.} {\bf
  D63} (2001) 126005}, [\href{http://arxiv.org/abs/hep-th/0101020}{{\tt
  hep-th/0101020}}].

\bibitem{Herzog:2004qw}
C.~P. Herzog, \emph{{Seiberg duality is an exceptional mutation}},
  \href{http://dx.doi.org/10.1088/1126-6708/2004/08/064}{\emph{JHEP} {\bf 08}
  (2004) 064}, [\href{http://arxiv.org/abs/hep-th/0405118}{{\tt
  hep-th/0405118}}].

\bibitem{Sharpe:2003dr}
E.~Sharpe, \emph{{Lectures on D-branes and sheaves}},  2003.
\newblock \href{http://arxiv.org/abs/hep-th/0307245}{{\tt hep-th/0307245}}.

\bibitem{10.2307/1969996}
R.~Bott, \emph{Homogeneous vector bundles}, {\emph{Annals of Mathematics} {\bf
  66} (1957) 203--248}.

\bibitem{10.2307/1970237}
B.~Kostant, \emph{Lie algebra cohomology and the generalized {Borel-Weil}
  theorem}, {\emph{Annals of Mathematics} {\bf 74} (1961) 329--387}.

\end{thebibliography}\endgroup
